\documentclass[12pt,preprint,longabstract]{aastex}

\usepackage{rotating}

\def\ltsima{$\; \buildrel < \over \sim \;$}
\def\simlt{\lower.5ex\hbox{\ltsima}}
\def\gtsima{$\; \buildrel > \over \sim \;$}
\def\simgt{\lower.5ex\hbox{\gtsima}}
\def\feh{\mathrm{[Fe/H]}}
\def\MH{\mathrm{[M/H]}}
\def\mh{\mathrm{[m/H]}}
\def\alp{\mathrm{[\alpha/Fe]}}
\def\logg{\log g}
\def\vrot{{V_{\rm rot}}}
\def\teff{{T_{\rm eff}}}

\def\snr{\hbox{S/N}}
\def\kms{\,\rm{km\,s}^{-1}}
\def\masyr{\mbox{mas yr$^{-1}$\,}}

\shorttitle{RAVE DR3}
\shortauthors{Siebert et al.}

\begin{document}


\title{The RAdial Velocity Experiment (RAVE): Third Data Release}
\author{
A. Siebert\altaffilmark{1},
M.~E.~K. Williams\altaffilmark{2},
A. Siviero\altaffilmark{2,3},
W. Reid\altaffilmark{4},
C. Boeche\altaffilmark{2},
M. Steinmetz\altaffilmark{2},
J. Fulbright\altaffilmark{5},
U. Munari\altaffilmark{3},
T. Zwitter\altaffilmark{6,7},
F.~G. Watson\altaffilmark{8},
R.~F.~G. Wyse\altaffilmark{5},
R.~S. de Jong\altaffilmark{2},
H. Enke\altaffilmark{2},
B. Anguiano\altaffilmark{2},
D. Burton\altaffilmark{8,9},
C.~J.~P. Cass\altaffilmark{8},
K. Fiegert\altaffilmark{8},
M. Hartley\altaffilmark{8},
A. Ritter\altaffilmark{4},
K.~S. Russel\altaffilmark{8},
M. Stupar\altaffilmark{8},
O. Bienaym\'e\altaffilmark{1},
K.~C. Freeman\altaffilmark{9},
G. Gilmore\altaffilmark{10},
E.~K. Grebel\altaffilmark{11},
A. Helmi\altaffilmark{12},
J.~F. Navarro\altaffilmark{13},
J. Binney\altaffilmark{14},
J. Bland-Hawthorn\altaffilmark{15},
R. Campbell\altaffilmark{16},
B. Famaey\altaffilmark{1},
O. Gerhard\altaffilmark{17},
B.~K. Gibson\altaffilmark{18},
G. Matijevi\v c\altaffilmark{6},
Q.~A. Parker\altaffilmark{4,8},
G.~M. Seabroke\altaffilmark{19},
S. Sharma\altaffilmark{15},
M.~C. Smith\altaffilmark{20,21},
E. Wylie-de Boer\altaffilmark{9}
}

\altaffiltext{1}{Observatoire astronomique de Strasbourg, Universit\'e
 de Strasbourg, CNRS, UMR 7550, 11 rue de l'universit\'e, 67000, Strasbourg, France}
\altaffiltext{2}{Leibniz-Institut f\"ur Astrophysik Potsdam (AIP), An der
 Sternwarte 16, D-14482, Potsdam, Germany}
\altaffiltext{3}{INAF Osservatorio Astronomico di Padova, 36012 Asiago
 (VI), Italy}
\altaffiltext{4}{Department of Physics and Astronomy, Faculty of
 Sciences, Macquarie University, NSW 2109, Sydney, Australia}
\altaffiltext{5}{Johns Hopkins University, Departement of Physics and
 Astronomy, 366 Bloomberg center, 3400 N. Charles St.,  Baltimore, MD
 21218, USA}
\altaffiltext{6}{University of Ljubljana, Faculty of Mathematics and
 Physics, Jadranska 19, 1000 Ljubljana, Slovenia}
\altaffiltext{7}{Center of excellence SPACE-SI, A\v sker\v ceva cesta
 12, 1000 Ljubljana, Slovenia}
\altaffiltext{8}{Australian Astronomical Observatory, P.O. box 296,
 Epping, NSW 1710, Australia}
\altaffiltext{9}{Research School of Astronomy and Astrophysics,
 Australian National University, Cotter Rd., ACT, Canberra, Australia}
\altaffiltext{10}{Institute of Astronomy, University of Cambridge,
 Madingley Road, Cambridge, CB3 OHA, UK}
\altaffiltext{11}{Astronomisches Rechen-Institut, Zentrum f\"ur Astronomie
 der Universit\"at Heidelberg, M\"onchhofstr. 12-14, D-69120, Heidelberg, Germany}
\altaffiltext{12}{Kapteyn Astronomical Institut, University of
 Groningen, Landleven 12, 9747 AD, Groningen, The Netherlands}
\altaffiltext{13}{Department of Physics and Astronomy, University of
 Victoria, P.O. box 3055, Victoria,BC V8W 3P6, Canada}
\altaffiltext{14}{Rudolf Peierls Center for Theoretical Physics,
 University of Oxford, 1 Keeble Road, Oxford, OX1 3NP, UK}
\altaffiltext{15}{Sydney Institute for Astronomy, School of Physics A28,
University of Sydney, NSW 2006, Australia}
\altaffiltext{16}{Western Kentucky University, Bowling Green, Kentucky, USA}
\altaffiltext{17}{Max-Planck-Institut f\"ur extraterrestrische
 Physick, Giessenbachstrasse, D-85748, Garching, Germany}
\altaffiltext{18}{Jeremiah Horrocks Institute, University of Central
 Lancashire,  Preston, PR1 2HE, UK}
\altaffiltext{19}{Mullard Space Science Laboratory, University College
 London, Holmbury St Mary, Dorking, RH5 6NT, UK}
\altaffiltext{20}{Kavli Institute for Astronomy and
 Astrophysics, Peking University, Beijing, China}
\altaffiltext{21}{National Astronomical Observatories, Chinese Academy
 of Sciences, Beijing, China}



\begin{abstract} 
We present the  third data release of the  RAdial Velocity Experiment (RAVE)
which is the  first milestone of the RAVE project,  releasing the full pilot
survey.   The catalog  contains $83\,072$  radial velocity  measurements for
$77\,461$ stars  in the  southern celestial hemisphere,  as well  as stellar
parameters for $39\,833$ stars.  This paper describes the content of the new
release, the new processing pipeline,  as well as an updated calibration for
the  metallicity   based  upon   the  observation  of   additional  standard
stars. Spectra will be made available  in a future release. The data release
can be accessed via the RAVE webpage: {\tt http://www.rave-survey.org}.
\end{abstract}



\section{Introduction}
\label{s:introduction}

A detailed understanding of the Milky Way, from its formation and subsequent
evolution,  to its  present-day structural  characteristics, remains  key to
understanding the cosmic  processes that shape galaxies.  To  achieve such a
goal, one needs access  to multi-dimensional phase space information, rather
than restricted  (projected) properties - for example,  the three components
of the  positions and  the three  components of the  velocity vectors  for a
given sample of stars.  Until a decade ago, only the position on the sky and
the proper motion  vector was known for most of the  local stars.  Thanks to
ESA's Hipparcos satellite \citep{hip},  the distance to more than $100\,000$
stars  within a  few  hundred parsecs  has  been measured,  allowing one  to
recover precise  positions in the local  volume (a sphere  roughly 100~pc in
radius  centered on the  Sun). However,  the $6^{\rm  th}$ dimension  of the
phase  space  was  still   missing  until  recently,  when  \citet{GCS}  and
\citet{famaey05} released  radial velocities for  subsamples of respectively
$14\,000$ dwarfs and $6\,000$ giants from the Hipparcos catalog.

In recent years, with the availability of multi-object spectrometers mounted
on  large field-of-view  telescopes, two  projects aiming  at  measuring the
missing dimension have  been initiated: RAVE and SEGUE,  the Sloan Extension
for Galactic  Understanding and Exploration.   SEGUE uses the  Sloan Digital
Sky Survey (SDSS) instrumentation  and acquired spectra for $240\,000$ faint
stars, $14<g<20.3$, in  212 regions sampling three quarters  of the sky. The
moderate resolution spectrograph (R$\sim  1\,800$) combined with coverage of
a large  spectral domain ($\lambda\lambda=3\,900-9\,000$~\AA)  allows one to
reach a radial velocity  accuracy of $\sigma_{\rm RV}\sim4\kms$ at $g\sim18$
and  $15\kms$  at $g=20$  as  well as  an  estimate  of stellar  atmospheric
parameters. The  SEGUE catalog was released  as part of the  SDSS-DR7 and is
described  in  \citet{yanni09}.   Altogether,  the SDSS-I  and  II  projects
provide spectra for  about $490\,000$ stars in the Milky  Way. As of January
2011, the SDSS Data Release 8 marks the first release of the SDSS-III survey
\citep{sdssIII}.   This  release  \citep{sdssdr8} provides  $135\,040$  more
spectra from the SEGUE-2 survey targeting stars in the Milky Way.

RAVE commenced observations in 2003 and has thus far released two catalogs :
DR1 in  2006 and  DR2 in  2008 \citep{dr1,dr2}, hereafter  Papers I  and II,
respectively.  The survey targets  bright stars compared to SEGUE, $9<I<12$,
in the southern celestial  hemisphere, making the two surveys complementary.
The RAVE catalogs contain  respectively $25\,000$ and $50\,000$ measurements
of radial  velocities plus  stellar parameter estimates  for about  half the
catalog  for  DR2.  RAVE  uses  the  6dF  facility on  the  Anglo-Australian
Observatory's   Schmidt  telescope  in   Siding  Spring,   Australia.   This
instrument  allows one to  collect up  to 150  spectra simultaneously  at an
effective resolution  of $R=7\,500$ in  a 385~\AA \, wide  spectral interval
around     the     near-infrared     calcium    triplet     ($\lambda\lambda
8\,410-8\,795$~\AA).   The CaII  triplet being  a strong  feature,  RAVE can
measure radial velocities with a median precision of about $2\kms$.


RAVE is designed to study the signatures of hierarchical galaxy formation in
the Milky Way and more specifically  the origin of phase space structures in
the    disk   and    inner   Galactic    halo.   Within    this   framework,
\citet{williams2011}     discovered    the     Aquarius     stream,    while
\citet{seabroke08}  studied the  net vertical  flux  of stars  at the  solar
radius and showed  that no dense streams with an  orbit perpendicular to the
Galactic plane exist in the solar neighborhood, supporting the revised orbit
of the Sagittarius dwarf galaxy  by \citet{fellhauer06}.  On the other hand,
\citet{klement08} looked directly at stellar streams in DR1 within 500~pc of
the Sun and  identified a stream candidate  on an extreme radial
orbit (the KFR08 stream), in  addition to three previously known phase space
structures    \citep[see   also][for   an    analysis   of    known   moving
 groups]{kiss2010}.   A later  analysis  of  the DR2  catalog  by the  same
authors,  using the newly  available stellar  atmospheric parameters  in the
catalog, revised their  detection of the KFR08 stream,  the stream being now
only marginally detected \citep{klement2010}.

If RAVE is designed to look  at cosmological signatures in the Milky Way, it
is  also  well-suited  to  address  more general  questions.   For  example,
\citet{smith07} used the  high velocity stars in the  RAVE catalog to revise
the local escape speed, refining the estimate of the total mass of the Milky
Way. \citet{coskunoglu2010} used  RAVE to revise the motion  of the Sun with
respect  to  the LSR,  while  \citet{siebert08}  measured  the tilt  of  the
velocity  ellipsoid  at 1~kpc  below  the  Galactic plane.   \citet{veltz08}
combined RAVE, UCAC2,  and 2MASS data towards the  Galactic poles to revisit
the thin-thick disk decomposition  and \citet{munari08} used RAVE spectra to
confirm the existence of  the $\lambda8\,648$\AA\, diffuse interstellar band
and its correlation with extinction.

RAVE, being a randomly-selected, magnitude-limited survey, possesses content
representative  of the  Milky Way  for the  specific magnitude  interval, in
addition to  peculiar and rare  objects within the same  interval. Together,
this makes  RAVE a particularly  useful catalog to  study the origin  of the
Milky Way's stellar populations. For example, \citet{ruchti2010} studied the
elemental abundances of a sample of  metal-poor stars from RAVE to show that
direct accretion of stars from dwarf  galaxies probably did not play a major
role in the formation of the thick disk, a finding corroborated by the study
of  the  eccentricity  distribution  of   a  thick  disc  sample  from  RAVE
\citep{wilson2010}.  Also,  \citet{gal2010} used RAVE to  study double lined
binaries using RAVE spectra  while \citet{fulbright2010} used RAVE to detect
very metal poor stars in the  Milky Way. It also happens that bright objects
from  nearby  Local  Group  galaxies  are  observed;  \citet{munari09},  for
example, identified eight luminous  blue variables from the Large Magellanic
Cloud in the RAVE sample.

So  far RAVE  has released  only radial  velocities and  stellar atmospheric
parameters. To really  gain access to the full 6D  phase space, the distance
to the stars remains a missing, yet important, parameter, unless one focuses
on  a particular  class of  stars, such  as red  clump stars  \citep[see for
 examples][]{veltz08,siebert08}.  Combining the photometric  magnitude from
2MASS  and RAVE stellar  atmospheric parameters,  \citet{breddels10} derived
the 6D coordinates  for 16,000 stars from the RAVE  DR2, allowing a detailed
investigation of the  structure of the Milky Way.   This effort of providing
distances for  RAVE targets was later  improved by \citet{zwitter_distance},
taking    advantage    of   stellar    evolution    constraints,   and    by
\citet{burnett_distance},  by  using  the  Bayesian  approach  described  in
\citet{burnett2010}.    The   distance   estimates   have   been   used   by
\citet{siebert10}  to  detect   non-axisymmetric  motions  in  the  Galactic
disk.  These  works will  be  extended to  DR3,  distributed  in a  separate
catalog,  and will  provide a  unique  sample to  study the  details of  the
formation of the Galaxy.  Moreover, for  the bright part of the RAVE sample,
the signal-to-noise ratio  per pixel allows one to  estimate fairly accurate
elemental abundances from the RAVE spectra. This catalog containing of order
10$^4$ stars (Boeche  et al., in prep) will provide  a unique opportunity to
combine dynamical and chemical analyses to understand our Galaxy.


In this paper we present the  3$^{\rm rd}$ data release of the RAVE project,
releasing the radial velocity data and stellar atmospheric parameters of the
pilot survey  program that  were collected during  the first three  years of
operation, therefore  DR3 includes the data  collected for DR1  and DR2. The
spectra are not part of this release.  These data were processed using a new
version of the processing pipeline.  This paper follows the first and second
data releases described in Papers I and II.  The pilot survey release is the
last release relying on the original input catalog, based on the Tycho-2 and
SuperCosmos  surveys. Subsequent  RAVE  releases will  be  based on  targets
selected from the  DENIS survey I-band.  The paper  is organized as follows:
Section~\ref{s:pipeline} presents the new version of the processing pipeline
which calculates the radial velocities and estimates the stellar atmospheric
parameters.  Section~\ref{s:validation}  presents the validation  of the new
data, as  well as  the updated calibration  relation for  metallicity, while
Section~\ref{s:catalog} describes the DR3 catalog.

\section{A revised pipeline for stellar parameters}
\label{s:pipeline}

In Papers  I and II we described  in detail the processing  pipeline used to
compute the radial velocities and the stellar atmospheric parameters, making
use of a best-matched template to  measure the radial velocities and set the
atmospheric  parameters reported  in  the catalog.   This pipeline  performs
adequately for  well-behaved spectra, permitting the  measurement of precise
radial velocities,  and we showed in  Paper II that  the stellar atmospheric
parameters $\teff$, $\logg$, and $\mh$ can be estimated. However, to compare
the  RAVE $\mh$  to high  resolution  measurements $\MH$\footnote{Throughout
 this  paper, $\mh$  refers  to  the metallicity  obtained  using the  RAVE
 pipeline  while  $\MH$  refers  to  metallicity  obtained  using  detailed
 analyses of high resolution echelle spectra.}, a calibration relation must
be  used. Also,  in the  case  where a  RAVE spectrum  suffers from  (small)
defects, the  stellar atmospheric parameters are  less well-constrained.  We
therefore  set out  to improve  the  pipeline, while  still maintaining  its
underlying computational techniques.  This section reviews the modifications
of the RAVE pipeline, which is otherwise fully described in Paper II.

\subsection{Stellar library}
\label{s:library}

The RAVE  pipeline for  DR1 and DR2  relied on the  \citet{munari} synthetic
spectra library  based on ATLAS~9  model atmospheres. This  library contains
spectra with three different values  for the micro-turbulence $\mu$ of 1, 2,
and  $4\kms$.    However,  the  library  is  well-populated   only  for  the
$\mu=2\kms$  value,  about  $3\,000$  spectra  having  $\mu=1$  or  $4\kms$,
compared to $\sim55\,000$ having $\mu=2\kms$.

For  this new  data release  (DR3), new  synthetic spectra  for intermediate
metallicities  were added  in  order  to provide  a  more realistic  spacing
towards the  densest region  of the observed  parameter space and  so remove
biases  towards  low metallicity.   The  new  grid  has $\mh=-2.5$,  $-2.0$,
$-1.5$,  $-1.0$, $-0.8$, $-0.6$,  $-0.4$, $-0.2$,  $0.0$, $0.2$,  $0.4$, and
$0.5\,$dex.

We  also  restricted  the  library  to  $\mu=2\kms$,  discarding  all  other
micro-turbulence values.  This  does not impact the quality  of the measured
stellar parameters as,  at our \snr\ level and resolution,  we are unable to
constrain the  micro-turbulence, and the  pipeline usually converges  on the
most common micro-turbulence value in the library ($\mu=2\kms$).

Furthermore, since  the nominal  resolution of the  6dF instrument  does not
allow us to measure precisely the  rotational velocity of the star, we chose
to restrict the  $V_{\rm rot}$ dimension, removing six  of the lower $V_{\rm
 rot}$ values ($0$, $2$, $5$, $15$, $20$, and $40\kms$), retaining only the
$10$, $30$, $50\kms$, and higher, velocities.

Removing one  dimension of the  parameter space and reducing  the rotational
velocity dimension  helps to stabilize the  solution and allows  us to lower
the number  of neighboring spectra  used for the  fit. We lower  this number
from  300  to  150.  As  for  Paper  II,  the  Laplace multipliers  for  the
penalisation terms were set using Monte Carlo simulations.  We increased the
Laplace  multiplier handling  the penalisation  on the  sum of  the weights,
which  constrains  the  level  of  the  continuum  to  unity  for  continuum
normalized spectra,  as a  $0.3$\% offset was  not uncommon in  the previous
pipeline.

\subsection{Signal-to-noise estimation}
\label{s:snr}

To date, the  processing pipeline used \snr\ estimates  as described in Paper
I. However, this  \snr\ estimate tends to underestimate the  true \snr\ and is
less dependent  on the true  noise than it  is on the weather  conditions or
spectrum defects, such as fringing (see  Paper II).  In Paper II, a new \snr\
estimate, S2N, was presented based on the best fit template but was not used
by the pipeline as it was an {\it a posteriori} estimate. We showed that S2N
is closer to the true \snr.

Because    of    the     new    continuum    correction    procedure    (see
Section~\ref{s:normalisation}), the \snr\  must be computed correctly before
the continuum  correction is applied. Therefore,  it must be  known prior to
the processing.   We thus developed an  algorithm to measure the  \snr\ of a
spectrum in which no flux information is used. This new \snr\ estimate, STN,
is obtained using the observed spectrum (no continuum normalization applied)
as follows:
\begin{itemize}
\item[1] Smooth the observed spectrum $s(i)$, with $i$ the pixel index, to
produce a smoothed spectrum $f(i)$. This smoothing is done with a smoothing
box three pixels long.
\item[2] Compute the residual vector $R(i)=f(i)-s(i)$ and its rms $\sigma$.
\item[3] Remove from $s$ pixels that diverge from $f$ by more than $2\sigma$.
\item[4] Smooth the clipped spectrum as above to form a new smoothed spectrum
$f$ and repeat the clipping process until convergence.
\item[5] Compute the local standard  deviation $\sigma_l(i)$ using  pixels
 $i-1,\,i$ and $i+1$.
\item[6] Compute STN$=\hbox{median}(s(i)/\sigma_l(i))/1.62$.
\end{itemize}

\begin{figure}[hbtp]
\centering
\includegraphics[width=15cm]{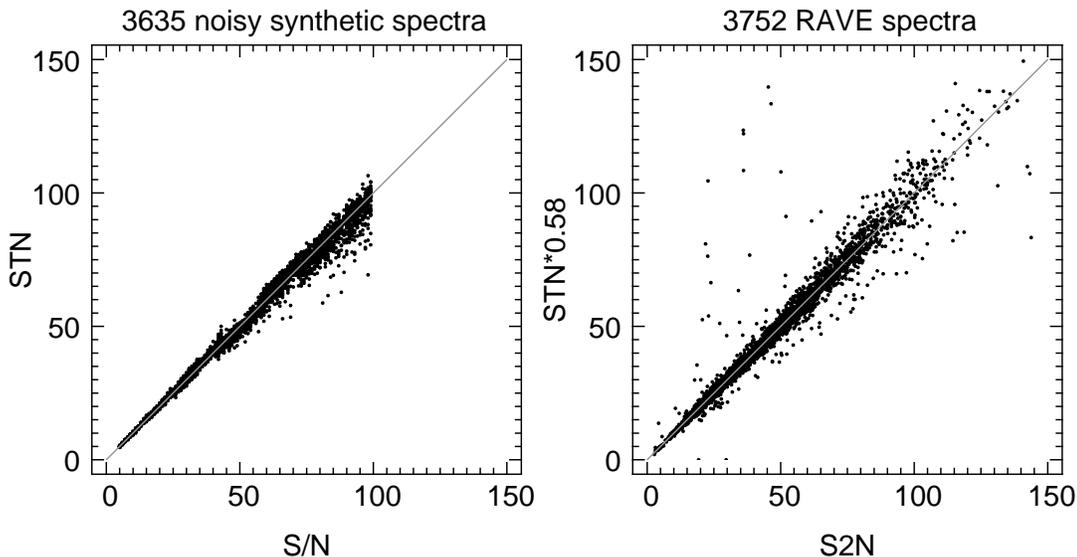}
\caption{Comparison of the various signal-to-noise estimates.  Left panel:
 signal-to-noise STN  compared to the  original RAVE \snr.   Right panel:
 comparison  of the  scaled  STN to  S2N,  the signal-to-noise  estimator
 constructed for DR2.}
\label{f:snr}
\end{figure}

\noindent 
The factor of  1.62 is set using numerical realizations  of a Poisson noise.
As shown in the left panel of Figure~\ref{f:snr}, \snr\ and STN are on a 1:1
relation.  However, in a real spectrum, instrument noise also contributes to
the residuals  and we  expect an additional  normalization factor.   The S2N
value as  computed in  Paper II  follows closely the  true \snr.   Hence, to
assess the  validity of the  STN measurement, we  compared it to the  S2N in
Paper II (Figure~\ref{f:snr} right panel).   A correction factor of 0.58 for
S2N  is found  to produce  a 1:1  relation between  the two  measurements, a
correction that we apply in the pipeline.

\subsection{Continuum normalization}
\label{s:normalisation}

In  the  low \snr\  regime  ($\snr<10$), the  metallic  lines  are no  longer
visible. In  this case, $\mh$  measurements converge to the  highest allowed
value  ($\mh=+0.5$~dex)  which gives  the  lowest  possible $\chi^2$  value,
i.e. the algorithm  fits the noise.  In this  regime, the stellar parameters
are not reliable and are therefore not published. In the intermediate regime
$10<\snr<50$, a correlation between $\mh$ and \snr\ is observed in
the RAVE data.

While some of the above correlation is understood and arises from the change
of the underlying  stellar content as one moves further  away from the plane
and  the  \snr\ simultaneously  decreases\footnote{The  exposure time  being
  fixed, a  lower \snr\ indicates a  fainter magnitude.}, some  part of this
correlation arises  from to the  continuum normalization failing  to recover
the proper continuum level.  The former pipeline uses the \textsc{IRAF} {\tt
  continuum} task with asymmetric rejection parameters ($1.5 \sigma$ for the
low rejection level  and $3.0 \sigma$ for the  high rejection level).  While
these parameters are  well-suited for the high \snr\  regime ($>60$), at low
\snr\ they tend to produce an estimated continuum that is too high.  This is
due to  the routine considering the  spikes below the  continuum as spectral
lines when, in fact, they are mainly due to noise.

\begin{figure}[hbtp]
\centering
\includegraphics[width=15cm]{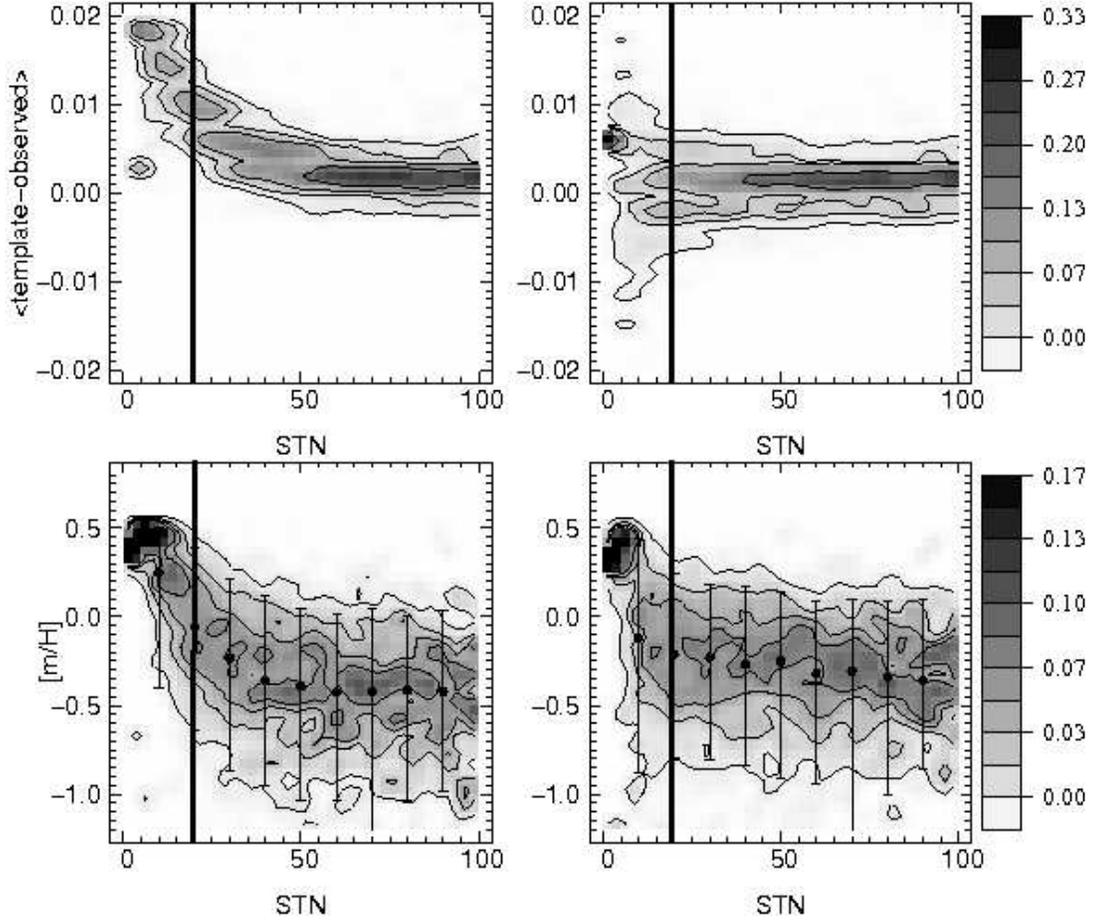}
\caption{Top panels: average  residuals best fit template$-$observed spectra
  for  4,684 RAVE  spectra  as a  function  of \snr.   Bottom panels:  $\mh$
  distributions as a function of \snr. The left columns are for the previous
  version of  the continuum normalization  algorithm while the  right column
  includes the low  rejection level being a function of  \snr. The gain from
  the  new  continuum  normalization   is  clear  from  these  figures:  the
  correlation between  metallicity and \snr\ is strongly  reduced, while the
  residuals do  not show  any correlation with  \snr.  The thick  black line
  represents  the  STN limit  below  which  atmospheric  parameters are  not
  published in the RAVE catalog.}
\label{f:continuum}
\end{figure}

We ameliorate this problem by using a low rejection value that is a function
of \snr.  This rejection level must be close to $1.5$ for high \snr\ spectra
and larger for  low S/N.  Numerical tests indicate  that using the following
formula

\begin{equation}
\mathrm{low}_\mathrm{rej}=1.5+0.2\exp\left(
-\frac{\mathrm{STN}^2}{2\sigma_{\mathrm{STN}}^2}\right)\,,
\end{equation}

\noindent
with    $\sigma_\mathrm{STN}=16$,    from   the    top    left   panel    of
Figure~\ref{f:continuum}, reduces  significantly the continuum normalization
problem. The  top panels in Figure~\ref{f:continuum} show  the mean residual
between the observed continuum-normalized spectra and best fit template as a
function of  \snr, before and after  the change in the  low rejection level,
while the  bottom panels present the  resulting distributions of  $\mh$ as a
function of \snr.

The  new  continuum  normalization  reduces  significantly  the  correlation
between metallicity and  \snr, while no trend in the  residual as a function
of  \snr\ remains.   This  indicates that  the  new continuum  normalization
algorithm performs adequately, although a  weak correlation is still seen in
the metallicity versus \snr\ ($\sim0.1\,$dex per 100 in \snr).

\subsection{Masking bad pixels}
\label{s:mask}

Approximately 20\% of  RAVE spectra suffer from defects  such as fringing or
residual cosmic rays, which cannot  be removed by the automatic procedure we
use  to reduce  our data.   While  residual cosmic  rays do  not affect  the
determination of  the stellar atmospheric  parameters (these are  similar to
emission lines, which  are not taken into account  in the template library),
fringing  results   in  poor  local  continuum   normalization,  leading  to
inaccurate parameter recovery.

Regions strongly affected  by fringing are difficult to  detect prior to the
processing, but we can make use of the best fit template to estimate whether
a spectrum  suffers from such  a continuum distortion and  therefore whether
the atmospheric parameter determination is likely to be in error.

\begin{figure}[hbtp]
\centering
\includegraphics[width=15cm]{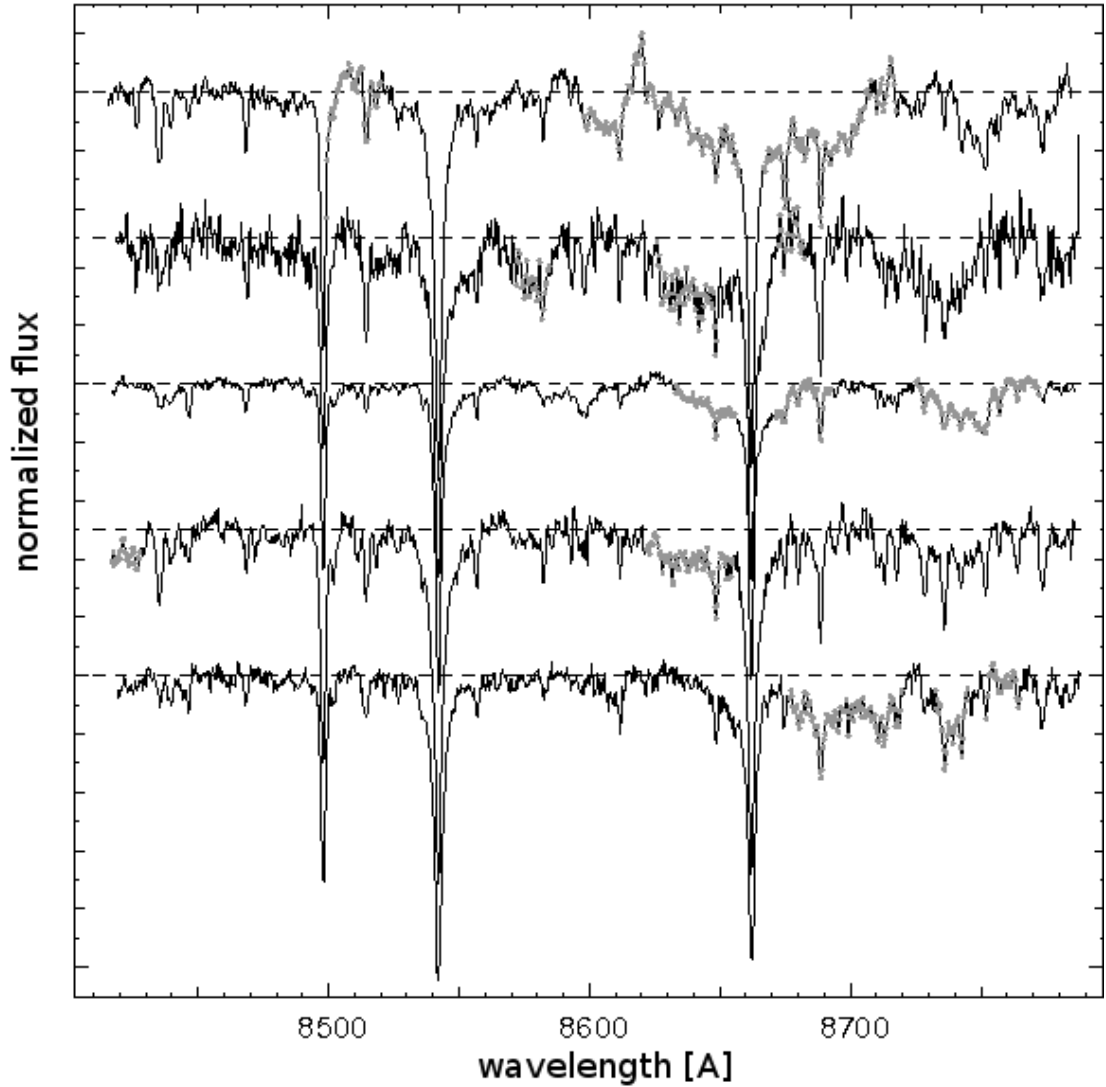}
\caption{Example of five RAVE spectra  with regions marked as problematic by
  the MASK code. The regions marked in grey are recognized as suffering from
  poor continuum normalization.  If more than 30\% of the spectrum is marked
  by the  code, the observation is  flagged as problematic  by the pipeline.
  The  normalized fluxes are  in arbitrary  units and  a vertical  offset is
  added between the spectra for clarity.}
\label{f:badcontinuum}
\end{figure}

To  estimate   the  fraction  of   a  spectrum  contaminated   by  continuum
distortions, we compute the reduced  $\chi^2(i)$ along the spectrum in a box
21 pixels  wide centered  on the pixel  $i$. We  then also compute  the mean
difference $S(i)$ between the best-fit template and the observed spectrum in
the  same box.   If  $\chi^2(i)>2$ and  $S(i)>2/\mathrm{STN}$, a  systematic
difference  between the  template  and the  observed  spectrum exists.   The
corresponding  region of  the spectrum  is then  flagged as  a  defect.  The
fraction of good  pixels in each spectrum is then recorded  and given in the
RAVE catalog (see MaskFlag in Table~\ref{t:A1}).  From visual inspection, we
find  that when  the  number of  bad pixels  is  larger than  30\% then  the
spectrum is  problematic and the  stellar parameters should be  treated with
caution.  Figure~\ref{f:badcontinuum} shows  different examples of real RAVE
spectra where a significant fraction of the spectrum is marked as defect.

\subsection{Improving the zero-point correction}
\label{s:zeropoint}

As explained  in previous papers  (e.g., Paper~I), thermal  instabilities in
the spectrograph  room induce zero  point shifts of the  wavelength solution
that  depend on  the  position along  the  CCD (e.g.,  fiber number).   This
results in instabilities of the radial velocity zero-point.

To  correct the  final radial  velocities  for this  effect, the  processing
pipeline uses  available sky lines in  the RAVE window and  fits a low-order
polynomial ($3^{\rm rd}$ order) to  the relation between sky radial velocity
and fiber number. This $3^{\rm  rd}$ order polynomial defines the mean trend
of zero point  offsets and provides the zero point  correction as a function
of  fiber number\footnote{The  zero-point correction  could in  principle be
  obtained directly from  the radial velocity of the  sky lines. However the
  radial  velocity measured  from  the sky  lines  suffers from  significant
  errors while the trend of the  zero-point offset with respect to the fiber
  number due to thermal changes is expected to be a smooth function of fiber
  number.  Therefore, using  a smooth function to recover  the mean trend is
  better suited  to correct  for zero point  offsets. Tests have  shown that
  using  a $3^{\rm rd}$  order polynomial  provides in  most cases  the best
  solution (see paper I).}.  However,  in some cases, a low-order polynomial
is not  the best solution and a  constant shift should be  used instead.  In
former releases, these cases were corrected by hand in the catalog.  In this
release, we introduced a new zero-point correction routine to the processing
pipeline  that  is  able  to  select which  correction  should  be  applied,
automatically.

The zero-point correction now computes  both the cubic correction, using the
$3^{\rm rd}$ order polynomial, and the constant correction. It then computes
the mean and  standard deviation between the measured  sky radial velocities
and the  corrections for  the entire  field and for  three regions  in fiber
number that are contiguous on  the CCD (fibers 1$-$50, 51$-$100, 101$-$150).
For each region,  the cubic fit is used unless any  of these four conditions
apply:
\begin{itemize}
\item[-] there are  less than two  sky fibers  in that region,  to avoid
 under-constrained fits,
\item[-] the mean in that region for the constant correction is better than
 the corresponding mean for the cubic fit,
\item[-] the standard deviation  for the cubic correction is greater than
 $5\kms$, which is the case for noisy data,
\item[-] the  maximum difference between the constant  correction and the
 cubic correction is larger than $7\kms$.
\end{itemize}

\begin{figure}[hbtp]
\centering
\includegraphics[width=8cm]{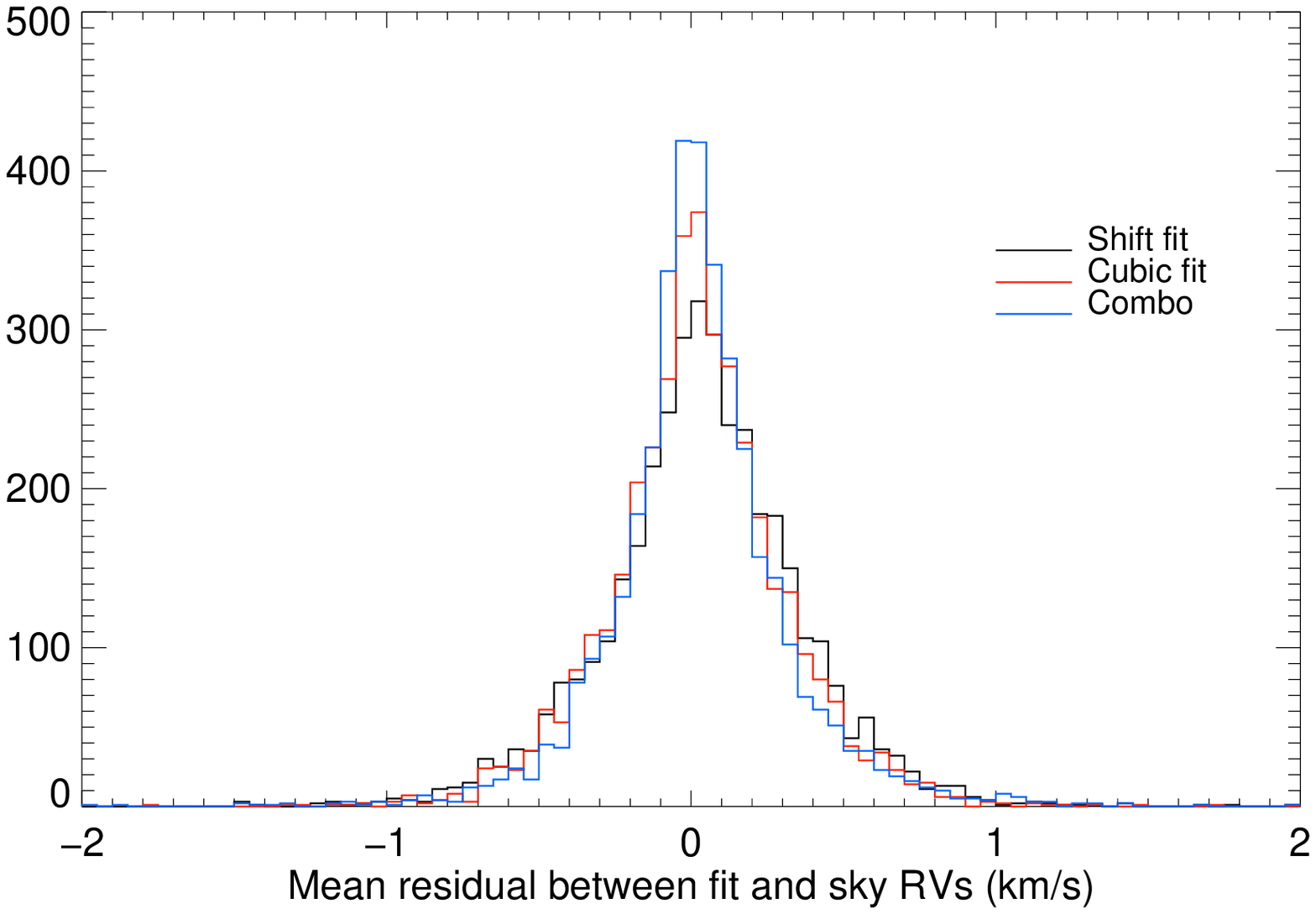}
\includegraphics[width=8cm]{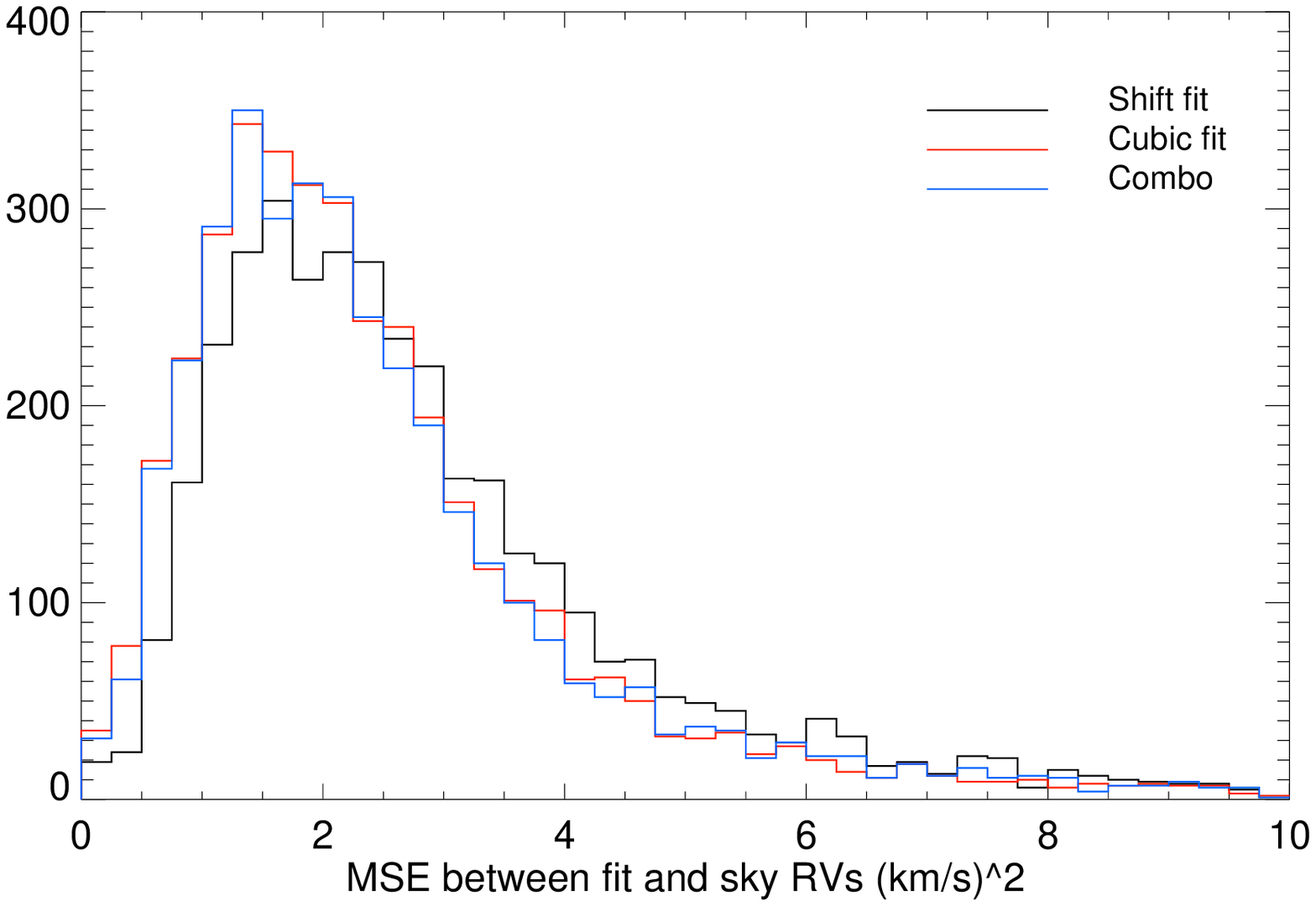}
\caption{Left: mean residual between the fit and the sky radial velocity for
 three different fitting functions. A constant shift (black histogram), the
 cubic  fit used  in  DR1 and  DR2  (red histogram),  and  the new  fitting
 procedure (blue histogram).  Right: associated mean-square-error.}
\label{f:zeropoint}
\end{figure}

We tested the  new procedure, together with other  options, against pairs of
repeat     observations.       The     results     are      presented     in
Table~\ref{t:zeropoint}. They  show clearly that the  new procedure performs
better  than the previous  version in  terms of  dispersion, while  the mean
difference is unchanged.  While  the constant term correction appears better
in  this table,  the left  panel  in Fig.~\ref{f:zeropoint}  shows that  the
distribution of the residuals is  less peaked than for the cubic correction.
In addition, the mean-square-error, defined as $MSE=E[(RV-RV_{\rm fit})^2]$,
shows a net  decrease with the new fitting procedure  compared to a constant
shift. This indicates  that for the general case,  a constant correction for
the  entire field  will result  in a  larger dispersion  and hence  a larger
zero-point offset residual.  This gives us confidence in the  use of the new
procedure.

\begin{table}[hbtp]
\centering
\caption{Radial  velocity difference  between pairs  of  repeat observations
 using different zero-point correction  solutions. The old correction is a
 combination of cubic  fit and corrections applied by  hand.  The number of
 pairs used is 25,172.}
\begin{tabular}{l c c c c c}
\hline
\hline
Method & $\mu~(\!\kms)$ & $\sigma~(\!\kms)$ & $N_{\rm reject}$ & 68\%
($\!\kms$) & 95\% ($\,\kms$)\\
\hline
No correction & 0.38 & 2.74 & 2\,572 & 3.0 & 18.9\\
Old correction & 0.23 & 2.49 & 2\,765 & 2.8 & 16.8\\
Cubic & 0.22 & 2.52 & 2\,958 & 2.9 & 21.3\\
Quadratic & -0.44 & 2.83 & 2\,645 & 3.2 & 20.2\\
Linear & 0.24 & 2.21 & 2\,850 & 2.5 & 16.7\\ 
Constant & 0.23 & 2.05 & 2\,990 & 2.3 & 16.5\\
New correction & 0.23 & 2.22 & 2\,817 & 2.5 & 16.6\\
\hline
\end{tabular}
\label{t:zeropoint}
\end{table}

\section{Calibration and validation}
\label{s:validation}

\subsection{Radial velocity}
\label{s:RV}

\subsubsection{Internal error distribution}
\label{s:internal_rv_dist}

\begin{figure}[hbtp]
\centering
\includegraphics[width=9cm]{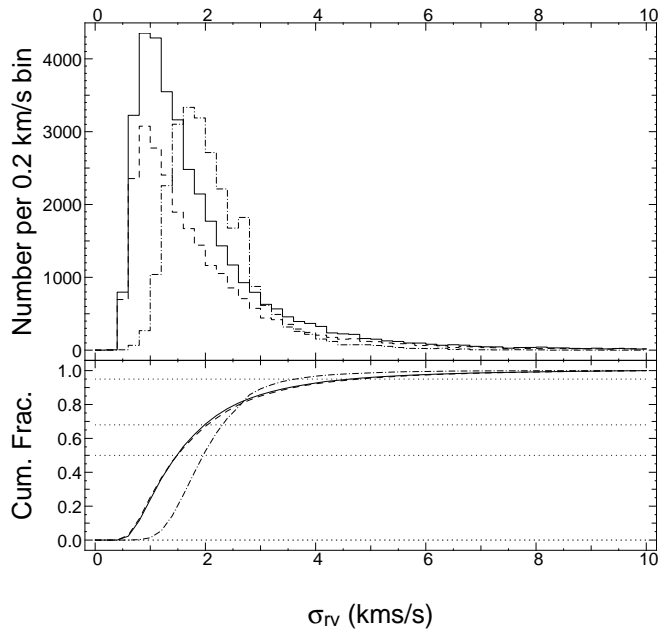}
\caption{Distribution of the radial velocity error (eRV) in the 3$^{\rm rd}$
 data  release.   Top: number  of  stars with  eRV  in  $0.2\kms$ bins  for
 first-year data  (dash-dotted line),  second-year data (dashed  line), and
 third-year data  (full line). Bottom: cumulative distribution  of the eRV.
 The dotted lines mark respectively 50, 68 and 95\% of the samples.}
\label{f:eRV}
\end{figure}

RAVE obtains its radial  velocity from a standard cross-correlation routine.
For each  radial velocity measurement  the associated error, eRV,  gives the
internal error  due to  the fitting procedure.   Figure~\ref{f:eRV} presents
the distribution  of eRV  per $0.2\kms$ bin  for the  data new to  each RAVE
release.  While first year data are of lower quality due to the second-order
contamination  of our  spectra,  second and  third  year data  are of  equal
quality  with  a  mode at  $0.8\kms$,  a  median  radial velocity  error  of
$1.2\kms$,  and  95\% of  the  sample  having  internal errors  better  than
$5\kms$.  Comparing these values to the old version of the pipeline used for
DR1  and  DR2  (see Table~2  and  Fig.~9  of  Paper  II), the  new  pipeline
marginally improves the  internal accuracy with a gain  of $\sim0.1\kms$ for
the mode and the median radial velocity error.

The aforementioned  error values represent the contribution  of the internal
errors to the  RAVE error budget.  External errors are  also present and are
partially due to the zero-point correction which corrects only a mean trend,
not  including  the  fiber-to-fiber  variations.  The  contribution  of  the
external  errors is  obtained using  external datasets  and is  discussed in
Section~\ref{s:external}.

\subsubsection{Zero-point error}

Our internal  error budget is the sum  of (i) the error  associated with the
evaluation of the maximum of  the Tonry-Davis correlation function, and (ii)
the contribution from the zero-point error.  The first contribution is given
by the pipeline (\S~\ref{s:internal_rv_dist}).   The magnitude of the second
term can be obtained from the  analysis of the re-observed targets as, for a
given  star  whose apparent  magnitude  is  fixed,  the radial  velocity  is
constant (if the star is not a  binary) and the internal errors are the main
source of uncertainties.

We therefore  use the re-observed stars  in the RAVE  DR3 catalog, selecting
only stars observed during the second and third year, as they share the same
global properties  in terms  of observing conditions.   Data from  the first
year  of   observing  are  discarded,  as  they   suffer  from  second-order
contamination  which  renders  the  internal  error  inhomogeneous  and  can
therefore bias  our estimate.   We also removed  from the sample  stars that
were observed on purpose to calibrate our stellar atmospheric parameters, as
these are  specific bright  targets with  high \snr\ that  do not  share the
random  selection function nor  the standard  observational protocol  of the
RAVE catalogue.

\begin{figure}[hbtp]
\centering
\includegraphics[width=7cm]{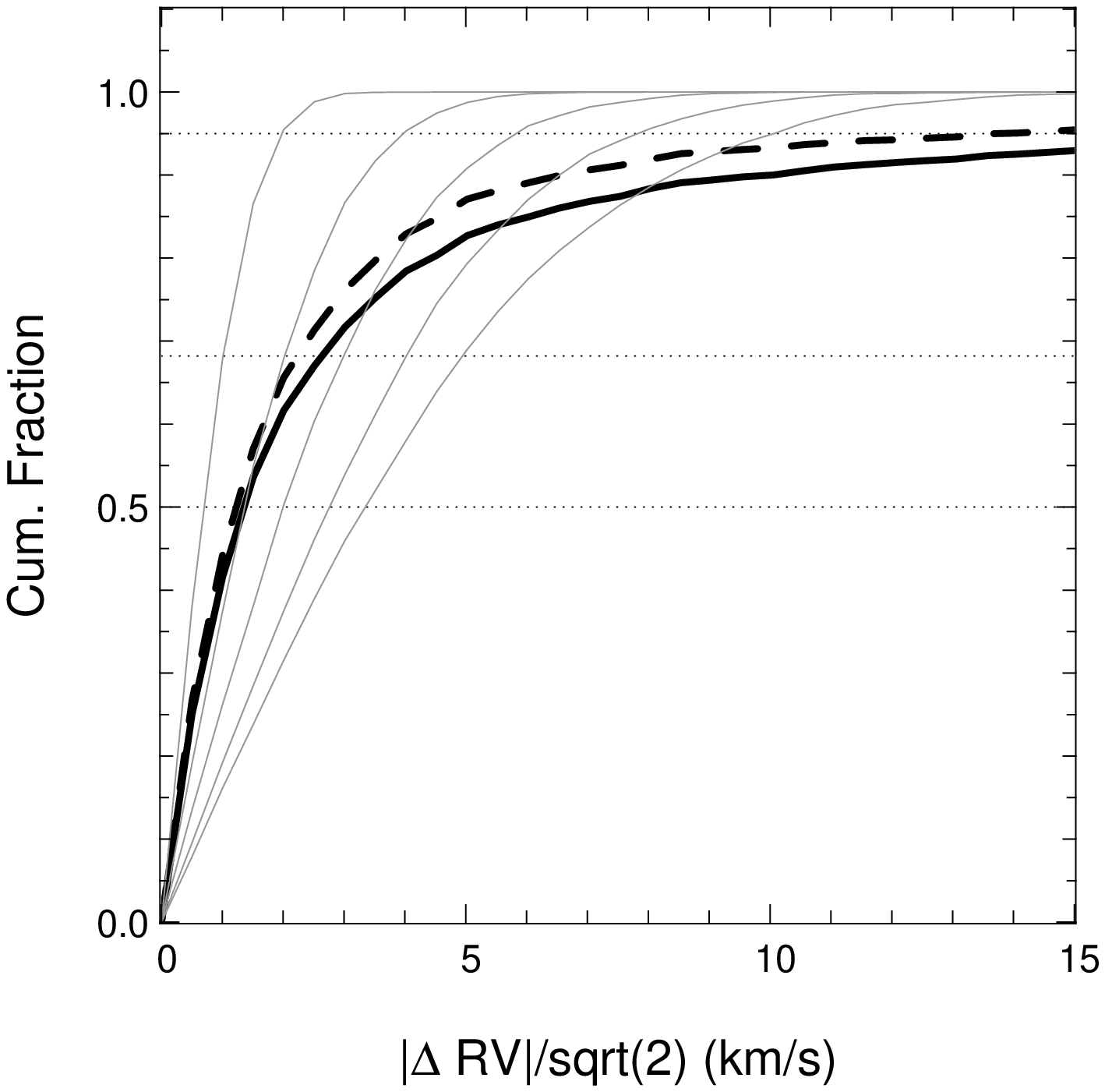}
\includegraphics[width=7cm]{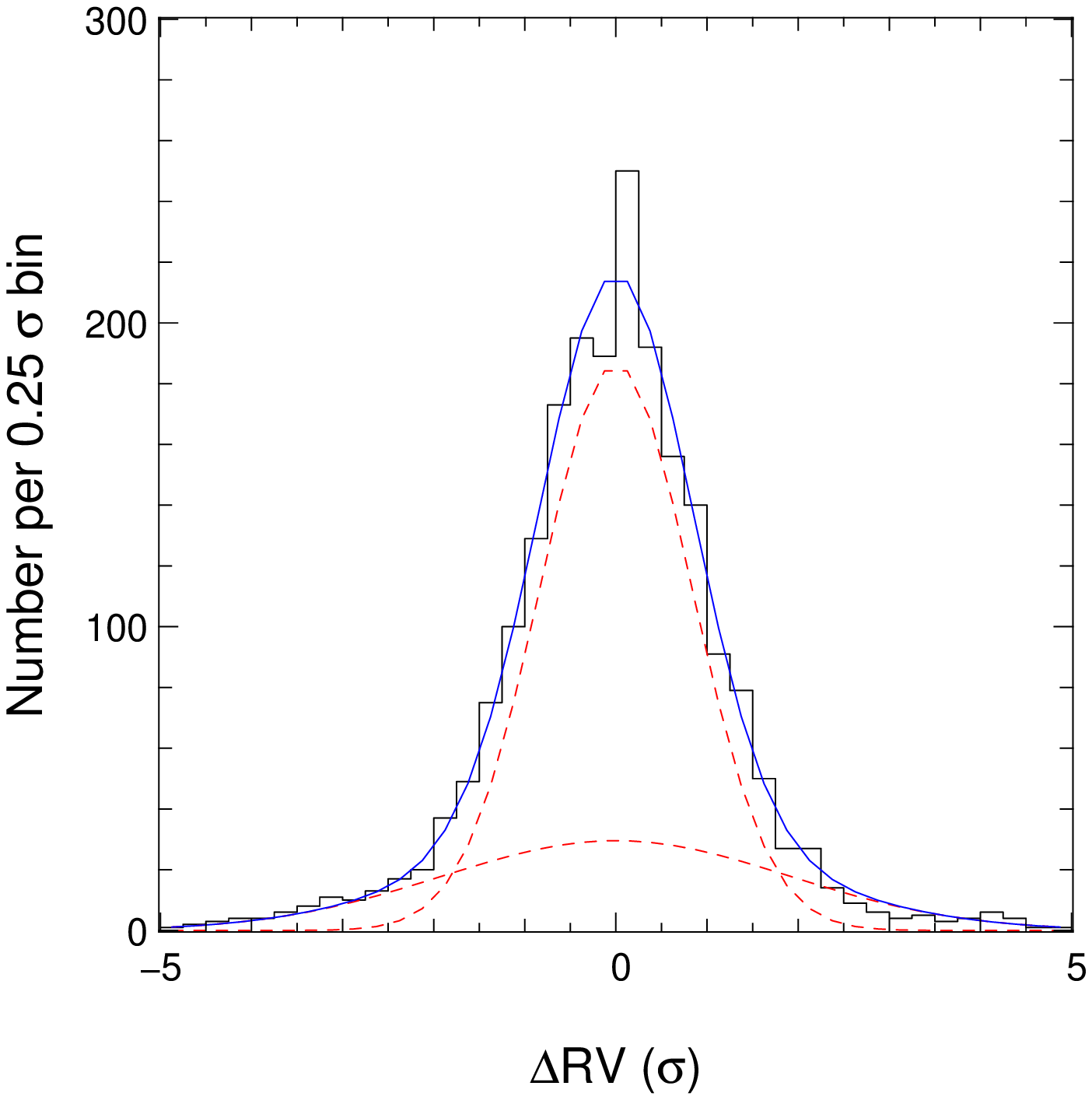}
\caption{Left:  Cumulative fraction  of the  radial velocity  difference for
 re-observed  RAVE targets  in  the  Third Data  Release.   The solid  line
 corresponds to the full sample, and  the dashed line relates to the sample
 restricted  to pairs  whose  individual measurement  differ  by less  than
 3$\sigma$  (hence rejecting  the spectroscopic  binaries with  the largest
 radial velocity difference).  The  horizontal lines indicate 50, 68.2, and
 95\% of the sample.  The grey  lines are the expected distributions of the
 radial velocity difference for Gaussian errors  of 1, 2, 3, 4, and $5\kms$
 from  inside out. Right:  distribution of  the radial  velocity difference
 $\Delta RV$ in  units of $\sigma$ for re-observed  targets.  The blue line
 corresponds  to our best-fit  double Gaussian  model to  the distribution.
 The red dashed lines show the respective contribution of each Gaussian.}
\label{f:RV_sigma}
\end{figure}

The cumulative  distribution of the radial velocity  difference is presented
in the left  panel of Fig.~\ref{f:RV_sigma} where the  solid line represents
the  full sample  of  re-observed targets  and  the dashed  line the  sample
restricted to individual measurements differing  by less than $3\sigma$ in a
pair.   Since our  sample is  contaminated by  spectroscopic  binaries, this
selection is compulsory  if one wants to address  the error distribution for
normal stars but is only a crude approximation when trying to remove all the
binaries in  the sample.   Applying this  cut rejects 6\%  of the  sample, a
value clearly below the  expected contamination (see below).  Therefore, the
errors estimated from the repeat observations are likely to overestimate the
true  errors.  With  this  limitation in  mind, from  Fig.~\ref{f:RV_sigma},
focusing on the dashed line, one  can conclude that 68.2\% of the sample has
an  error below  $2.2\kms$ while  $\sim93$\% of  the sample  lies  below the
$5\kms$ accuracy limit.

To  estimate  the contribution  from  the  zero-point  errors to  the  total
internal error budget, we computed  the distribution of the {\it normalized}
radial  velocity  difference, the  relative  difference  in radial  velocity
between two observations divided by the  square root of the quadratic sum of
the errors on radial velocity. If our measurements were affected only by the
random errors for (i), then the distribution of this {\it normalized} radial
velocity difference  would follow a  Gaussian distribution of zero  mean and
unit standard deviation. An additional  contribution to the error budget due
to  a random  zero-point  error  would broaden  the  distribution and  hence
enhance the  dispersion of the  resulting distribution.  The result  of this
test is  presented in the  right panel of Figure~\ref{f:RV_sigma},  where we
fitted the sum of two Gaussians to the observed distribution.

The  dominant Gaussian distribution  corresponds to  stars stable  in radial
velocity.  The width  of the associated Gaussian function  is $0.83 \sigma$,
narrower  than a normal  distribution, indicating  that the  internal errors
quoted in  the catalog are  likely overestimated. Our quoted  internal error
can therefore be  assumed to be an upper bound on  the true internal errors,
including the contribution of the zero-point error.

{\bf Spectroscopic  binary contamination:} subsidiarily,  the broad Gaussian
comprises spectra with defects (or  where the zero-point solution could have
diverged)  as well  as the  contribution from  spectroscopic  binaries.  The
fraction of spectra with defects is small in this sample, as the catalog has
been   cleaned   of  fields   where   the   zero-point   solution  did   not
converge. Hence, the relative weight  of the two Gaussian functions gives an
estimate,  in  reality  an  upper  limit,  of  the  contamination  level  by
spectroscopic binaries  with radial velocity  variation between observations
larger than  $1\sigma$ in the RAVE  catalog.  Our best-fit  solution gives a
relative contribution for this second  population of 26\% which allows us to
conclude that  the fraction of  spectroscopic binaries with  radial velocity
variations larger than $2\kms$ in the  RAVE catalog is less than or equal to
26\%.  A  more detailed analysis  of repeated observations based  on 20\,000
RAVE stars by \citet{sb1} gives a  lower limit of 10-15\% of the RAVE sample
being affected  by binarity \citep[see also][]{gal2010}.   However, the time
span between repeat observations being biased towards short periods (days to
weeks), long period  variations are not detected. The  previous estimates do
not take into  account this population and a more  detailed analysis will be
required  to estimate  the  contribution  of long  period  variables to  our
survey.

\subsubsection{Validation using external datasets}
\label{s:external}

\begin{figure}[hbtp]
\centering
\includegraphics[width=7cm]{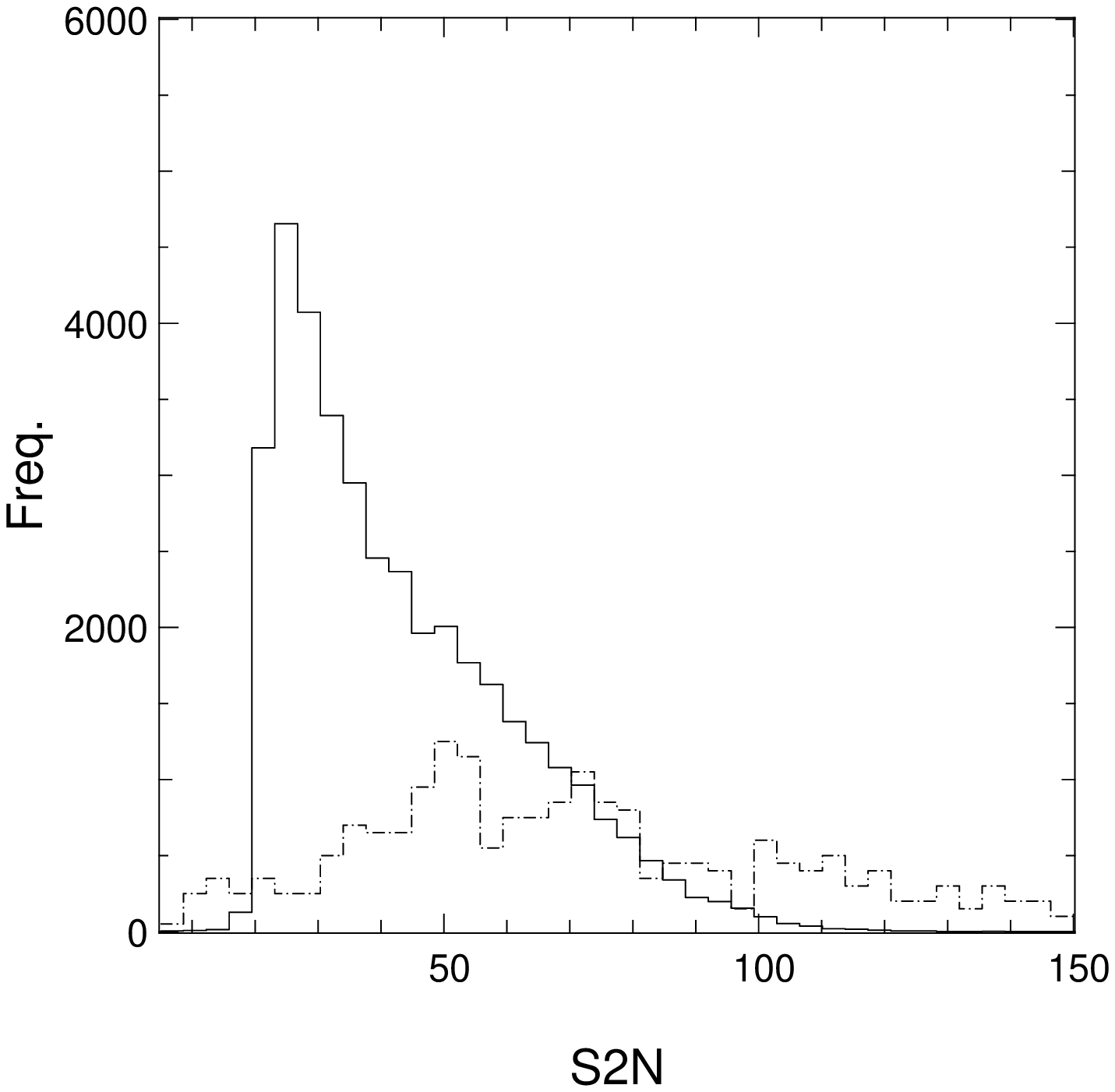}
\includegraphics[width=7cm]{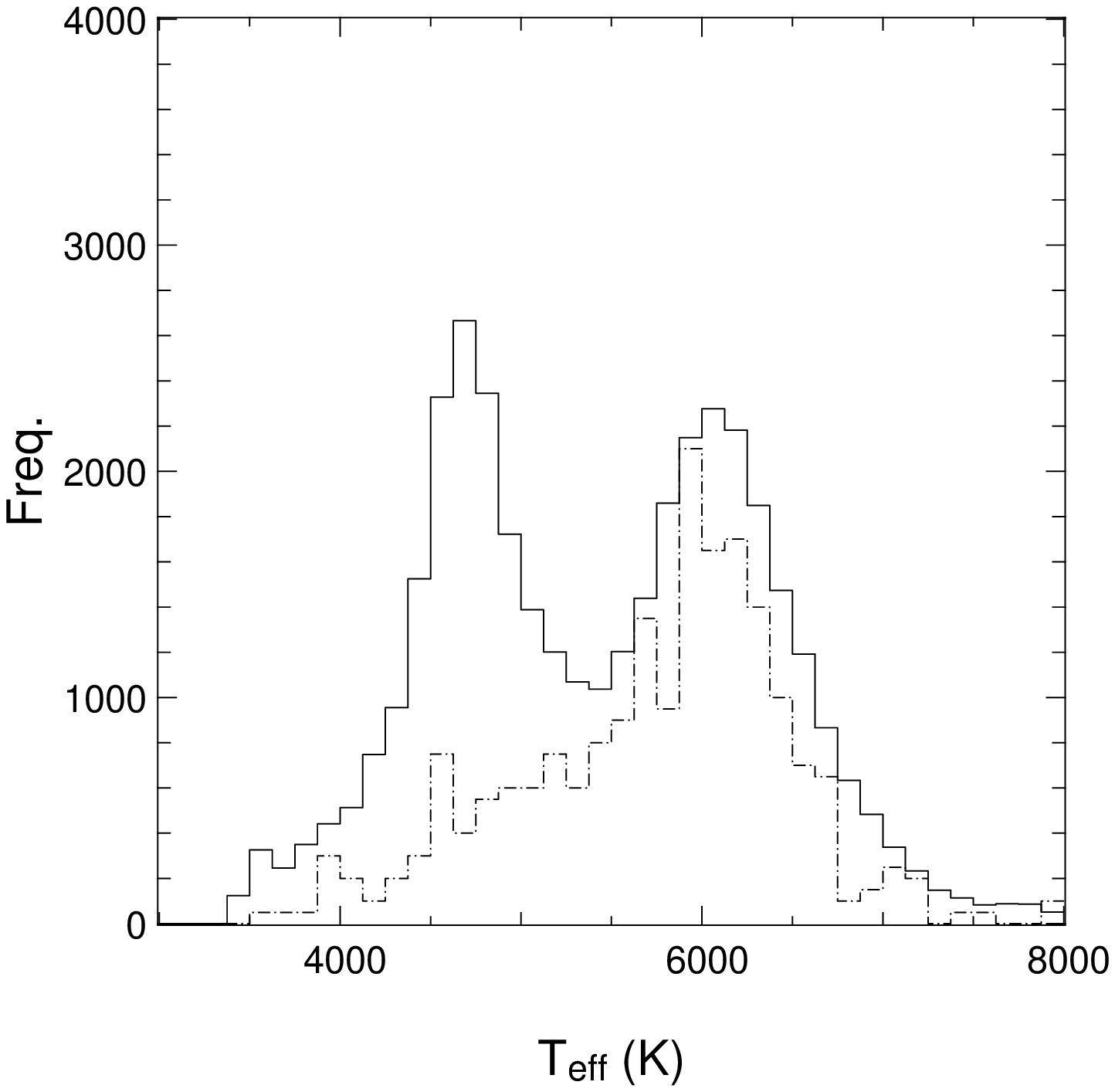}
\includegraphics[width=7cm]{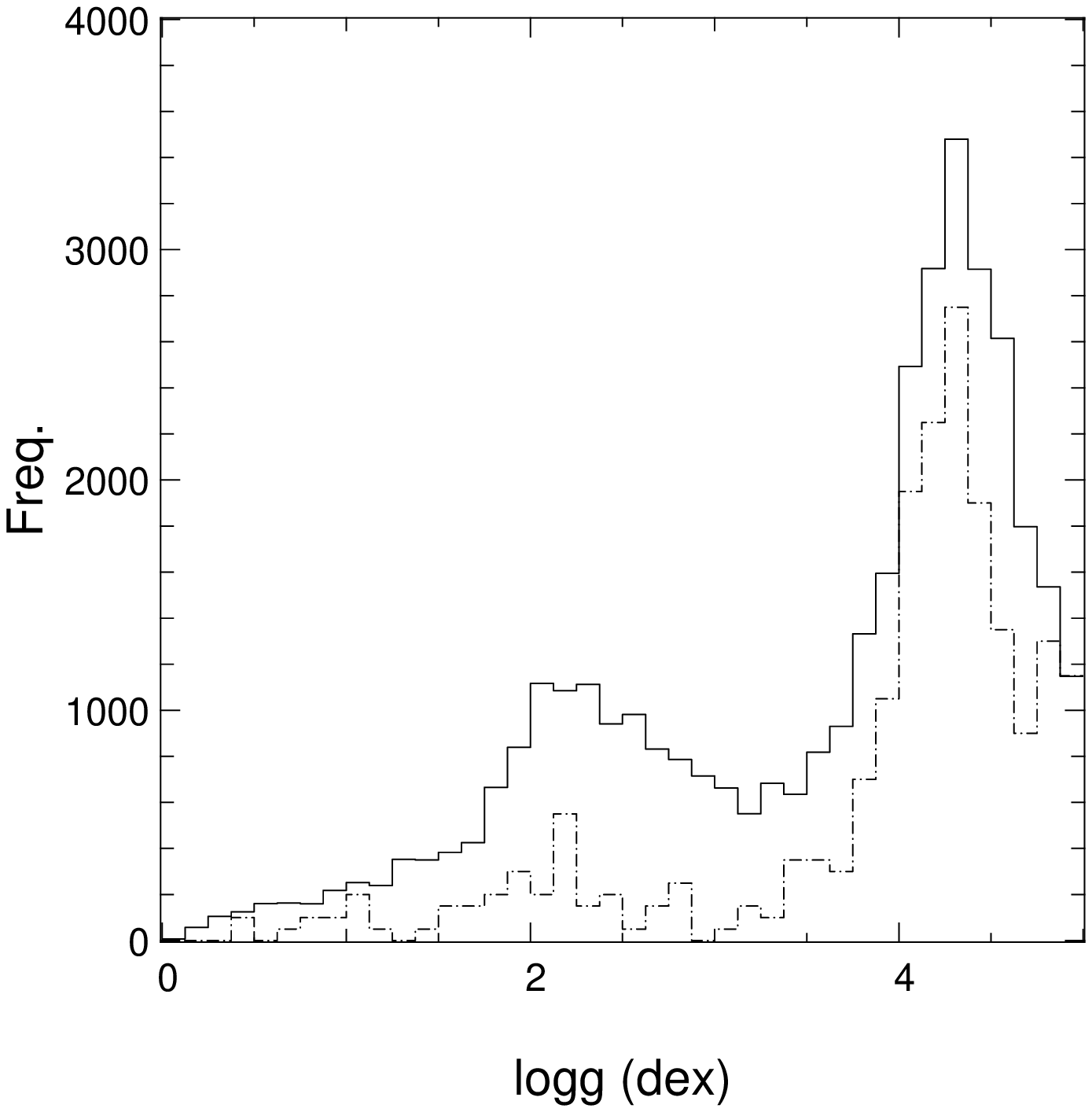}
\includegraphics[width=7cm]{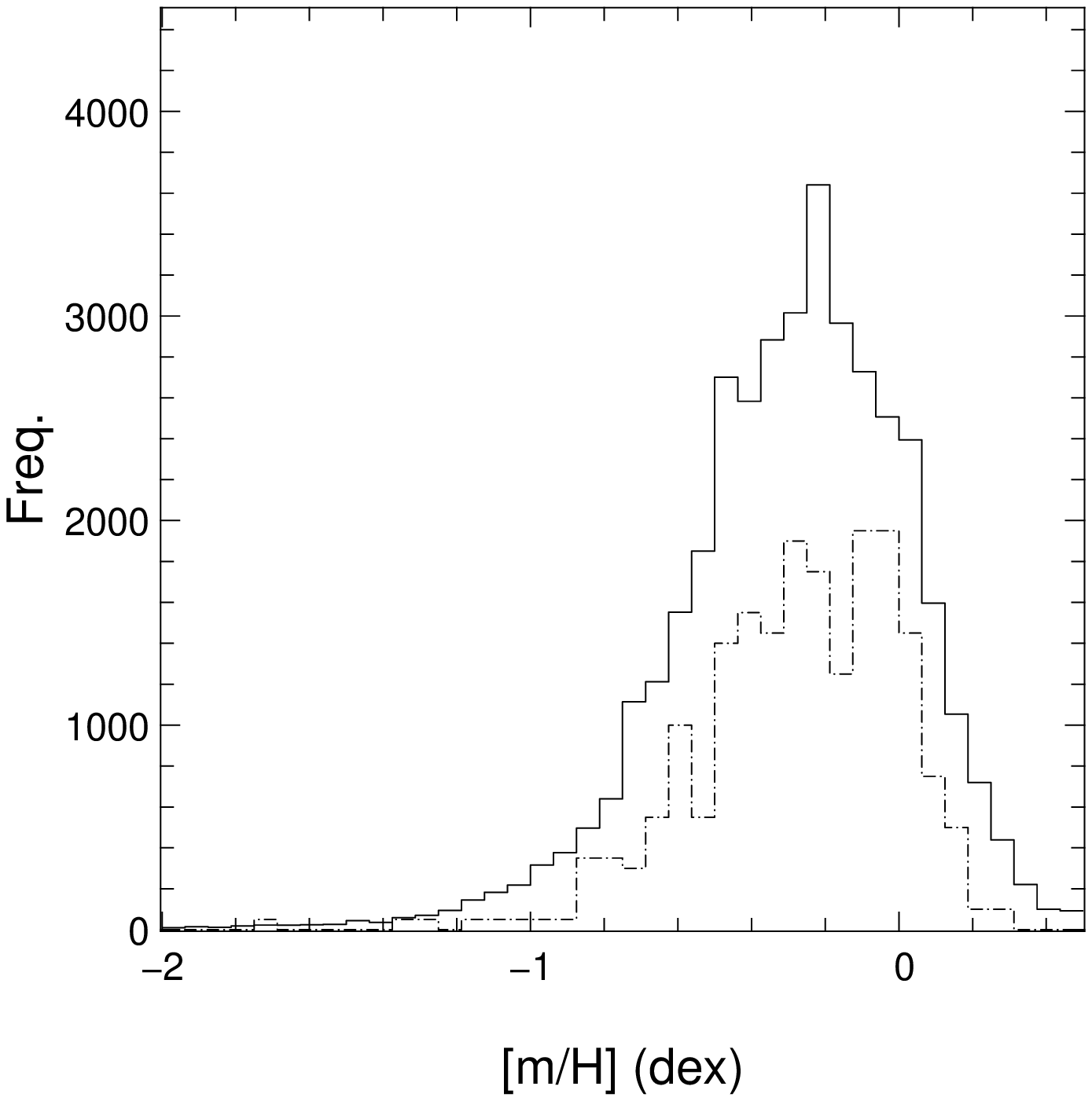}
\caption{Histograms of the distribution of the reference sample (dash-dotted
 histograms)  and  the  RAVE DR3  sample  (full  lines)  as a  function  of
 signal-to-noise,  $\teff$,   $\logg$  and  RAVE   $\mh$.  The  dash-dotted
 histograms are multiplied by a factor 50 to enhance their visibility. }
\label{f:sampledist}
\end{figure}

Our  external datasets  (or, `reference'  datasets) comprise  data  from the
Geneva-Copenhagen  Survey \citep[][hereafter  GCS]{GCS},  Elodie and  Sophie
high resolution observations from the Observatoire de Haute Provence, Asiago
echelle observations,  and spectra  obtained with the  ANU 2.3m  facility in
Siding Spring.  The targeted stars are chosen to cover the possible range of
signal-to-noise    conditions    and    stellar   atmospheric    conditions.
Figure~\ref{f:sampledist} presents the  distributions of the reference stars
as a function of signal-to-noise S2N, $\teff$ $\logg$ and $\mh$ compared the
the RAVE DR3 distributions.  While for $\mh$ the distribution ressembles the
distribution of the  data release, the distribution of  $\logg$ shows a lack
of  giant stars  that  translates to  a  reduced peak  at temperature  below
$5\,000$~K compared to  the full DR3 sample. This is due  to the GCS sample,
our primary source  of reference stars, that contains F and  G dwarfs and no
giants.  For the  S2N distribution, we chose to  sample almost uniformly the
RAVE S2N interval, top  left panel of Fig.~\ref{f:sampledist}, which enables
us to verify that signal-to-noise does  not impact the quality of our radial
velocities (see below).

\begin{figure}[hbtp]
\centering
\includegraphics[width=7cm]{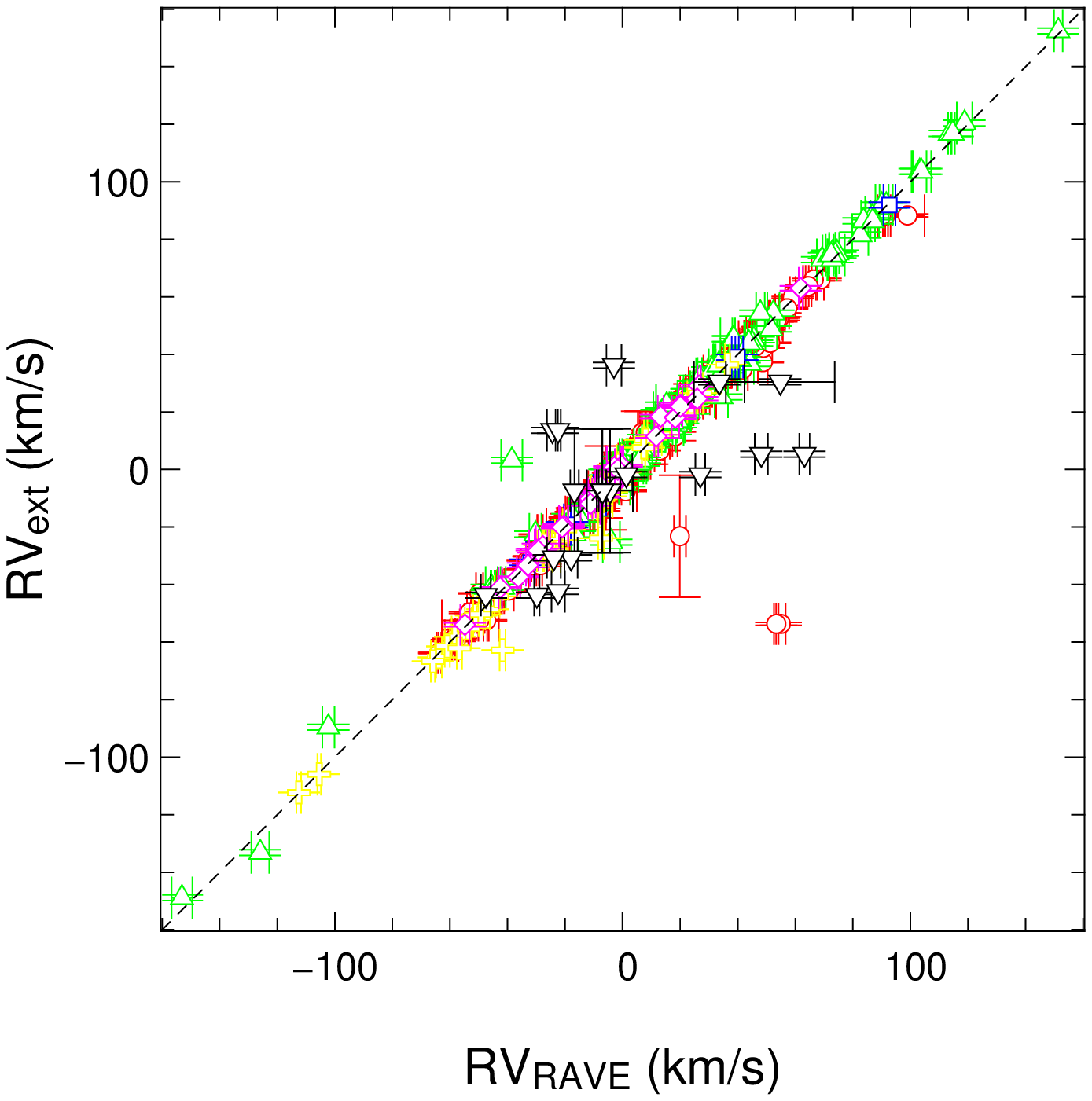}
\includegraphics[width=7cm]{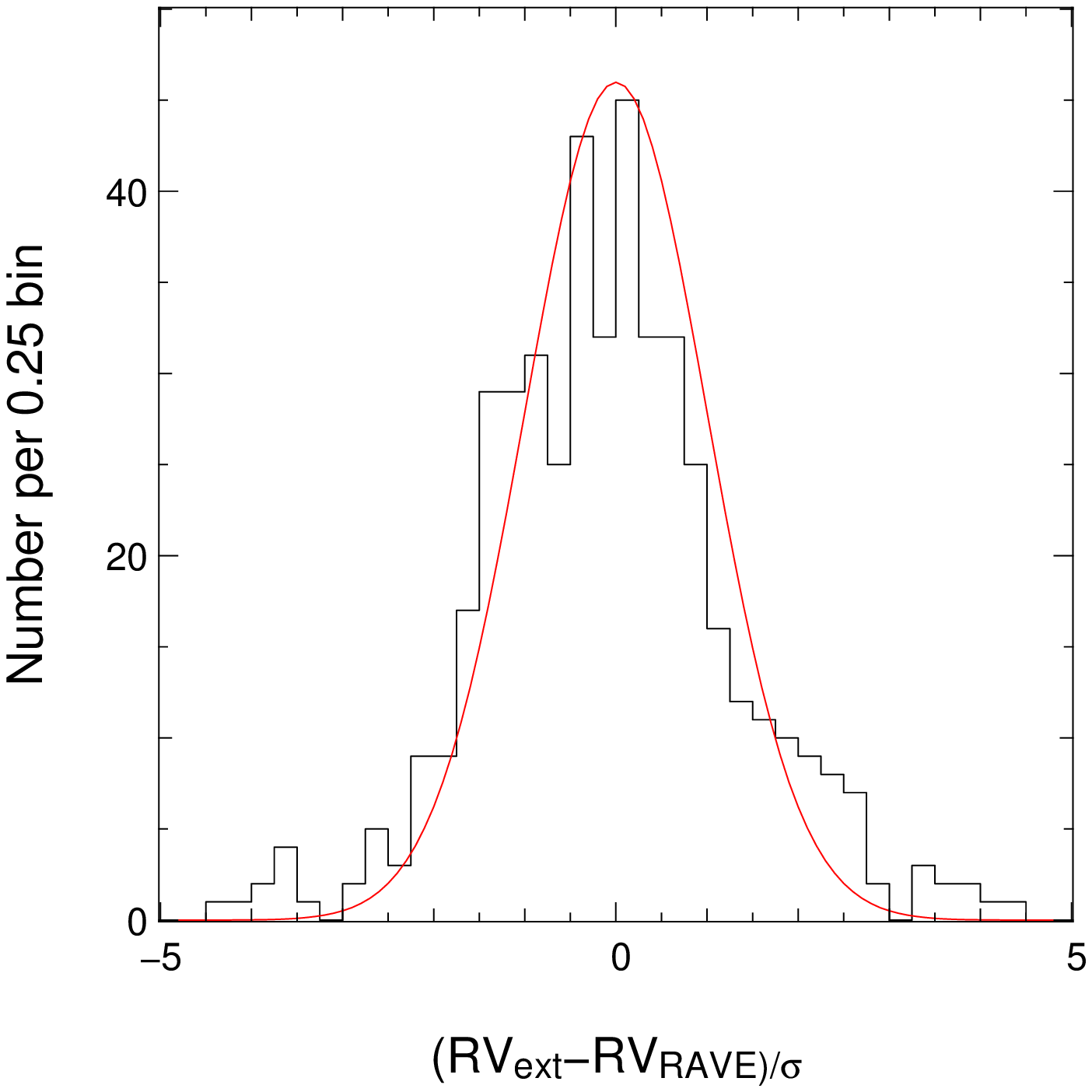}
\caption{Comparison of  RAVE radial velocities to external  sources.  Left :
 $RV_{\rm RAVE}$  vs.  $RV_{\rm  ext}$ for all  the different  sources: GCS
 (red circles),  ANU 2.3m (green triangles), Elodie  (blue squares), Sophie
 (yellow  crosses), and  Asiago  echelle spectra  (magenta diamonds).   The
 black  downwards  triangles  are  stars identified  as  binaries.   Right:
 distribution of the radial  velocity differences divided by the associated
 errors.   The red  curve is  a Gaussian  distribution with  zero  mean and
 $\sigma=1$.}
\label{f:RVcomp}
\end{figure}

A  comparison of the  radial velocities  obtained by  RAVE and  the external
datasets is  presented in  Figure~\ref{f:RVcomp}, while the  detailed values
for the comparison for each sample can be found in Table~\ref{t:external}.

\begin{table}[hbtp]
\centering
\caption{Global properties  of the comparison  of RAVE radial  velocities to
 external datasets for  stars observed during the second  and third year of
 the  program. $\Delta RV$  is defined  as $\Delta  RV=RV_{\rm ext}-RV_{\rm
   RAVE}$.  The mean deviations  and standard deviations are computed using
 a sigma  clipping algorithm.  The second column  gives the number  of data
 points  used  to  compute the  mean  and  $\sigma$  while the  numbers  in
 parenthesis are  the total number of  stars in the sample  ($N_1$) and the
 number of unique  objects ($N_2$).  The last two  lines are obtained after
 correcting each dataset for the mean deviation.}
\begin{tabular}{l r l c c}
\hline
\hline
Reference & $N$ & ($N_1$,$N_2$) & $\langle\Delta RV\rangle$ & $\sigma(\Delta RV)$\\
dataset   &    & & $\!\kms$ & $\!\kms$\\
\hline
GCS & 224 &(285,162) & -0.28 & 1.76 \\
Sophie & 35& (37,34) & -0.77&  1.62  \\
Asiago & 30 &(30,25)& 1.08 & 1.45 \\
Elodie & 6 &(9,9) & -0.63 & 0.36\\
2.3m & 76 &(125,74)& 0.87 & 2.39 \\
All & 373 &(486,304) & -0.22 & 2.72\\
All but GCS & 142 &(201,142) & 0.50 & 2.16\\
\hline
\multicolumn{5}{c}{mean deviation corrected}\\
\hline
All & 142 &(486,304) & -0.18 & 2.66 \\
All but GCS & 127 &(201,142) & 0.10 & 1.96\\
\hline
\end{tabular}
\label{t:external}
\end{table}

With the new version of the  pipeline, we find no significant difference for
the mean radial velocity difference compared to DR2. The values for the mean
difference   and   its  dispersion   are   consistent   between  these   two
releases. From  the right panel  of Figure~\ref{f:RVcomp} one sees  that the
distribution  of the  radial  velocity difference  divided  by the  internal
errors is wider than a normal distribution : its dispersion is 1.37$\sigma$.
We can then  estimate the upper limit to the  external error contribution as
$\sigma_{\rm ext}\leq  0.9\kms$.  This is  an upper limit as  the zero-point
errors  of the  other  sources of  radial  velocity also  contribute to  the
measured $\sigma_{\rm ext}$ and are unknown.

\begin{figure*}[hbtp]
\centering
\includegraphics[width=7.5cm]{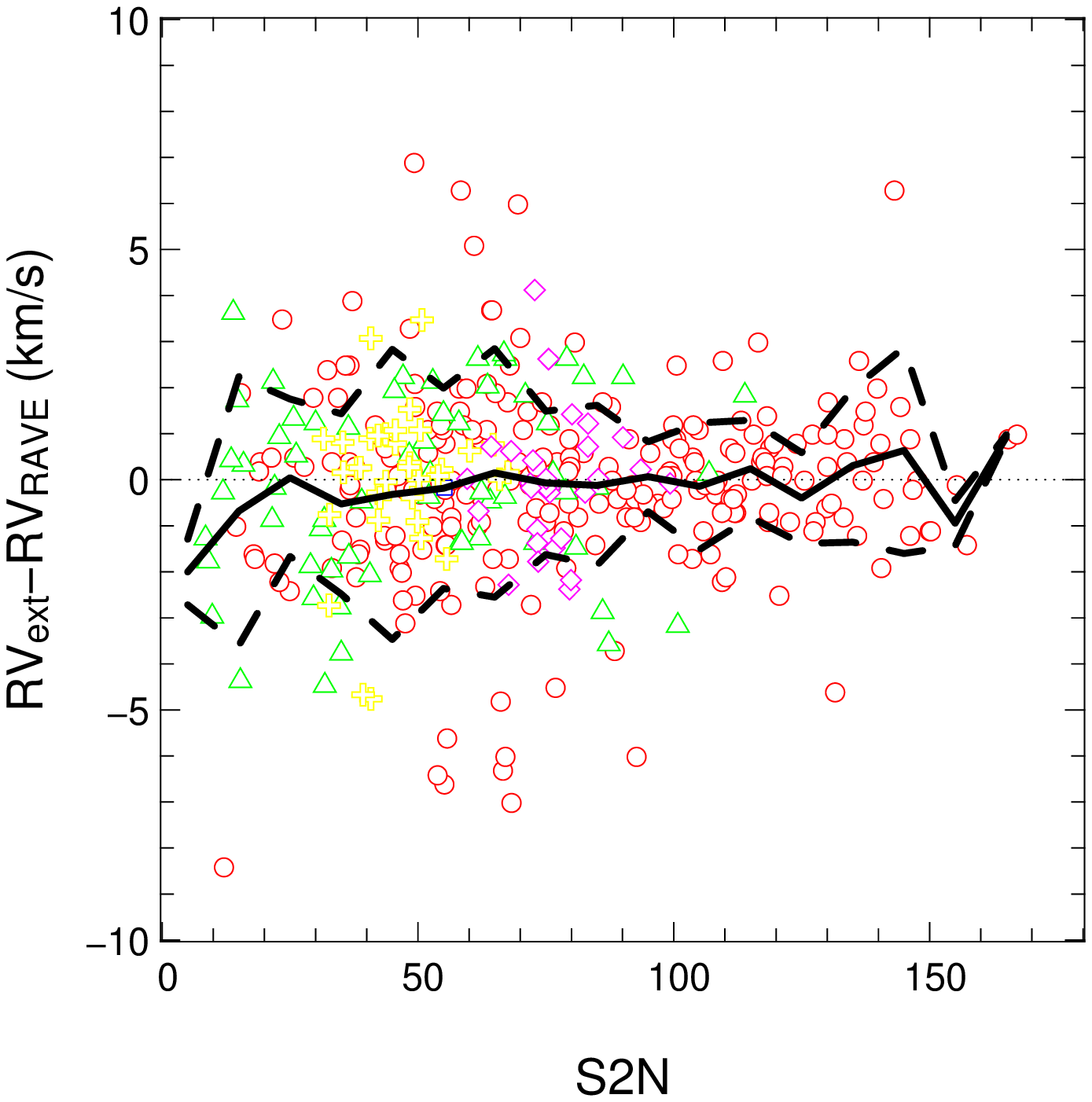}
\includegraphics[width=7.5cm]{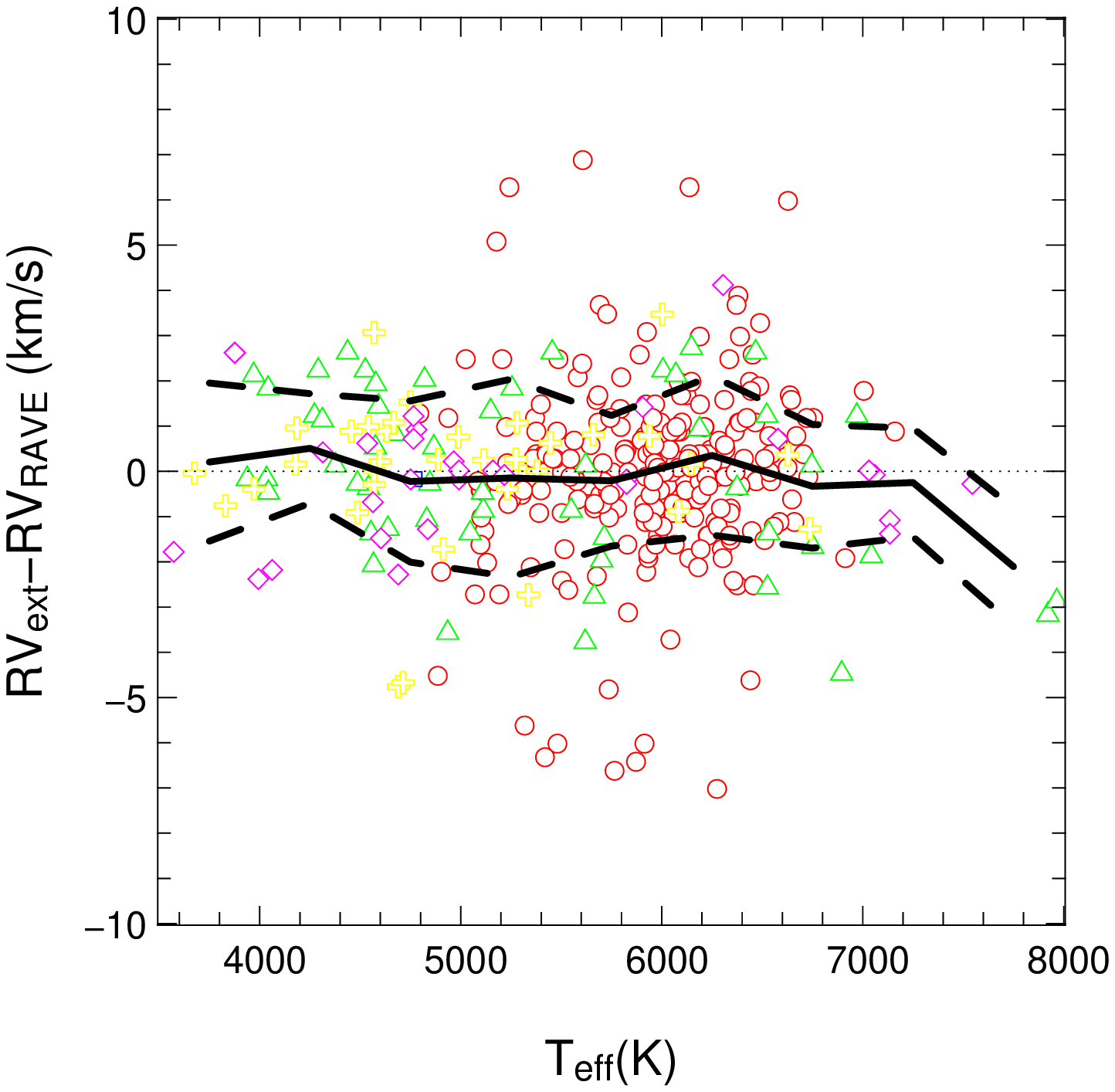}
\includegraphics[width=7.5cm]{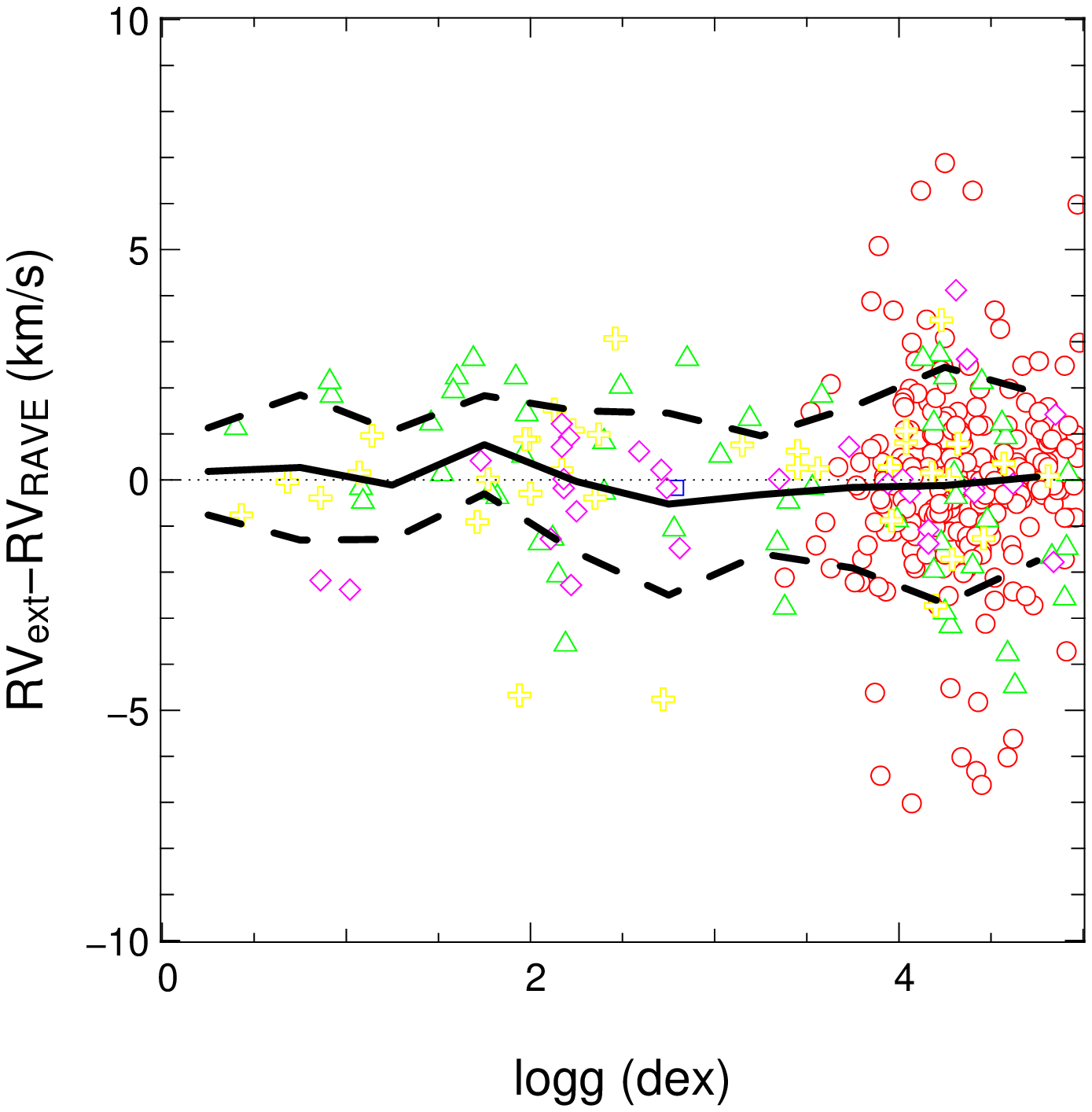}
\includegraphics[width=7.5cm]{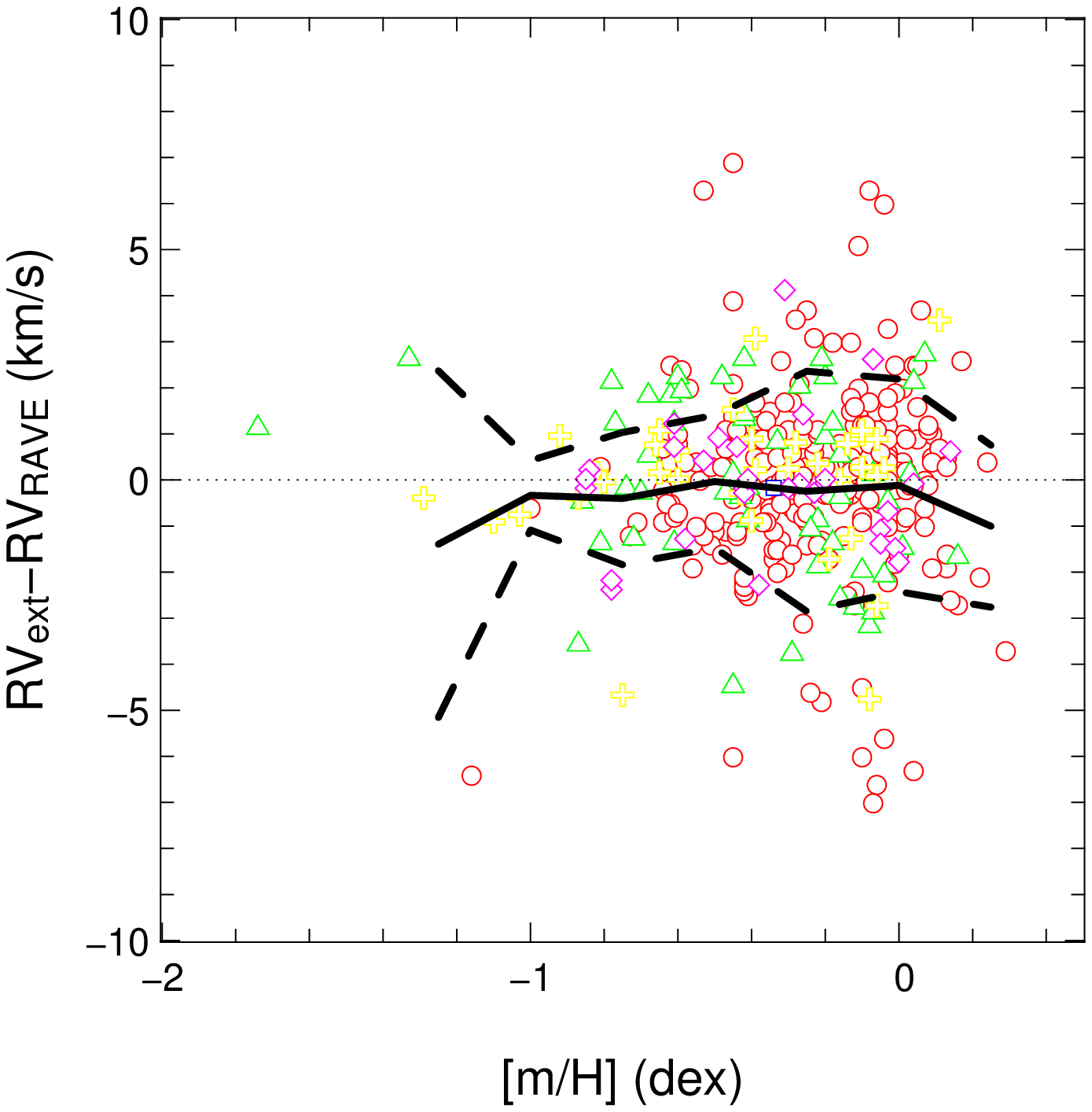}
\caption{Radial velocity  difference between  the RAVE observations  and the
 external  sources as  a function  of  the signal-to-noise  ratio S2N  (top
 left), effective temperature (top right), $\logg$ (bottom left), and $\mh$
 (bottom   right)   of   the   RAVE  observation.    The   symbols   follow
 Figure~\ref{f:RVcomp} while the full  and dashed thick lines represent the
 mean and dispersion  about the mean of the  radial velocity difference per
 interval of 10 in S2N, 500~K  in $\teff$, 0.5~dex in $\log g$, or 0.25~dex
 in $\mh$.}
\label{f:RVsnr}
\end{figure*}

The dependency of the radial velocity difference on signal-to-noise ratio is
weak, as  can be seen from  Figure~\ref{f:RVsnr} (top left  panel). The mean
difference is  consistent with  no offset,  at all S2N  levels.  There  is a
slight tendency for an increase in dispersion at low S2N, but the dispersion
values  remain  very  well-behaved  ($\sigma\sim1.2\kms$  at  S2N$>$100  and
$\sigma\sim2.0\kms$ for  S2N$<$40).  In  addition, no strong  variation with
$\logg$,  $\teff$, or  $\mh$ is  seen, indicating  that our  radial velocity
solution is stable as a function of stellar type.

\subsection{Stellar atmospheric parameters}
\label{s:parameters}

During the  second and  third years of  its program, RAVE  observed $2\,266$
stars  more  than once;  $1\,917$  stars  were  observed twice,  $256$  were
observed three times, and $93$  were observed four times.  $1\,391$ of these
stars have  more than one measurement  of stellar parameters.   We use these
re-observations to estimate the stability and error budget for our estimated
stellar atmospheric parameters. These parameters are the parameters from the
synthetic template spectrum used to compute the final radial velocity.  This
template  is constructed using  a penalized  chi-square algorithm  where the
template spectrum is a weighted sum  of the synthetic spectra of the library
of  \citet{munari}.   The  weights  of   the  best-match  are   obtained  by
minimization  of a  $\chi^2$ plus  additional constraints  (weights  must be
positive and smoothly distributed  in the atmospheric parameters space). The
algorithm is described in Paper II.

\subsubsection{Internal stability from repeat observations}
\label{s:repeat_param}

\begin{figure*}[hbtp]
\centering
\includegraphics[height=4.5cm]{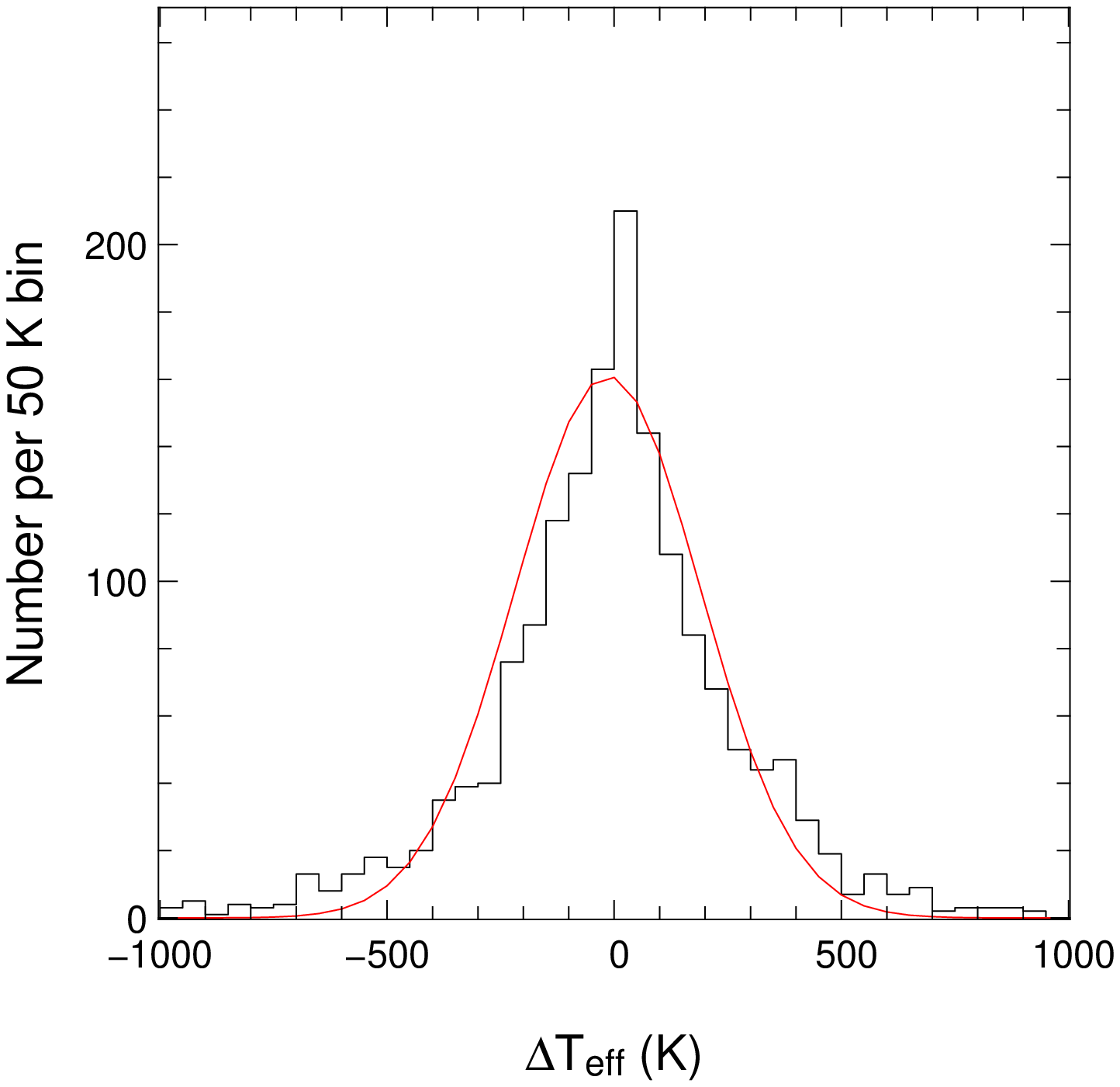}
\includegraphics[height=4.5cm]{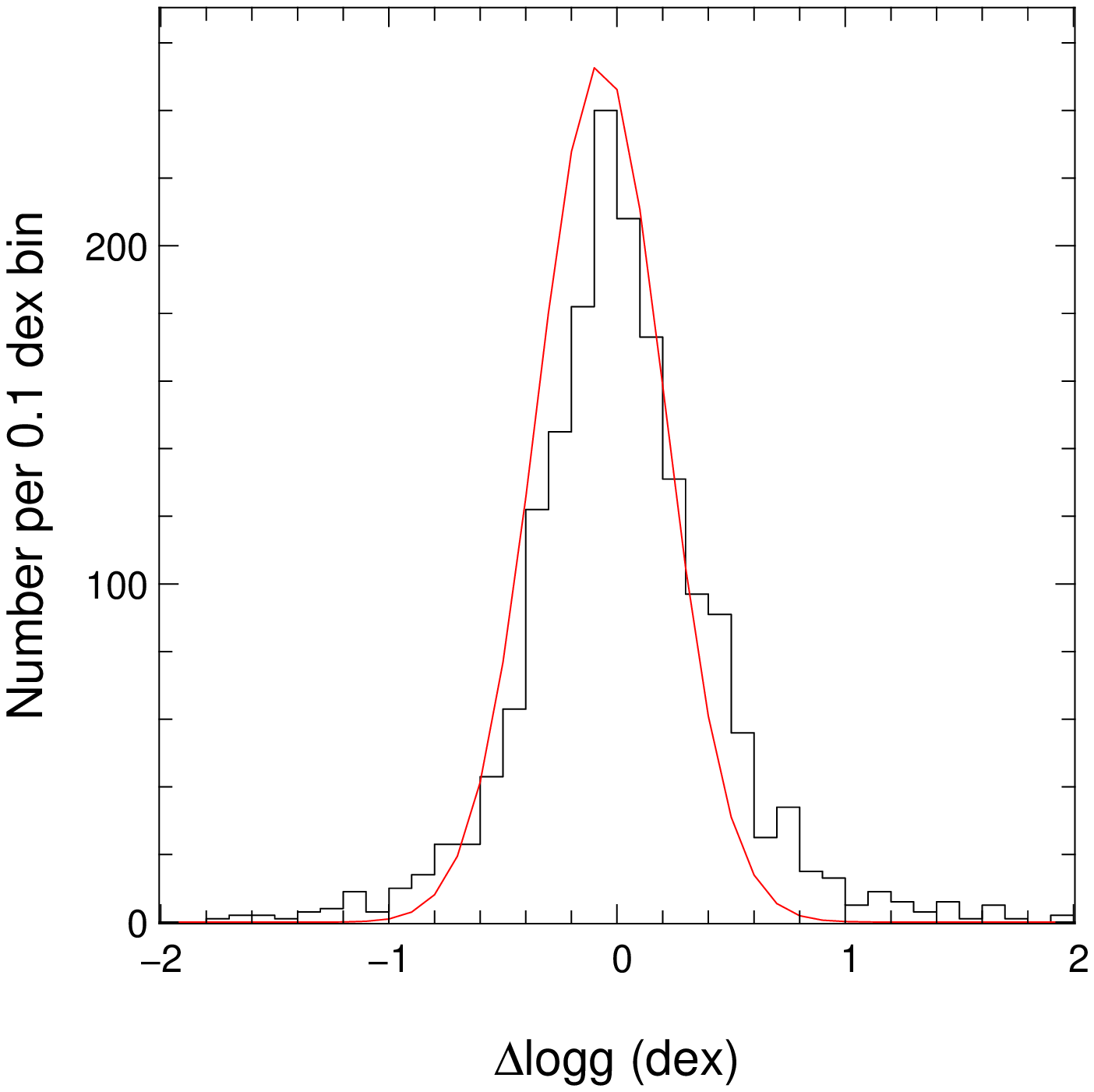}
\includegraphics[height=4.5cm]{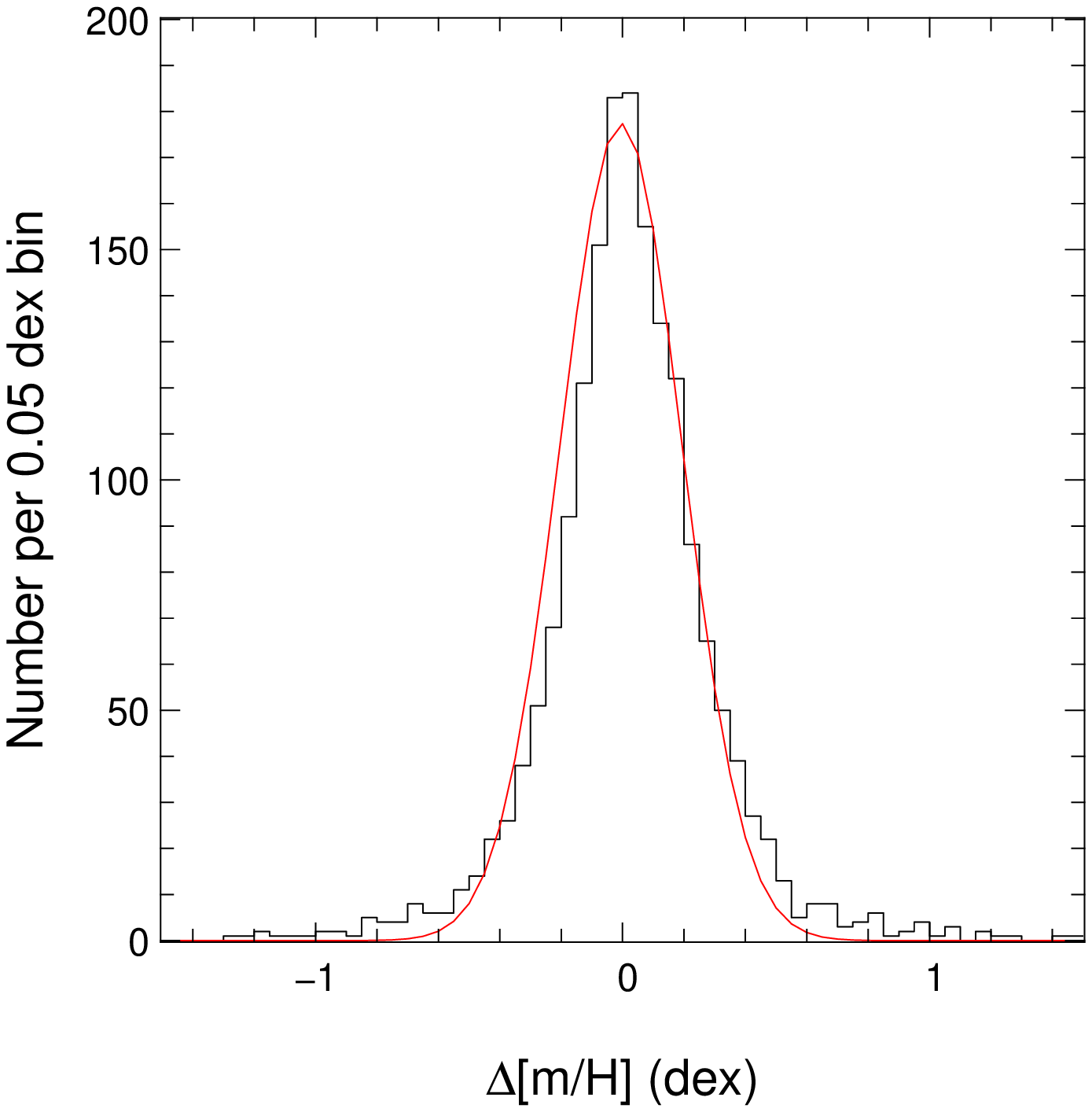}
\includegraphics[height=4.5cm]{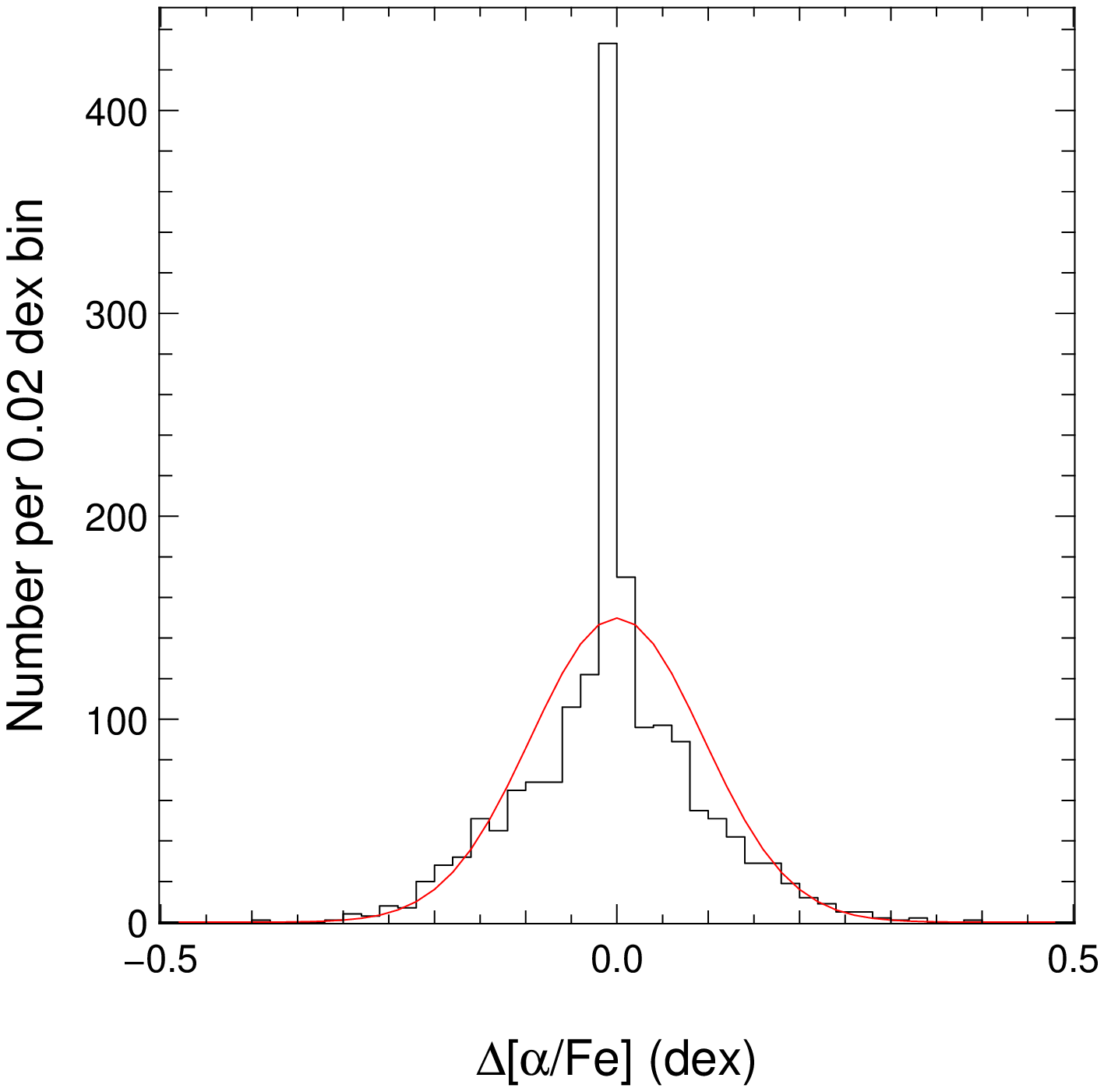}
\includegraphics[height=4.5cm]{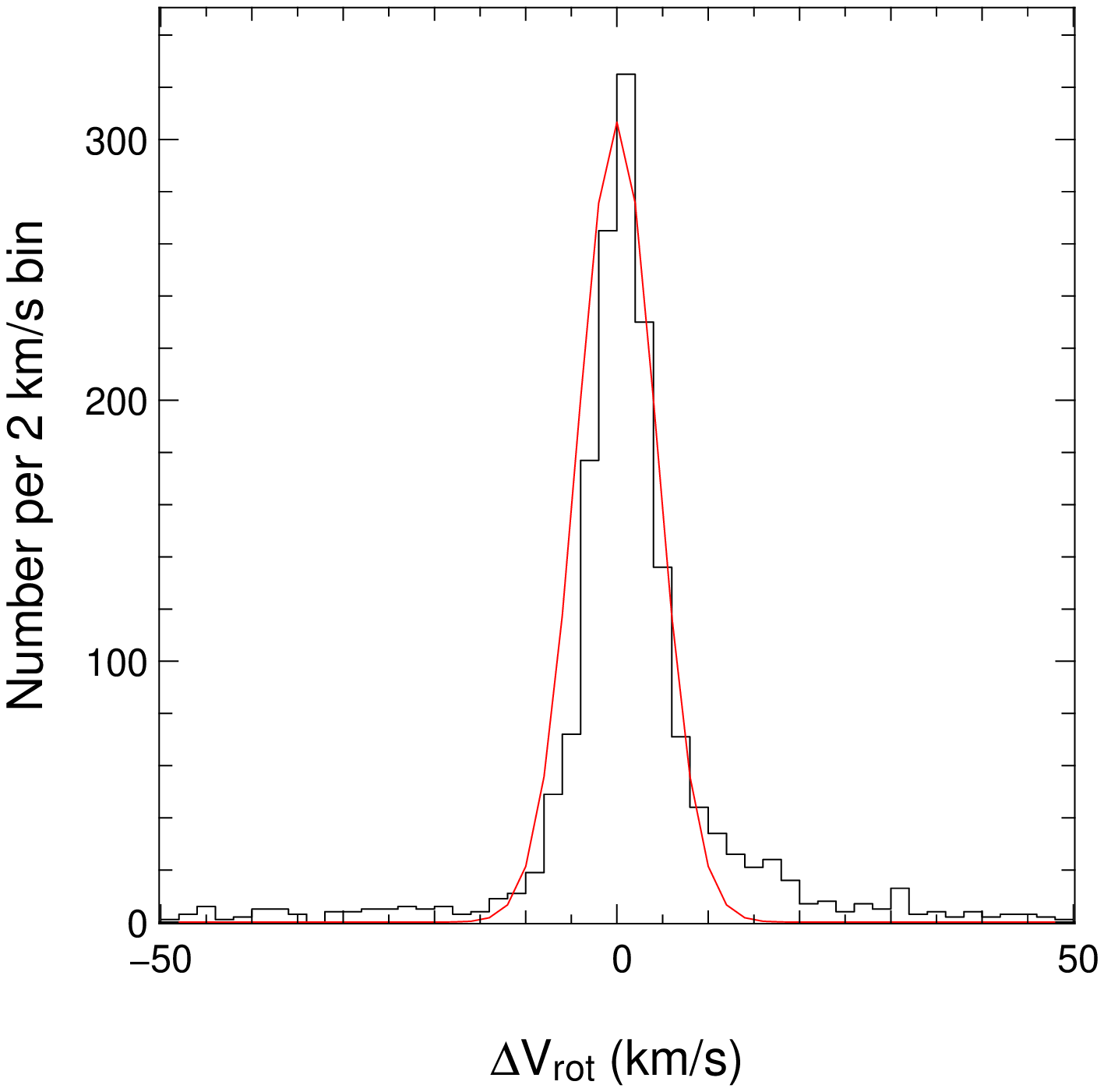}
\caption{Distributions of the difference in the measured stellar atmospheric
 parameters in re-observed  targets. The spectrum with highest  \snr\ for a
 given star  is used as  reference. The red  lines in the  different panels
 correspond to  a Gaussian function whose parameters  (mean and dispersion)
 are   obtained   using  an   iterative   sigma-clipping  algorithm   (see
 Table~\ref{t:reobs_param}).}
\label{f:reobs_param}
\end{figure*}

\begin{figure*}[hbtp]
\centering
\includegraphics[height=4.5cm]{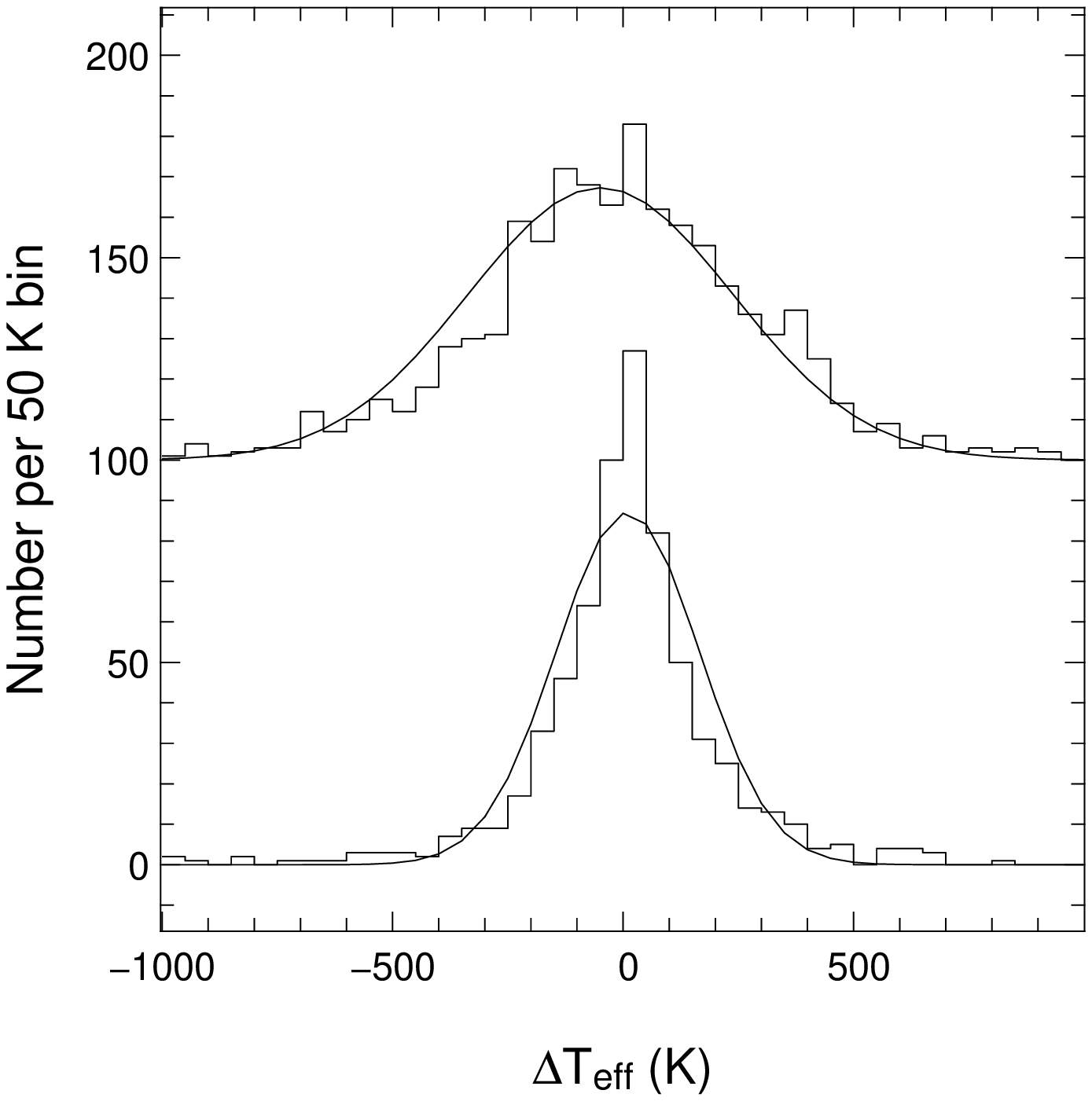}
\includegraphics[height=4.5cm]{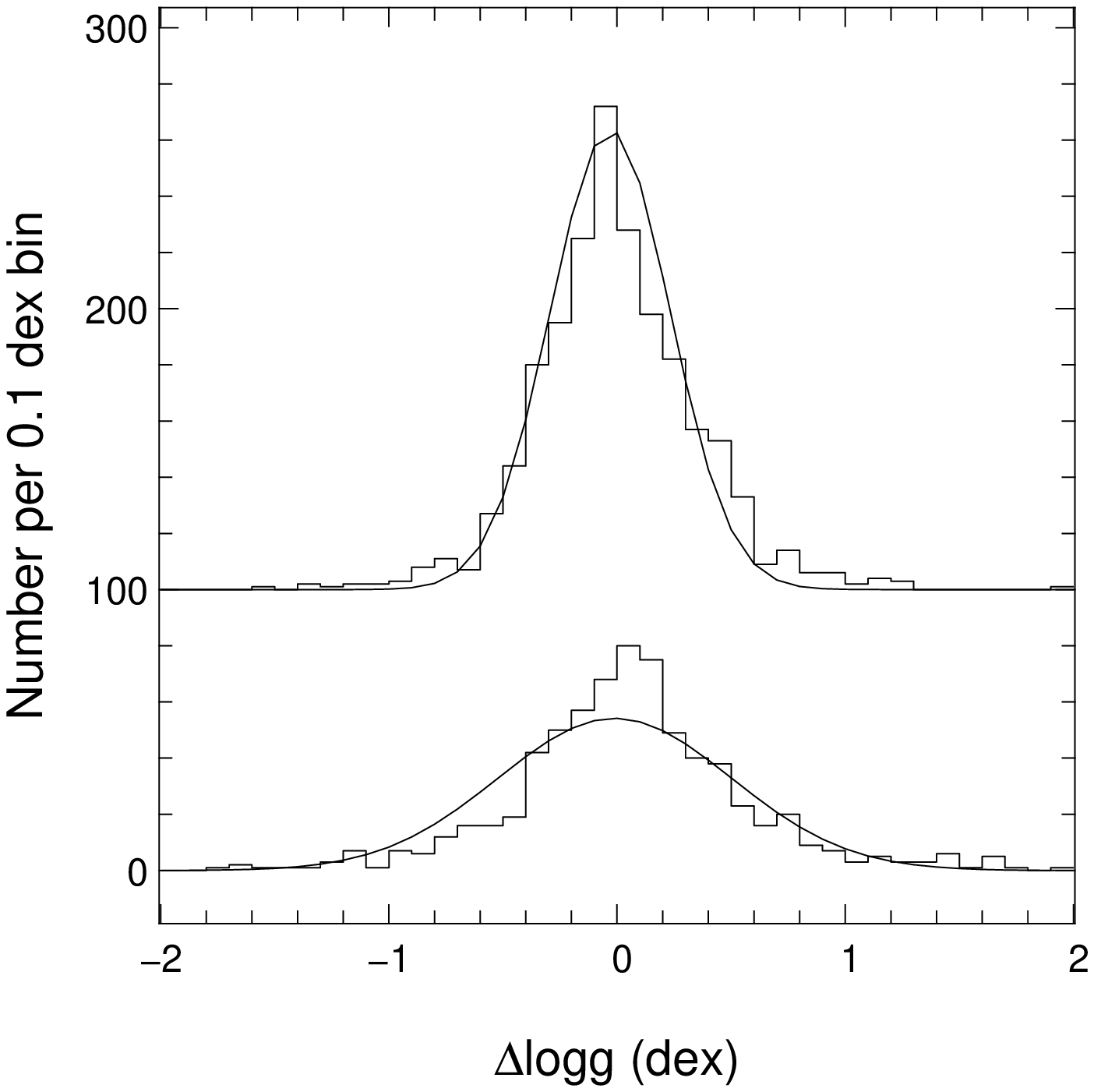}
\includegraphics[height=4.5cm]{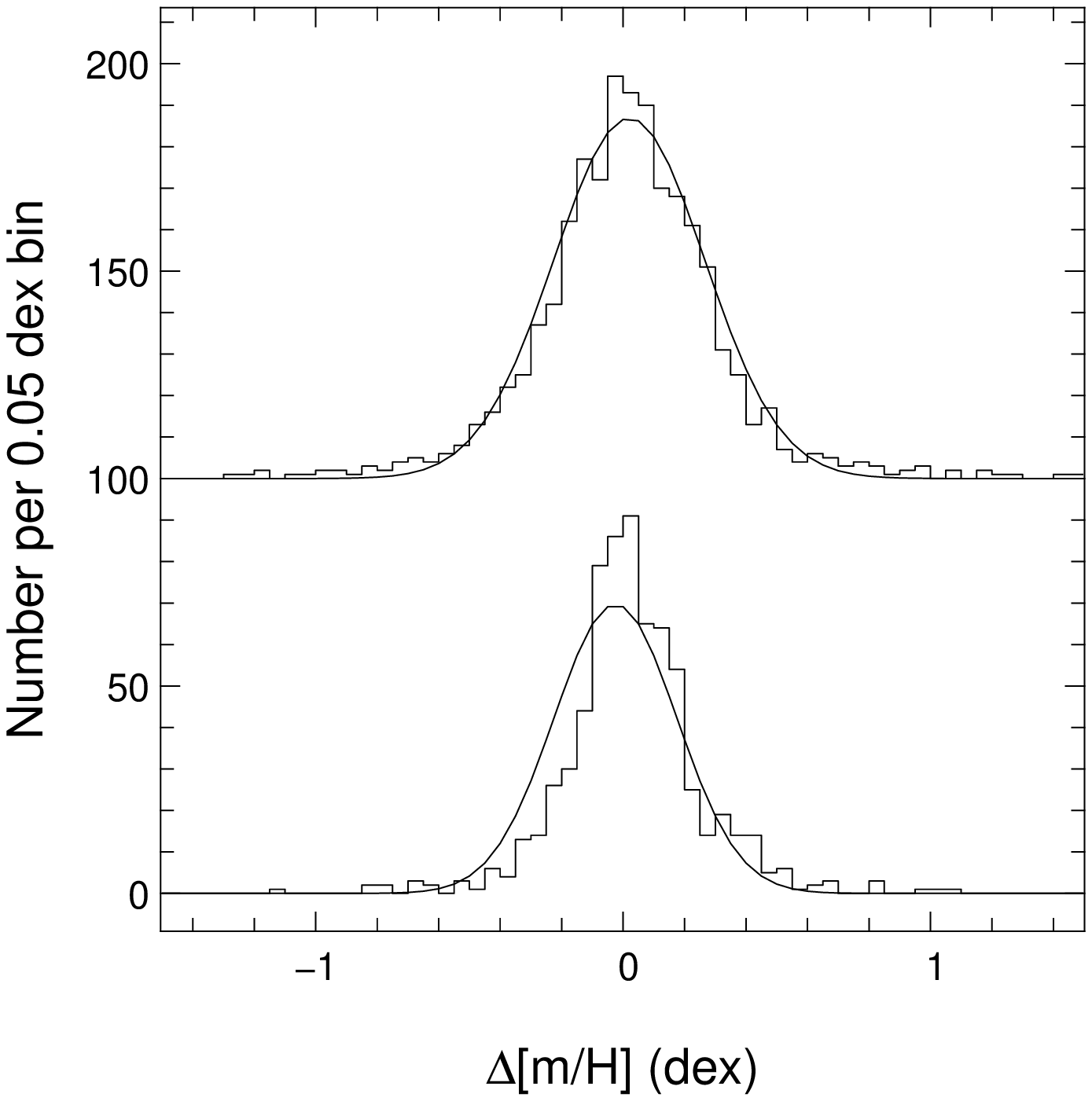}
\includegraphics[height=4.5cm]{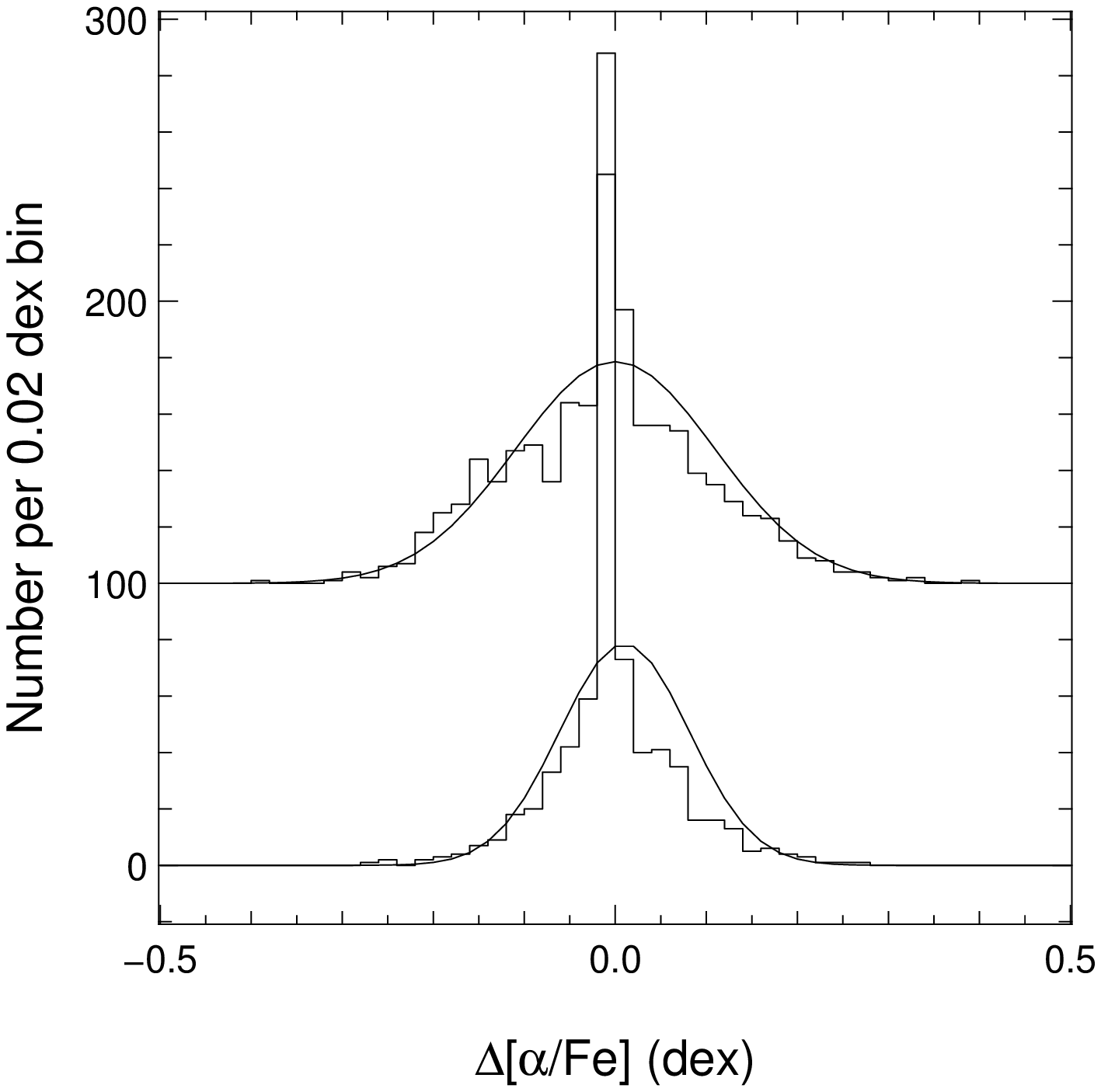}
\includegraphics[height=4.5cm]{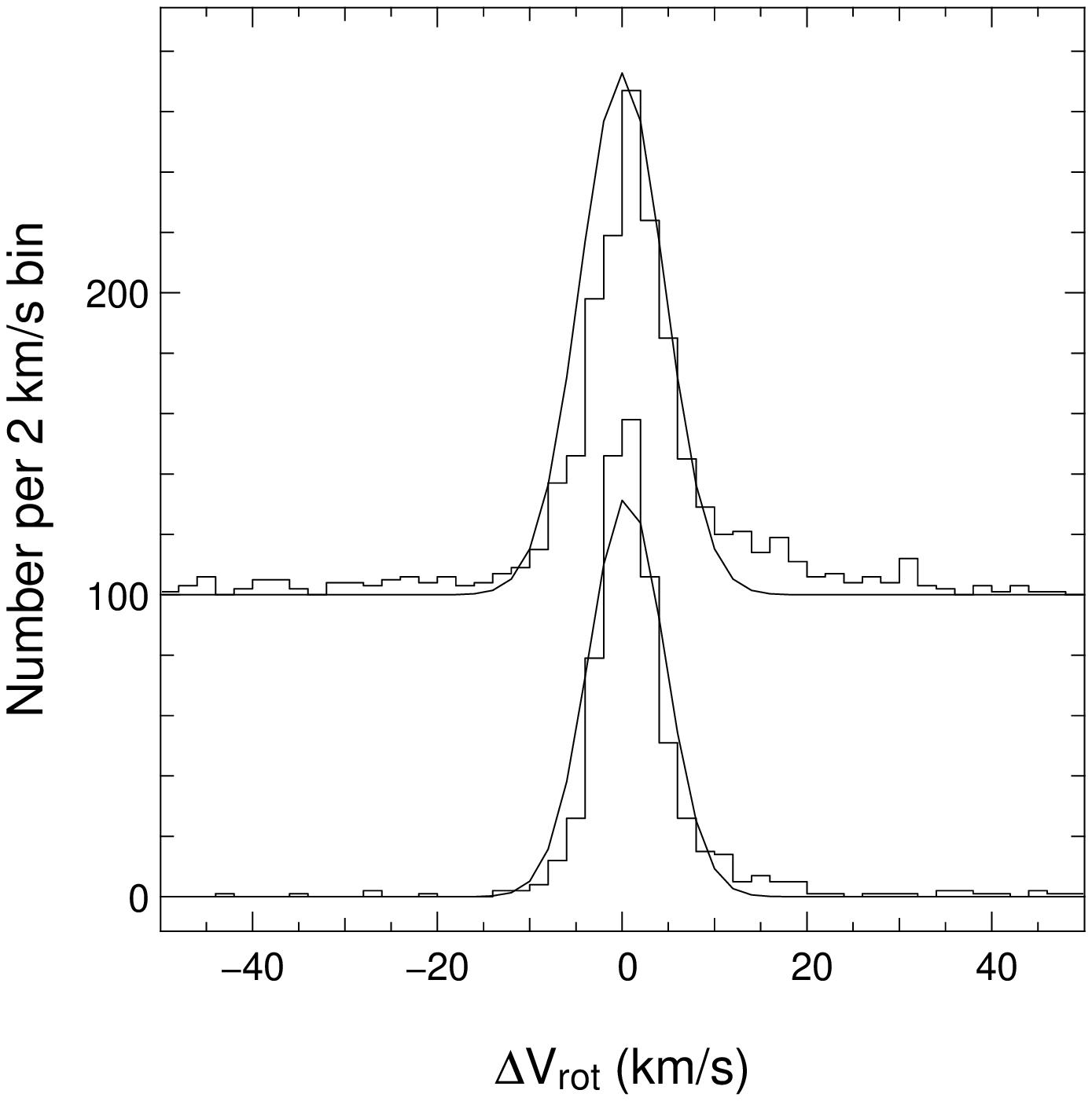}
\caption{Same as  Fig.~\ref{f:reobs_param} but for the  sub-samples of dwarf
 stars  (top curves)  and giant  stars  (bottom curves).   The samples  are
 selected according  to $\logg$ using the  separating line $\logg=3.5$~dex.
 The histograms for  dwarf stars are shifted upwards by  100 counts per bin
 for clarity.}
\label{f:reobs_param_split}
\end{figure*}

As  a first  step, we  estimate  the stability  from the  difference in  the
measured parameters using,  for a given star, the  spectrum with the highest
\snr\  as  the  reference  measurement.   The distribution  of  the  stellar
parameter differences $\Delta P$, where $P$ may stand for any of the stellar
atmospheric  parameters considered,  is shown  in Figure~\ref{f:reobs_param}
while Figure~\ref{f:reobs_param_split} presents the distributions for dwarfs
and giants  stars respectively.  The red  curves in each  panel are Gaussian
functions whose parameters (mean  and standard deviation) are obtained using
an iterative sigma-clipping algorithm.   The corresponding mean and standard
deviation for each parameter  are reported in Table~\ref{t:reobs_param}. For
all  parameters, the  mode of  the  distributions is  consistent with  zero,
indicating good  stability of  our atmospheric parameter  measurements.  The
average internal error for the  atmospheric parameters can be estimated from
the  standard deviation.   For $\teff$  one obtains  200~K and  0.3  dex for
$\logg$, while the  $\mh$ and $\alp$ distributions show  a dispersion of 0.2
and 0.1~dex  respectively.  These values must be  regarded as underestimates
of  the true  errors as  they do  not include  external errors  such  as the
inadequacy  of  the  template   library  in  representing  real  spectra  or
variations in  the abundances  of the chemical  species with respect  to the
solar abundances (using but one value of the $\alpha$-enhancement).

In Fig.~\ref{f:reobs_param}  the distributions of $\teff$,  $\mh$ and $\alp$
are relatively symmetric although  not Gaussian. The distribution of $\logg$
is less  symmetric and that  of $\vrot$ is  very skew.  Since  our reference
measurements are the spectra with  the highest \snr, symmetry indicates that
there  is no  strong bias  in the  atmospheric parameter  estimation  as one
reduces the signal-to-noise ratio: a  systematic effect with the \snr\ would
imply that as  one lowers the \snr\ the measured  parameters would be either
higher or lower than the reference value.

For $\vrot$,  a systematic  effect is  likely. As one  lowers the  \snr, the
wings of  the spectral lines become  more affected by the  noise, making the
lines appear  narrower, hence  mimicking a lower  $\vrot$.  The  same effect
applies to $\logg$.

\begin{table}[hbtp]
\centering
\caption{Standard RAVE errors on  stellar atmospheric parameters from repeat
  observations  for the  full sample  of  re-observed stars.   The mean  and
  standard  deviations  are   computed  using  an  iterative  sigma-clipping
  algorithm and $\Delta P=P_{\rm ref}-P_{\rm star}$.}
\begin{tabular}{r l c c}
\hline
\hline
\multicolumn{2}{c}{$P$} & $\langle \Delta P\rangle$ & $\sigma_{P}$ \\
\hline
$\teff$ & (K) & -7 & 204 \\
$\logg$ & (dex) & 0.0 &  0.3 \\
$\mh$ & (dex) & 0.0 & 0.2 \\
$\alp$ & (dex) & 0.0 & 0.1 \\
$\vrot$ & ($\,\kms$) & 0.3 & 4.3 \\
\hline
\end{tabular}
\label{t:reobs_param}
\end{table}

Internal  errors  on  the  atmospheric  parameters depend  on  the  physical
condition  of the  star, $\logg$  being  better constrained  for giants  and
$\teff$ for cool stars. The internal  errors, as defined in Paper II, depend
mostly on the  algorithm used and the grid spacing  of the synthetic spectra
for these two  parameters.  Neither has been modified in  the new version of
the  pipeline.  Hence,  the  internal errors  for  the different  parameters
remain unchanged and upper limits  for these errors are presented in Fig.~19
in Paper  II. However, using re-observed  RAVE stars, one is  able to refine
this estimate based on the scatter of the atmospheric parameter measurements
in  various  $\teff$ and  $\logg$  intervals.  These  refined estimates  are
presented in Table~\ref{t:reobs_refined}  where a smooth-averaging procedure
is used  to compute the dispersion at  a given grid point.  Only grid points
with three or more repeated observations are given in the table.

\begin{table}[hbtp]
\centering
\caption{Dispersion  in $\teff$  (K), $\logg$  (dex)  and $\mh$  (dex) as  a
 function  of  $\teff$  and  $\logg$.   The  dispersions  are  computed  by
 smooth-averaging sigmas in individual  grid points. Only grid points where
 three or more repeated objects are present are quoted.}
\begin{tabular}{r l p{0.8cm} p{0.8cm} p{0.8cm} p{0.8cm} p{0.8cm} p{0.8cm} p{0.8cm} p{0.8cm} p{0.8cm} p{0.8cm} p{0.8cm}}
\hline
\hline
\multicolumn{2}{c}{$\teff $(K)$ \backslash \logg$~(dex)} & 0 & 0.5 & 1 & 1.5 & 2 & 2.5 & 3 & 3.5 & 4 & 4.5 & 5\\
\hline
 & ($\teff$)& 30  & 40 & 50 & 50 & 80 & 180 & 500 & 200 & 100 & 100 & 110\\
4000 & ($\logg$) & 0.07 & 0.16 & 0.19 & 0.21 & 0.43 & 0.29 & 0.81 & 0.62 & 0.17 & 0.16 & 0.051\\
& ($\mh$) & 0.06 & 0.08 & 0.08 & 0.09 &0.14 & 0.42 & 0.13 & 0.26 & 0.10 & 0.07 & 0.11\\
\hline
& ($\teff$)& & 50 & 60 & 60 & 50 & 50 & 60 & 160 & 120 & 70 & 50\\
4500 & ($\logg$) & & 0.12 & 0.18 & 0.20 & 0.18& 0.19 & 0.22 & 0.34 & 0.16 & 0.17 &
0.06\\
& ($\mh$)& & 0.07 & 0.09 & 0.08 & 0.07 & 0.07& 0.08 & 0.09 & 0.09 & 0.09 & 0.06\\
\hline
& ($\teff$)& & & & 180 & 70 & 70 & 90 & 110 & 100 & 80 & 50\\
5000 & ($\logg$) & & & & 0.58 & 0.17 & 0.21 & 0.24 & 0.20 & 0.16 & 0.14 & 0.06\\
& ($\mh$)&  & & & 0.28 & 0.07 & 0.09 & 0.09 & 0.09 & 0.08 & 0.07 & 0.05\\
\hline
& ($\teff$)& & & & & 600 & 200 & 180 & 190 & 130 & 120 & 90\\
5500 & ($\logg$) & & & & & 0.39 & 0.29 & 0.38 & 0.27 & 0.14 & 0.12 & 0.08\\
& ($\mh$)& & & & & 0.13 & 0.17 & 0.17 & 0.12 & 0.10 & 0.08 & 0.05\\
\hline
& ($\teff$)& & & & & & 850 & 300 & 180 & 110 & 120 & 100\\
6000 & ($\logg$) & & & & & & 0.98 & 0.89 & 0.28 & 0.15 & 0.14 & 0.08\\
& ($\mh$)& & & & & & 0.36 & 0.43 & 0.11 & 0.10 & 0.09 & 0.07\\
\hline
& ($\teff$)& & & & & & & 400 & 140 & 110 & 130 & 130\\
6500 & ($\logg$) & & & & & & & 1.13 & 0.18 & 0.14 & 0.16 & 0.08\\
& ($\mh$)& & & & & & & 0.35 & 0.13 & 0.10 & 0.10 & 0.08\\
\hline
& ($\teff$)& & & & & & & & 160 & 150 & 150 & 140\\
7000 & ($\logg$)& & & & & & & & 0.13 & 0.12 & 0.17 & 0.07\\
& ($\mh$)& & & & & & & & 0.08 & 0.10 & 0.12 & 0.12\\
\hline
& ($\teff$)& & & & & & & & 200 & 110 & 200 & 500\\
7500 & ($\logg$) & & & & & & & & 0.16 & 0.10 & 0.15 & 0.19\\
& ($\mh$)& & & & & & & & 0.11 & 0.08 & 0.11 & 0.21\\
\hline
\end{tabular}
\label{t:reobs_refined}
\end{table}

\subsubsection{Effect of the correlations between atmospheric parameters}
\label{s:correl}

\begin{figure*}[hbtp]
\centering
\begin{tabular}{p{1cm} c c c c}
& $\Delta \teff$ (K) & $\Delta \logg$ (dex) & $\Delta \mh$ (dex) &
 $\Delta \alp$ (dex)\\
\begin{sideways} \hspace{0.5cm} $\Delta V_{\rm rot}\ (\kms)$  \end{sideways}&
\includegraphics[width=3.5cm,height=3.5cm]{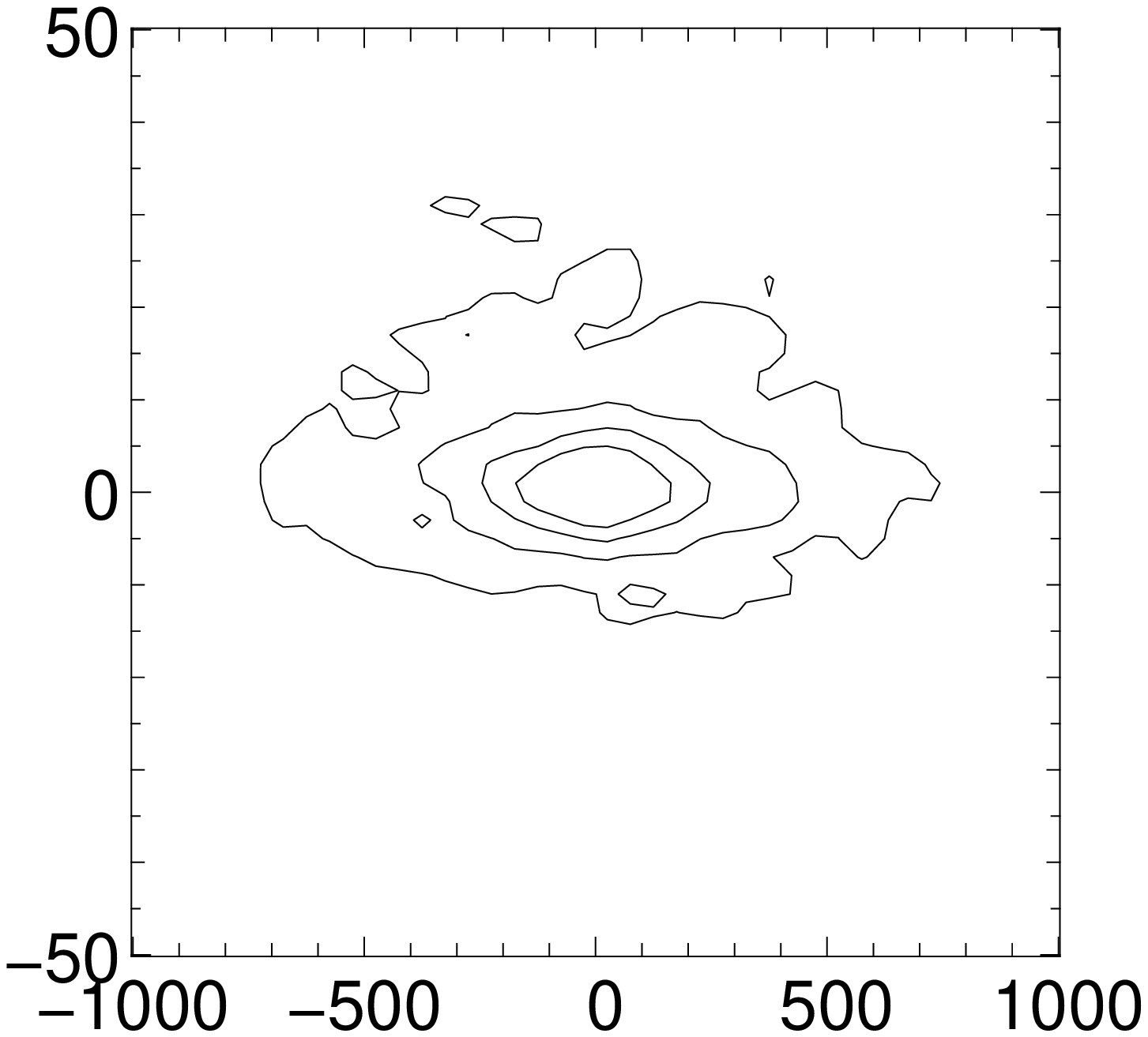}&
\includegraphics[width=3.5cm,height=3.5cm]{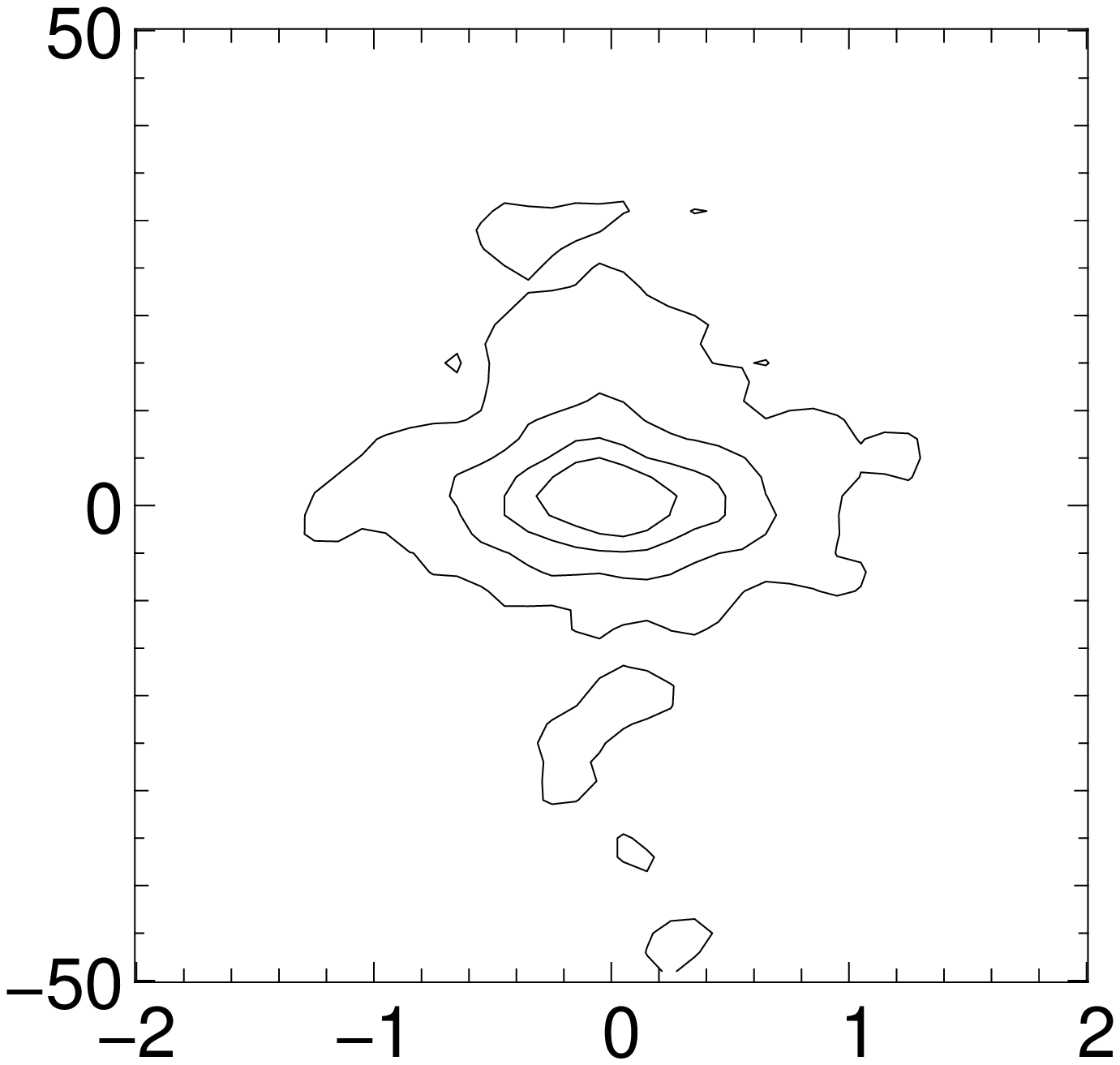}&
\includegraphics[width=3.5cm,height=3.5cm]{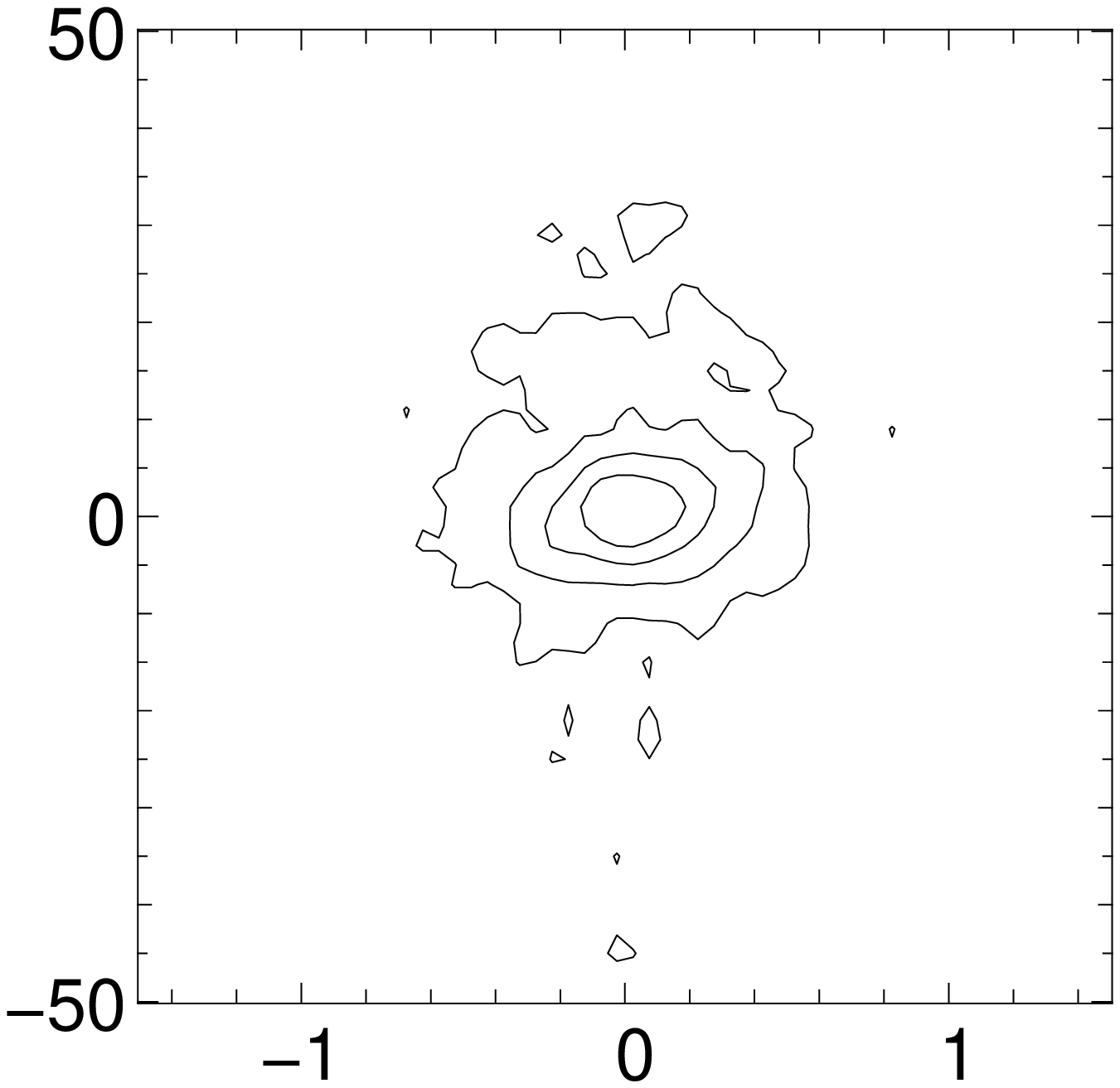}&
\includegraphics[width=3.5cm,height=3.5cm]{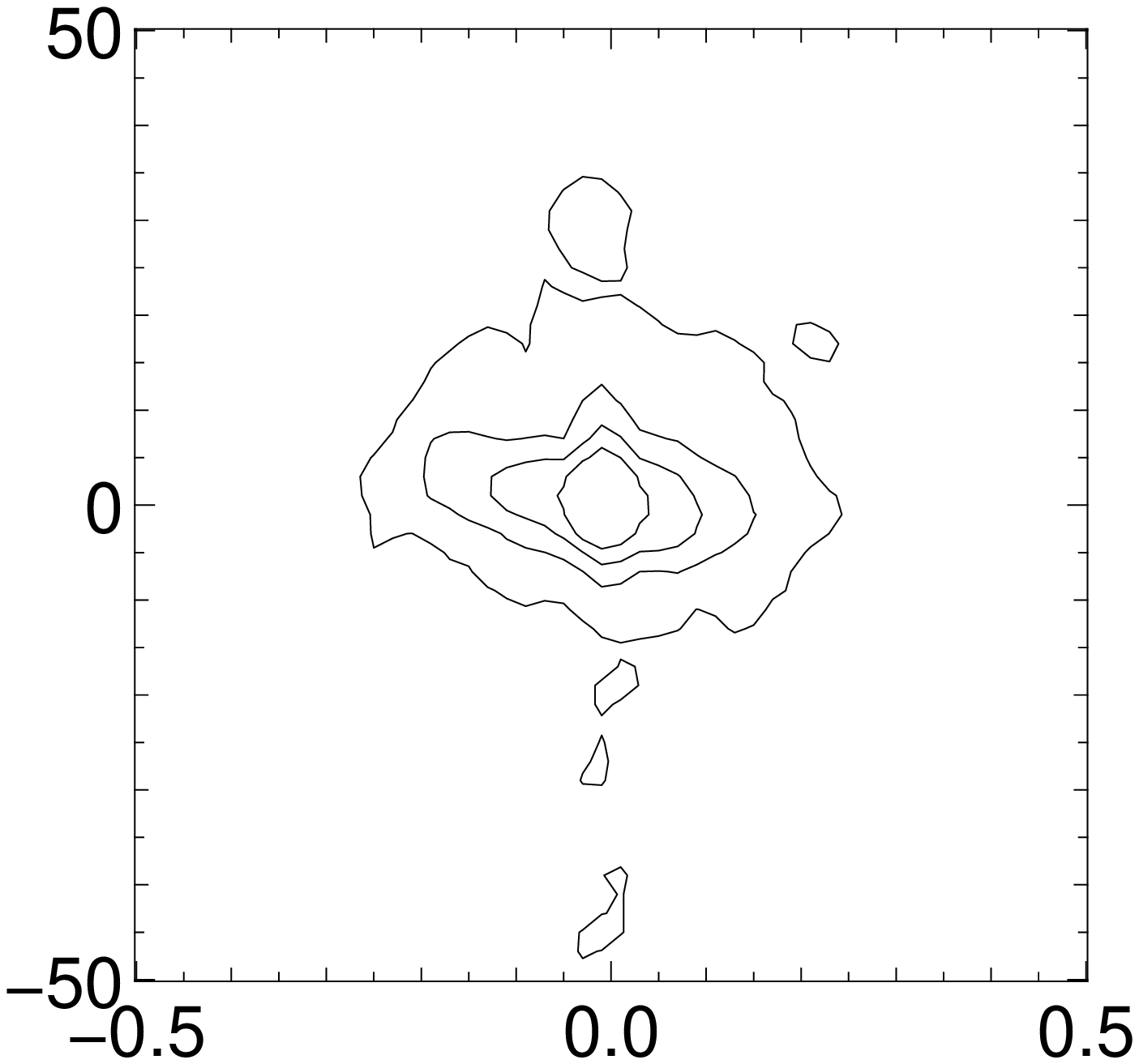}
\\
\begin{sideways} \hspace{0.5cm} $\Delta \alp$ (dex) \end{sideways}&
\includegraphics[width=3.5cm,height=3.5cm]{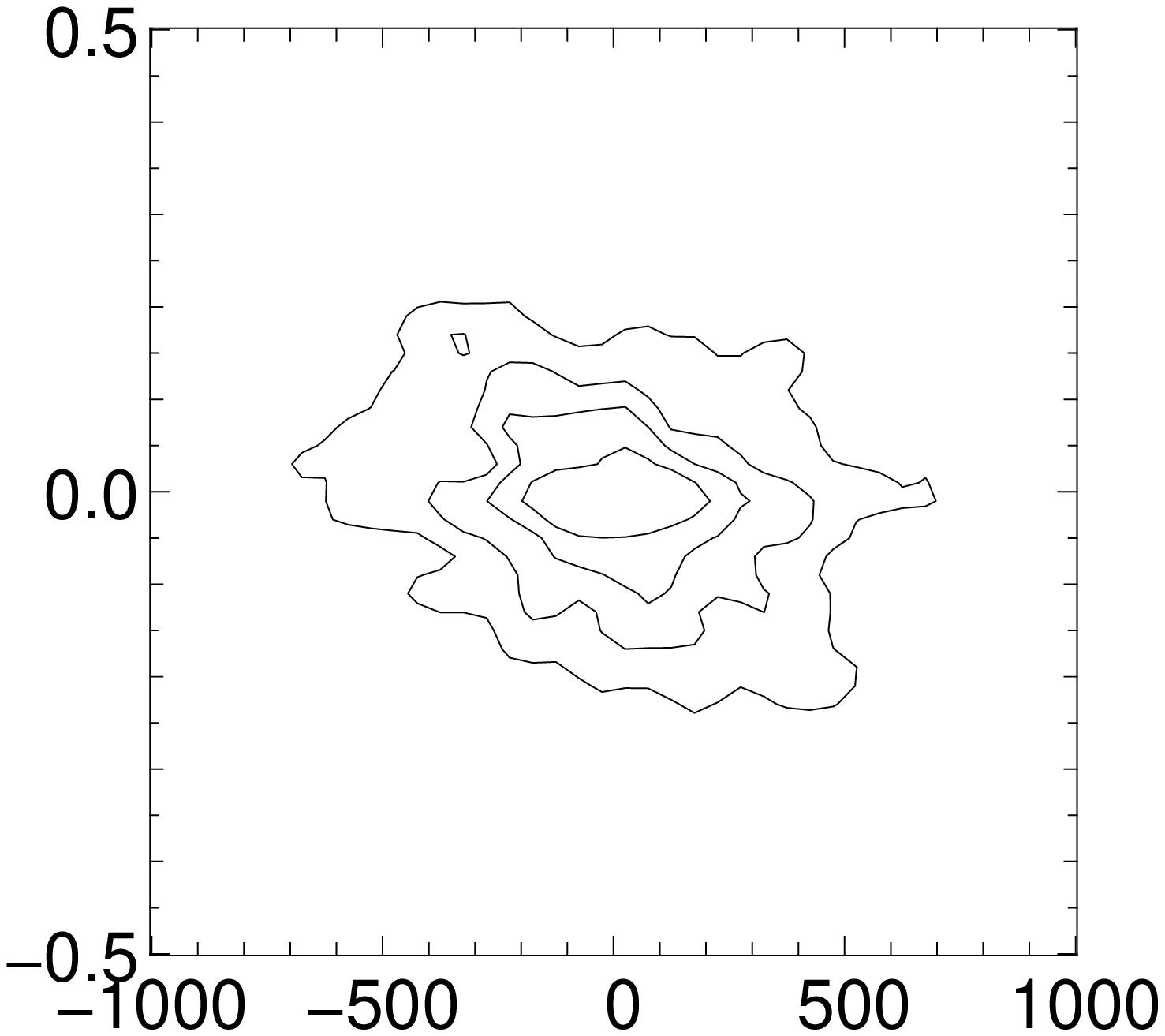}&
\includegraphics[width=3.5cm,height=3.5cm]{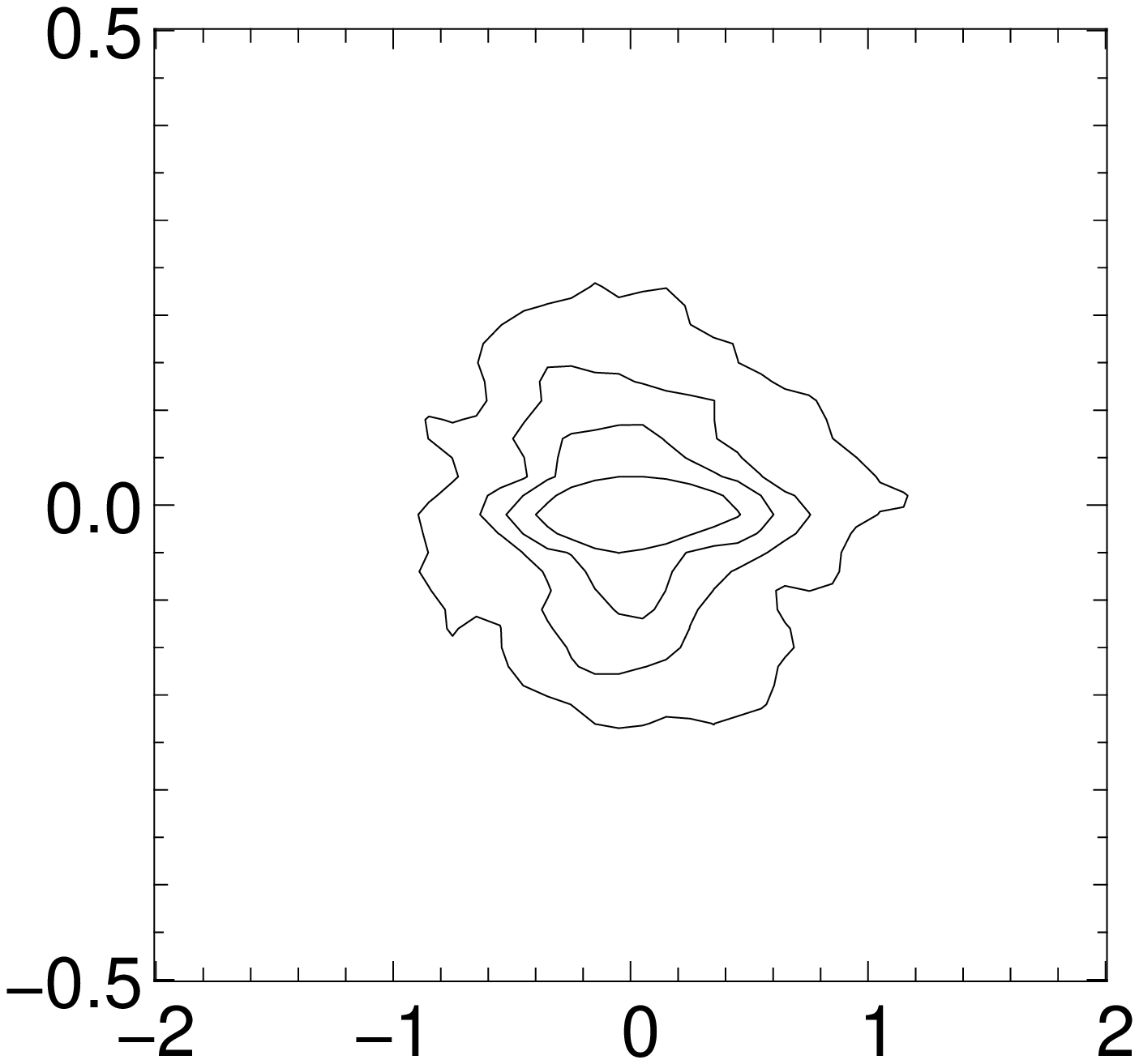}&
\includegraphics[width=3.5cm,height=3.5cm]{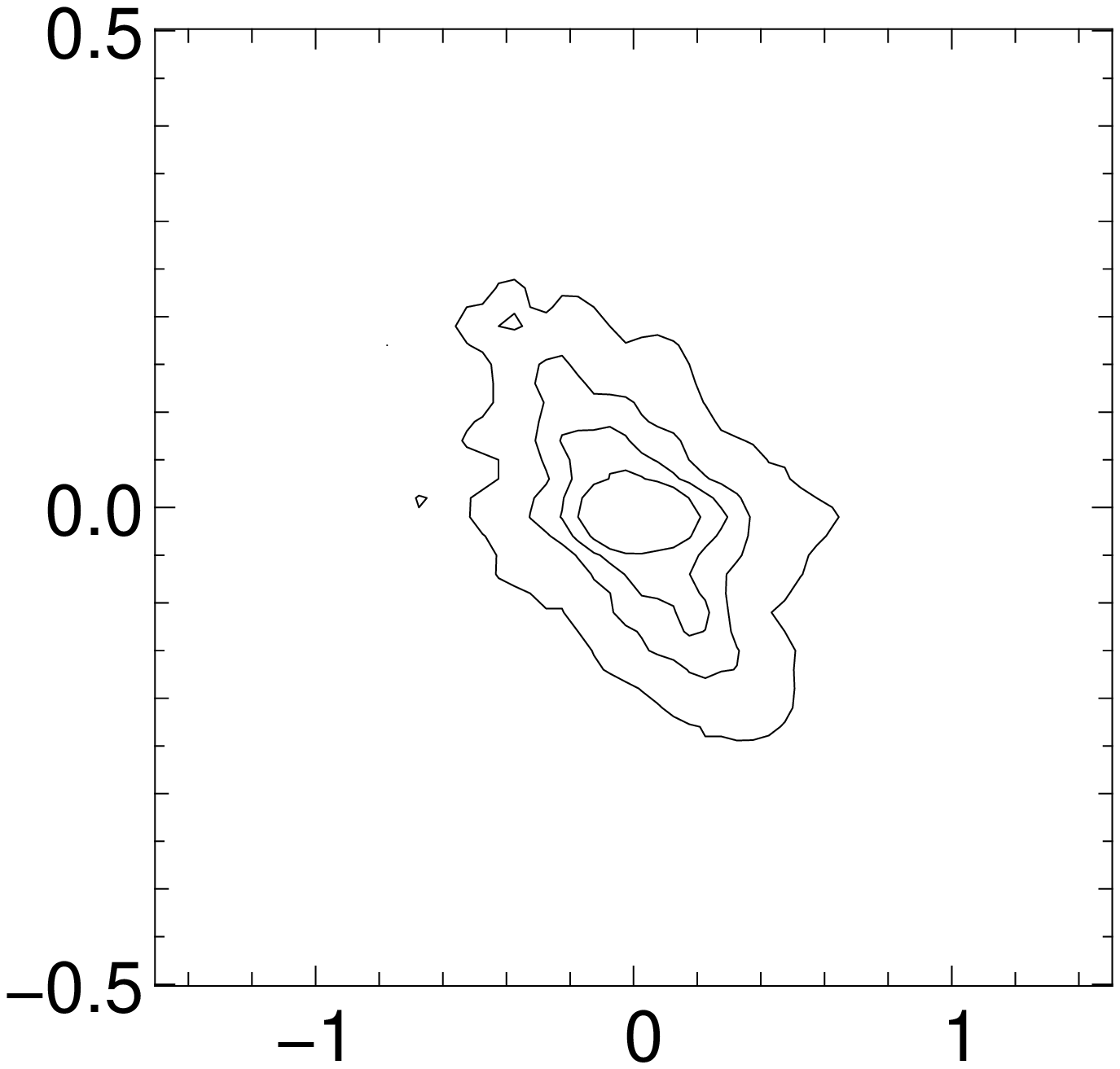}
\\
\begin{sideways} \hspace{0.5cm} $\Delta \mh$ (dex) \end{sideways}&
\includegraphics[width=3.5cm,height=3.5cm]{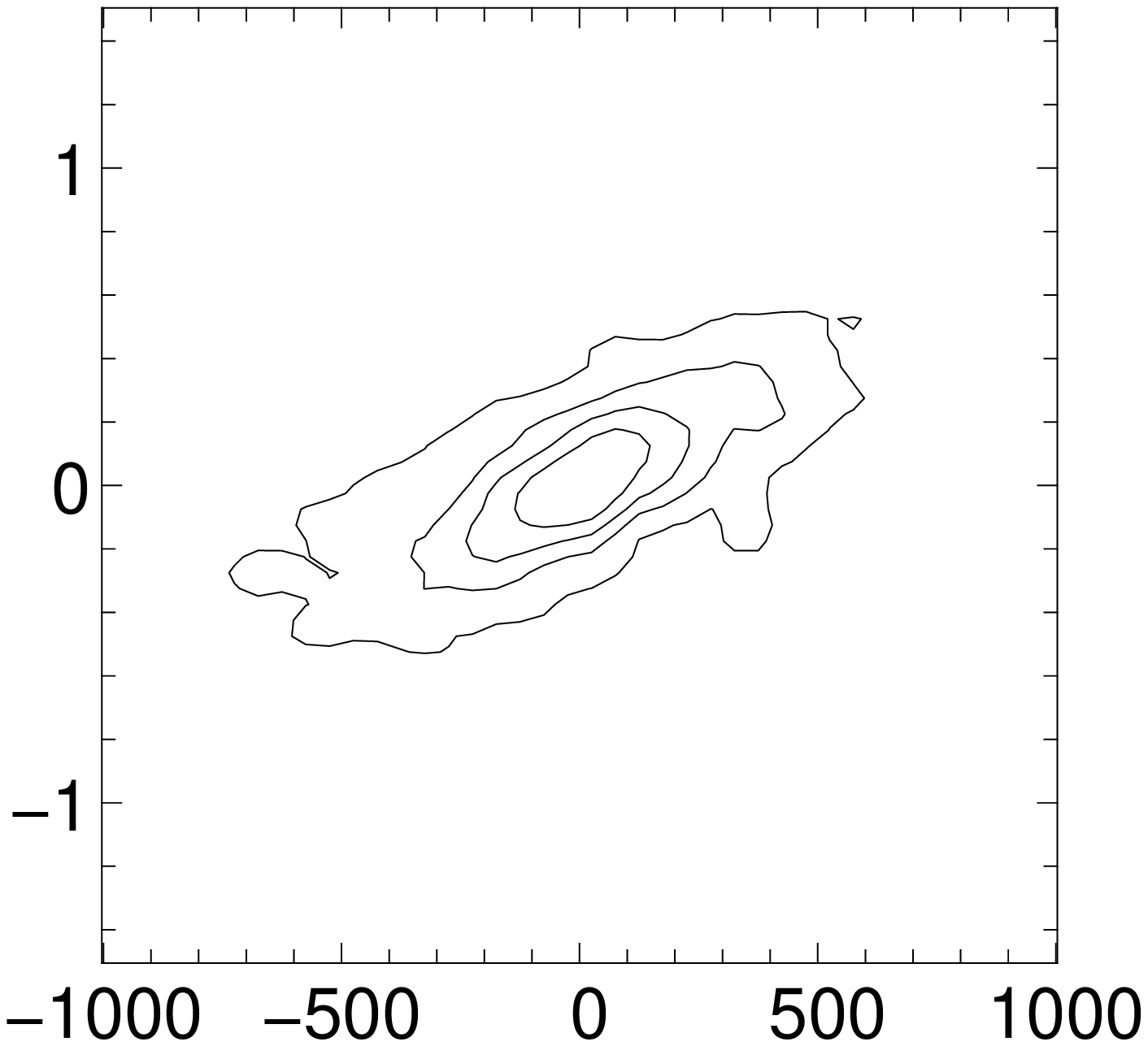}&
\includegraphics[width=3.5cm,height=3.5cm]{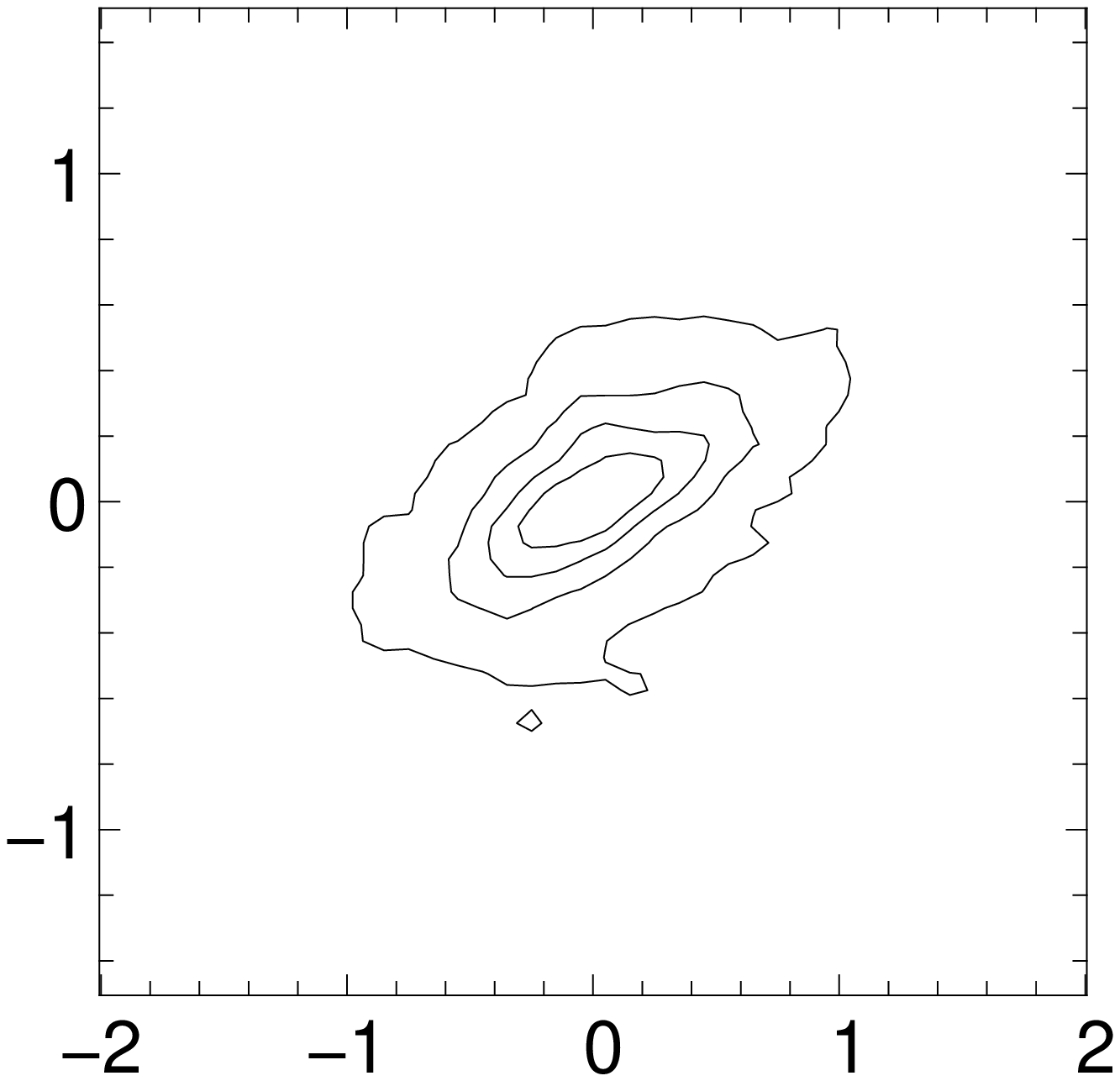}&
\\
\begin{sideways} \hspace{0.5cm} $\Delta \logg$ (dex) \end{sideways}&
\includegraphics[width=3.5cm,height=3.5cm]{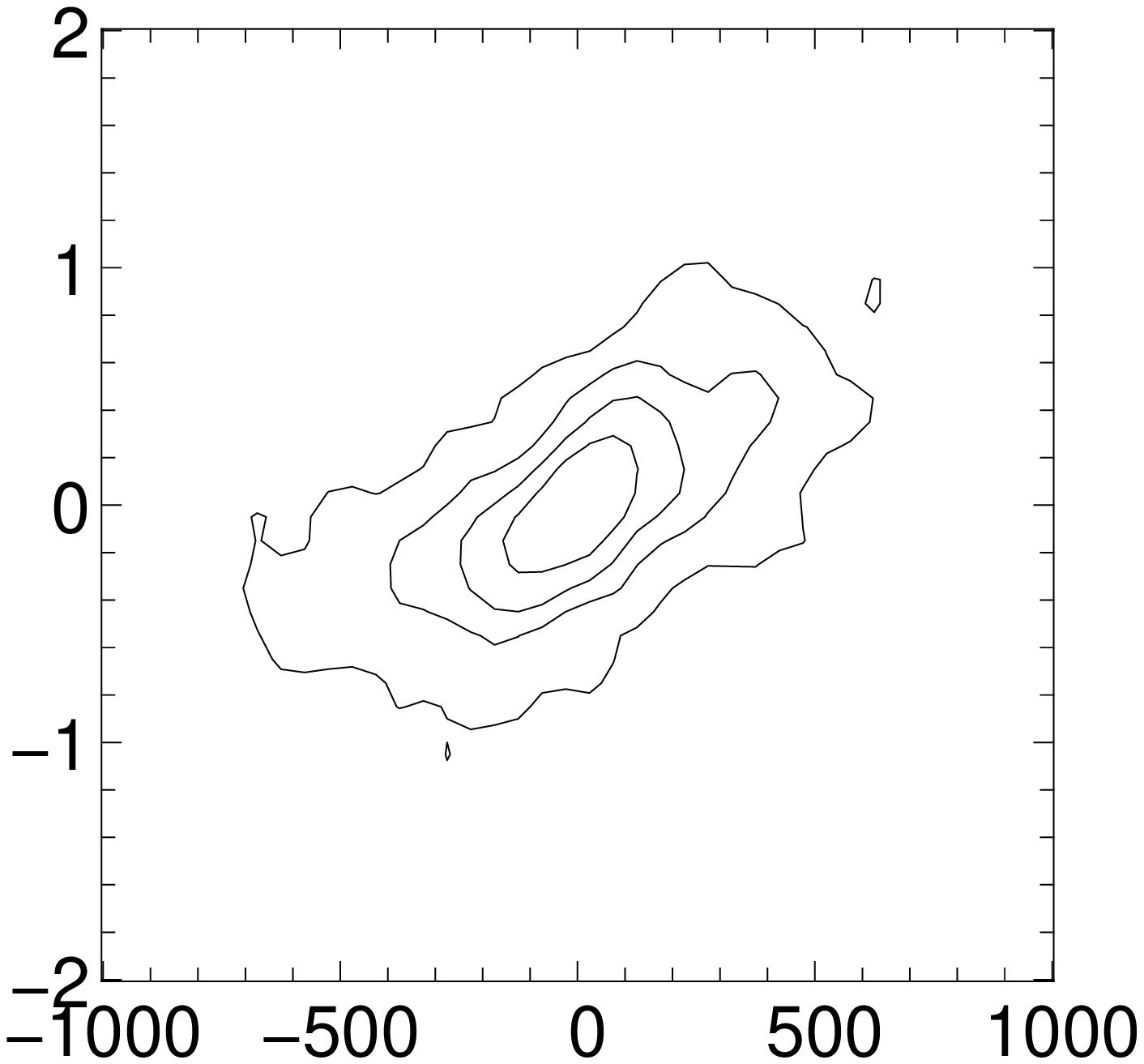}&
\\
\end{tabular}
\caption{Correlation  between the  stellar atmospheric  parameters  based on
 re-observed RAVE targets. The contours contain  30, 50, 70, and 90\% of the
 data  respectively.  A  correlation between  the error  in  two parameters
 indicates that a  systematic error in one parameter  influences the result
 in the other.}
\label{f:reobs_param_correl}
\end{figure*}

In  Paper II,  we showed  that the  method we  use to  estimate  the stellar
atmospheric  parameters  introduces  correlations   in  the  errors  of  the
recovered  parameters. Here,  we use  the re-observations  of  standard RAVE
program stars to  estimate the amplitude of these  correlations. The results
of these tests are  presented in Figure~\ref{f:reobs_param_correl} where the
contours  in  each  panel  contain  30,  50,  70,  and  90\%  of  the  total
sample. Looking  at the  different panels, a  clear correlation  is observed
between the  deviations in $\teff$,  $\logg$ ,and $\mh$ while  deviations in
$\alp$ are  only correlated with deviations  in $\mh$. $\vrot$  on the other
hand does not show any  correlation, regardless of the atmospheric parameter
considered. Since the correlation between  $\logg$ and $\mh$ is broader than
between $\logg$  and $\teff$, it  is likely that  errors on $\teff$  are the
primary  source of  errors, and  that these  errors propagate  to  the other
atmospheric parameters.

These correlations  indicate that the true  $\MH$ will be a  function of all
the  parameters, except  for  $\vrot$. The  correlation  with $\logg$  being
weaker than that with $\teff$ and $\mh$, the true calibration relation might
be independent of $\logg$ or at least, we expect $\logg$ to play a secondary
role in the estimation of the  true $\MH$.  This will be studied more deeply
in the next paragraph.

\subsubsection{Comparison to external data}
\label{s:calib_param}

In  the  previous  paragraphs,  we  checked  the  consistency  of  the  RAVE
atmospheric-parameter  solutions  and the  correlations  that exist  between
these  parameters.   The  consistency   of  the  atmospheric  parameters  is
satisfactory  given  our medium  resolution  ($R\sim7\,500$)  and our  small
wavelength  interval.   The  dispersions  around the  reference  values  are
$\sim200$~K for $\teff$, 0.3~dex for $\logg$, 0.2~dex for $\mh$, and 0.1~dex
for $\alp$, with no significant centroid offset.

The  next  step is  to  compare  our  measured atmospheric  parameters  with
independent measurements. As for DR2,  RAVE stars are generally too faint to
have been observed  in other studies from the  literature. We therefore used
custom RAVE observations of  bright stars from the literature\footnote{These
  stars are  not part  of the original  input catalog  but are added  to the
  observing  queue  to  permit   the  validation  of  the  RAVE  atmospheric
  parameters.}  as  well  as  high-resolution observations  of  bright  RAVE
targets  to construct our  calibration sample.   This sample  comprises four
different sources of atmospheric parameters:
\begin{itemize}
\item[-] RAVE observations of \citet{soubiran} stars,
\item[-] Asiago echelle observations of RAVE targets ($R\sim20\,000$),
\item[-] AAT 3.9m UCLES echelle observations of RAVE targets,
\item[-] APO     ARC    echelle    observations     of    RAVE    targets
 ($R\sim35\,000$).
\end{itemize}
The  last three sources  of calibration  data make  the bright  RAVE targets
sample  and were  all reduced  and processed  within the  RAVE collaboration
using  the same  technique and  are therefore  merged in  the  following and
referred to  as ``echelle data''.   We follow a standard  analysis procedure
using Castelli  ODFNEW atmosphere models. The  gf values for  iron lines are
taken from three different sources
\begin{itemize}
\item[-] the list from \citet{ful00} for metal poor stars based
\item[-] a list of differential $\log gf$ from Acturus \citep{ful06} best
 suited for metal rich giants
\item[-] a list of differential $\log gf$ from the Sun best suited for dwarf
 stars.
\end{itemize}
The  three line lists  give reasonable  agreement ($\Delta\teff<50\,{\mathrm
 K}$  and  $\Delta\feh<0.1\,{\mathrm   dex}$)  in  the  parameter  boundary
regions.  The  alpha- and heavy-element  line list is  basedon \citet{ful00}
for metal-poor stars and \citet{ful07} for metal-rich stars.  $\teff$ values
are obtained using the excitation balance, forcing the distribution of $\log
\epsilon({\mathrm  Fe})$\footnote{$\epsilon(X)$ is the  ratio of  the number
 density of atoms of element $X$  to the number density of hydrogen atoms.}
vs.   excitation  potential  for  individual  Fe\,I lines  to  have  a  flat
slope. $\logg$  is obtained  via the ionisation  balance, forcing  the $\log
\epsilon(\mathrm{Fe})$ values derived from  Fe\,I and Fe\,II lines to agree.
Both  methods are  fully independent  from the  technique used  by  the RAVE
pipeline to estimate atmospheric parameters from medium-resolution spectra.

For the  RAVE observation of stars  studied in the literature,  we chosed to
build our  sample upon the  \citet{soubiran} catalog. This  catalog contains
abundances measurements from the literature paying a particular attention at
reducing the systematics between the various studies.  It makes this catalog
particularly suited for calibration purposes.

\begin{figure}
\centering
\includegraphics[width=7cm]{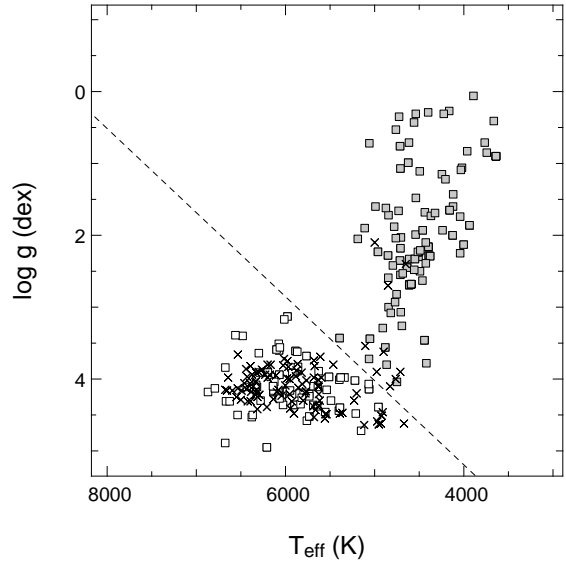}
\caption{Location   of  the   reference  stars   in   the  ($\teff$,$\logg$)
 plane.  Squares  are  echelle  data,  the  dashed  line  representing  our
 separation between dwarfs (open symbols) and giants (grey symbols) for the
 calibration relation. Crosses are stars in \citet{soubiran}.}
\label{f:calib_HR}
\end{figure}

Table~\ref{t:datasource}  summarizes  the   content  of  each  sample  while
Figure~\ref{f:calib_HR} presents the distribution  in $\logg$ and $\teff$ of
stars in the  calibration sample. The GCS also  provides photometric $\teff$
measurements but as for DR2, we choose not to include photometric $\teff$ in
our analysis.

\begin{table}
\centering
\caption{Samples  used to  calibrate the  RAVE atmospheric  parameters.  The
 echelle  sample covers  the  data  obtained using  UCLES,  ARC, and  Asiago
 spectrographs and were processed and analysed consistently.}
\label{t:datasource}
\begin{tabular}{c c c c c c c}
\hline 
\hline 
Sample & $N_{\rm star}$  & $N_{\rm obs}$ & $\teff$ & $\logg$ &
$\MH$ & $\alp$\\ 
\hline 
Echelle & 162  & 228 & $\surd$ & $\surd$ & $\surd$ & $\surd$\\ 
Soubiran \& Girard & 102 & 107 & $\surd$ & $\surd$ & $\surd^{(1)}$
&  $\surd$\\ 
\hline  
\multicolumn{7}{p{12cm}}{(1):  \citet{soubiran} do  not
 report metallicity $\MH$, so their  values are derived from a weighted sum
 of the quoted element abundances of Fe, O, Na, Mg, Al, Si, Ca, Ti, and Ni,
 assuming the solar abundance ratio from \citet{AG89}.}\\
\end{tabular}
\end{table}

In the following, we separate the analysis of $\teff$ and $\logg$ from 
$\MH$, the latter requiring a specific calibration.

\noindent $\bullet$ $\teff$ and $\logg$ :

Table~\ref{t:teff_logg} presents  the results of the comparison  of the RAVE
pipeline outputs with the reference datasets. Since outliers are present, we
use  a standard iterative  (sigma-clipping) procedure  to estimate  the mean
offset  and standard  deviation  for each  atmospheric  parameter.  The  new
version of the  pipeline shows a slight tendency  to overestimate $\teff$ by
$\sim50-60$~K  compared to  the previous  version, with  an increase  of the
standard  deviation  from  188~K  to  250~K. For  $\logg$  the  results  are
consistent between the two versions of  the pipeline.  We note here that the
reference samples used for the new release have increased considerably, with
the number of  \citet{soubiran} stars increasing by a factor  of two and the
number of echelle observations by a factor of four.

\begin{table}
\centering
\caption{Mean offset and standard  deviation for $\teff$ and $\logg$ between
  the reference  datasets and RAVE DR3  values.  $N_{\rm tot}$  is the total
  number of  observations in the reference datasets,  and $N_{\rm rej,Teff}$
  and $N_{\rm  rej,\logg}$ are  the number of  observations rejected  by the
  iterative procedure for estimating  the mean difference and dispersion for
  $\teff$ and $\logg$ respectively. }
\label{t:teff_logg}
\begin{tabular}{c c c c c c c c}
\hline
\hline
Sample & $N_{\rm tot}$ & $\Delta \teff$ & $\sigma_\teff$ &$N_{\rm rej,Teff} $& $\Delta
\logg$ & $\sigma_{\logg}$ & $N_{\rm rej,\logg}$\\
\hline
Echelle & 227 & $-85\pm14$ & 209 & 11 & $-0.12\pm0.03$ & 0.43 & 6\\
Soubiran \& Girard & 107 &$-63\pm26$ & 262 & 7 & $-0.05\pm0.03$ & 0.35 & 2\\
\hline
All & 334 & $-72\pm14$ & 251 & 12 & $-0.10\pm0.02$ & 0.40 & 9\\
\hline
\end{tabular}
\end{table}

\begin{figure}
\centering
\includegraphics[width=5cm]{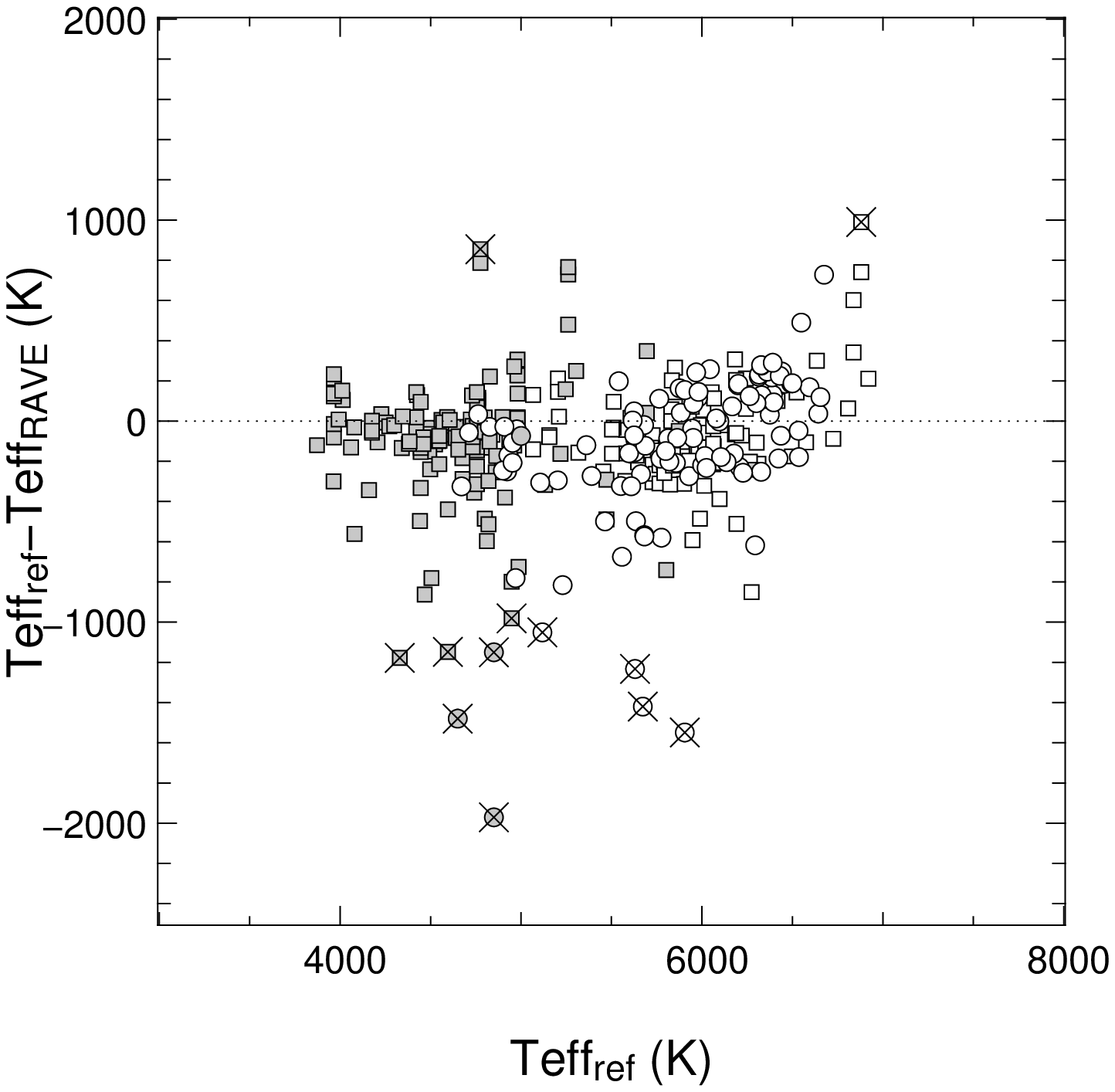}
\includegraphics[width=5cm]{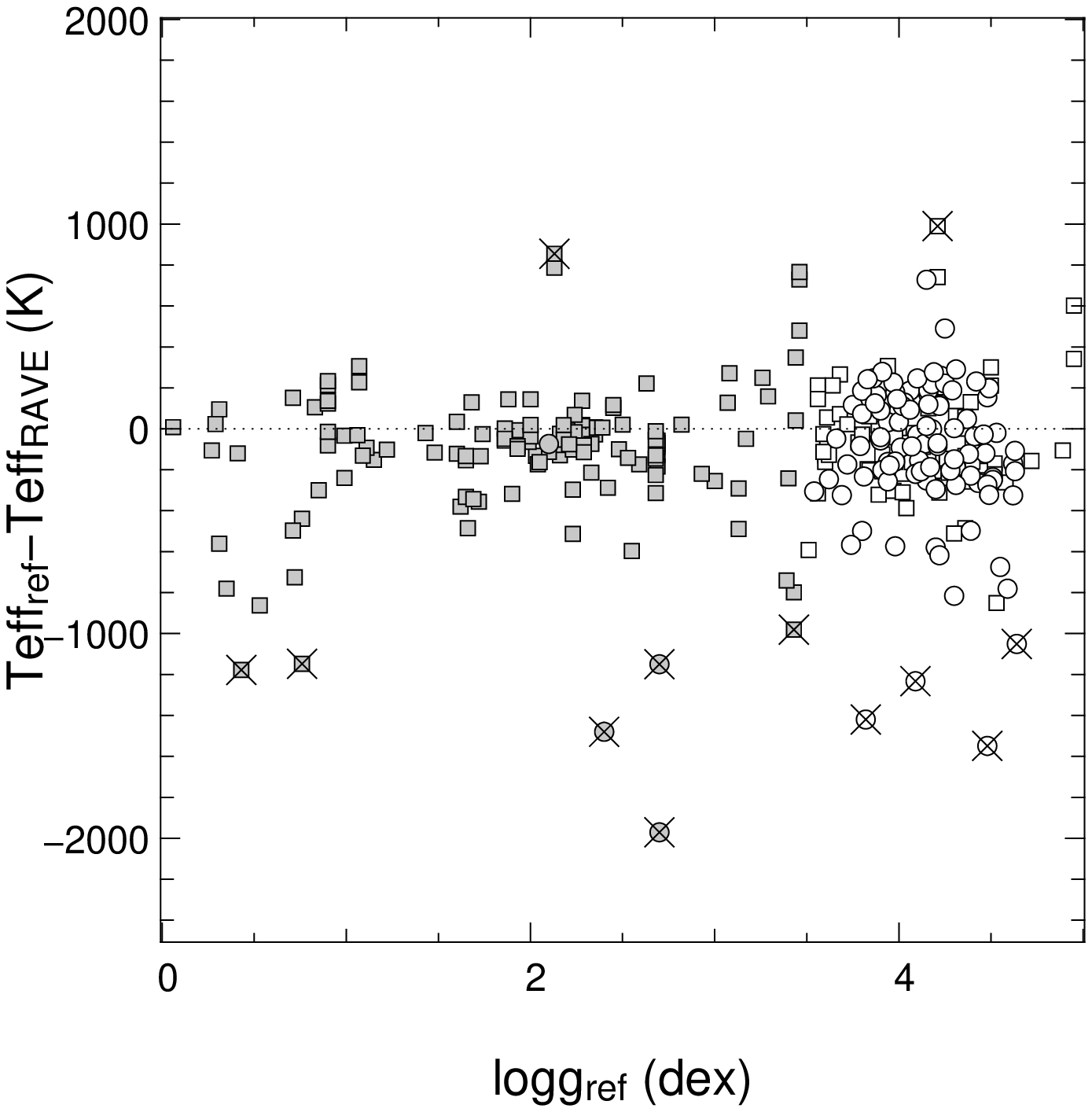}
\includegraphics[width=5cm]{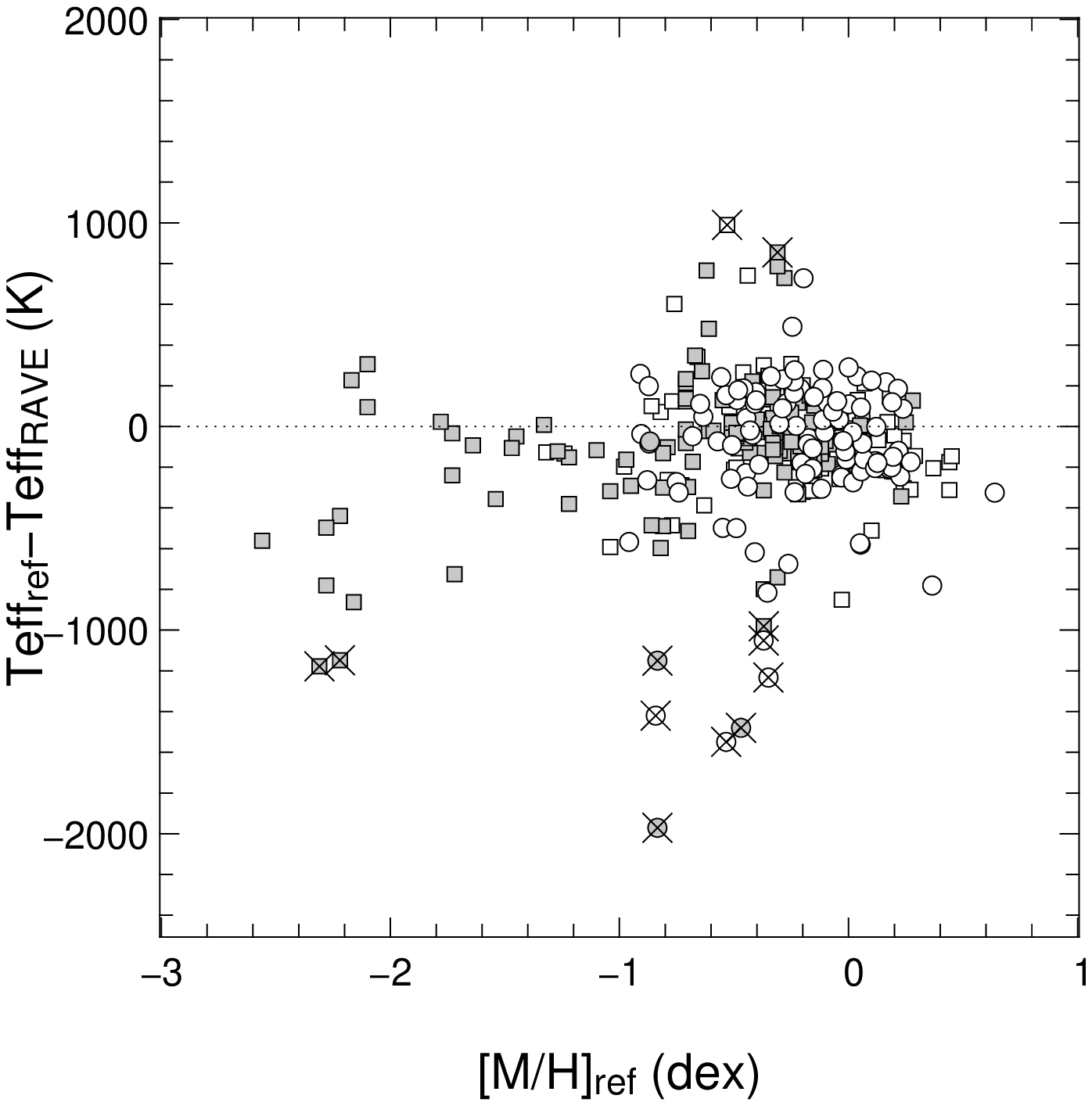}
\includegraphics[width=5cm]{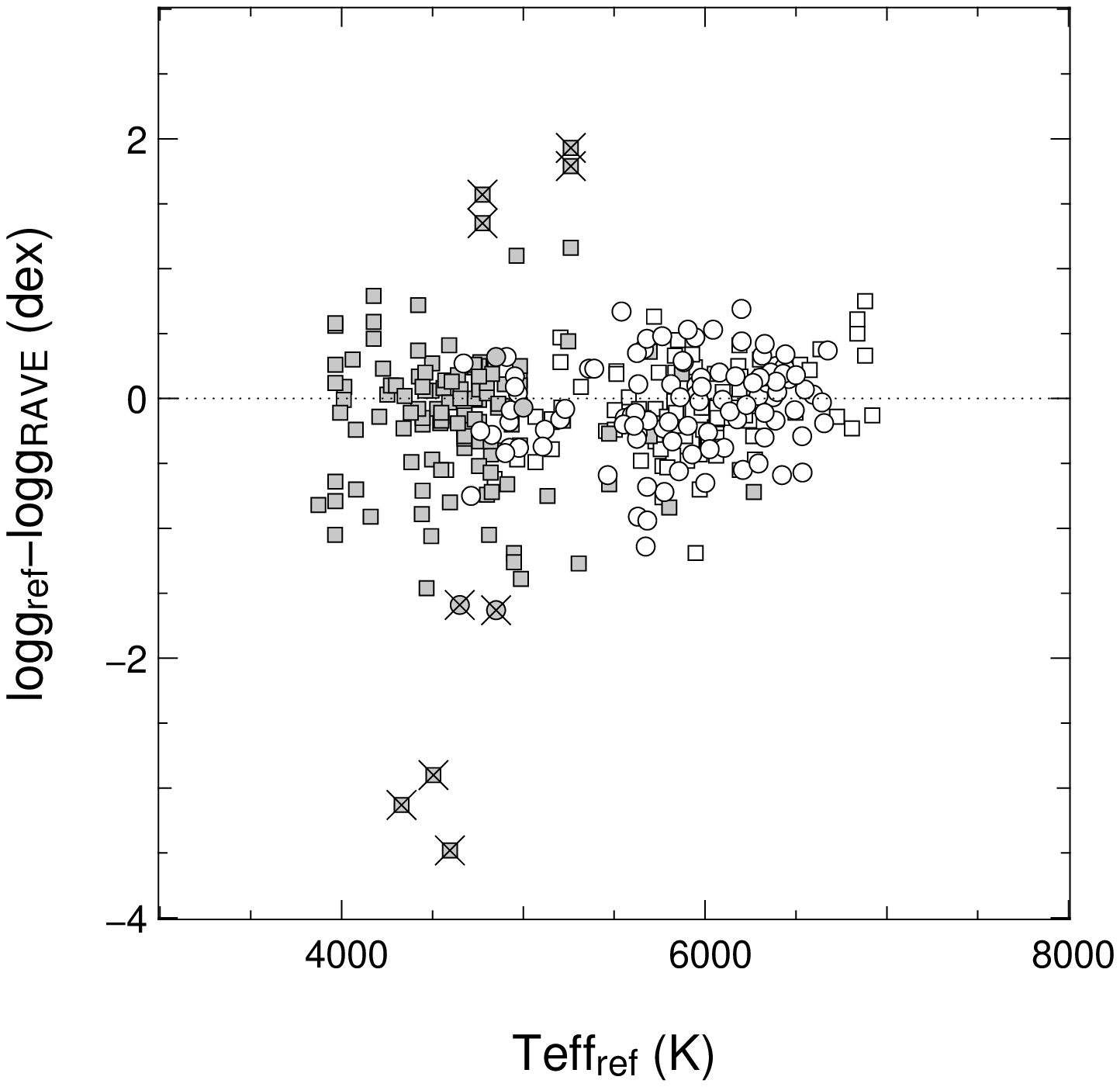}
\includegraphics[width=5cm]{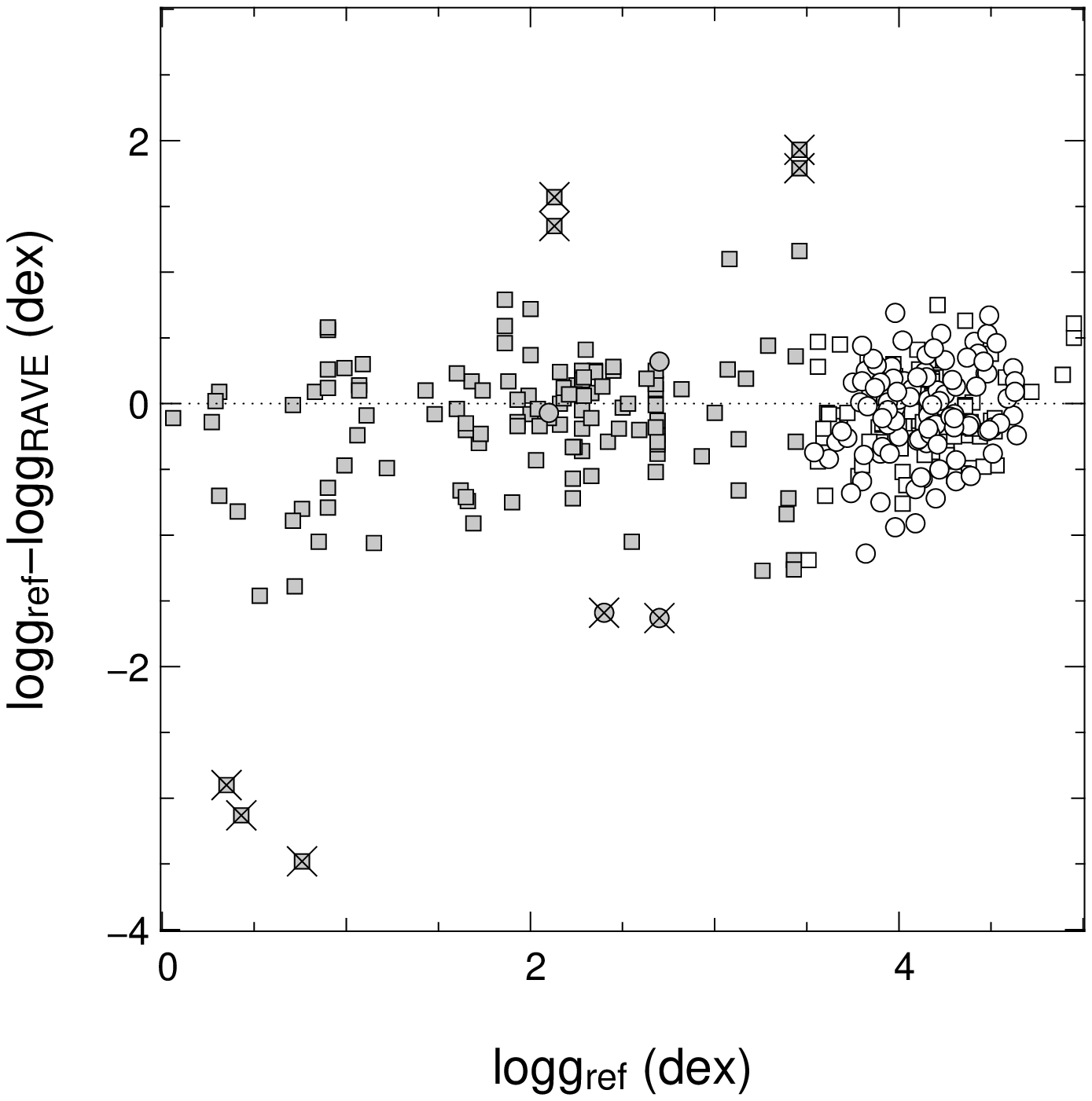}
\includegraphics[width=5cm]{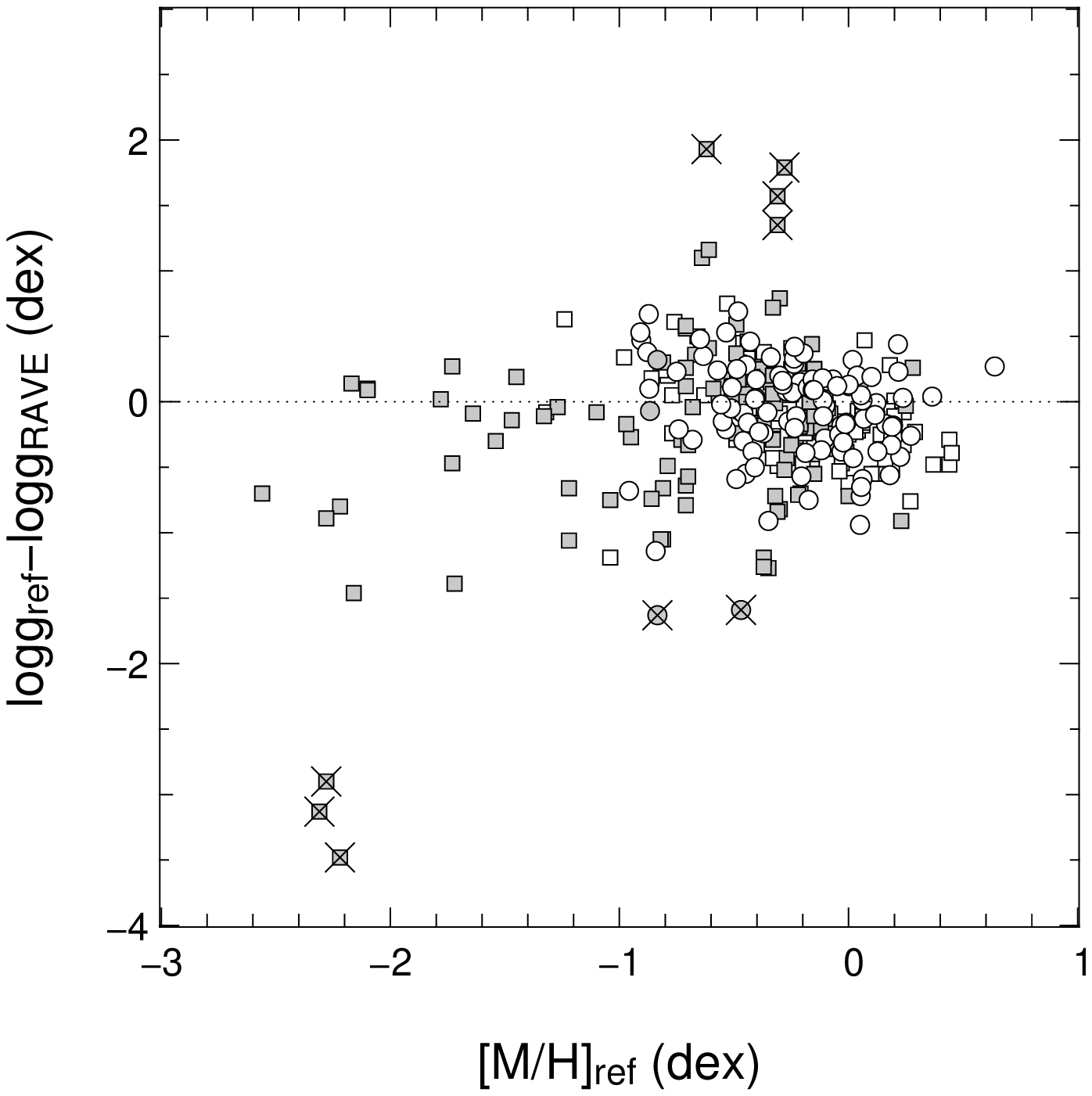}
\caption{Difference  between  the atmospheric  parameters  of the  reference
 datasets and  of the RAVE  DR3 parameters as  a function of  the reference
 $\teff$,  $\logg$  and $\MH$  for  $\teff$  (top),  and $\logg$  (bottom).
 Circles stand  for stars in \citet{soubiran} while  squares denote echelle
 data.  Grey symbols  represent the giants, open symbols  mark the location
 of the dwarfs.  Crosses indicate  data rejected by the iterative procedure
 used for Table~\ref{t:teff_logg}.}
\label{f:teff_logg_calib}
\end{figure}

To  further  validate our  atmospheric  parameters,  we  compare the  offset
between the reference atmospheric parameters  with the RAVE values.  This is
presented  in Figure~\ref{f:teff_logg_calib}  for $\teff$  (top  panels) and
$\logg$ (bottom panels)  as a function of reference  $\teff$ (left), $\logg$
(middle), and $\mh$ (right). The  crosses indicate the data discarded by the
iterative procedure as being outliers.

For $\teff$, no  correlation is observed either as a  function of $\teff$ or
$\mh$.  Considering  the echelle  data alone (open  squares) a  tendency for
$\teff$ to be overestimated as  $\logg$ increases is observed, producing the
$-85\,$K  offset  reported  in  Table~\ref{t:teff_logg}.   However,  at  low
$\logg$  the  discrepancy vanishes.   This  tendency  is  not seen  for  the
\citet{soubiran} stars.  Since this effect is not systematic, it leads us to
conclude that the  apparent trend in $\teff$ with $\logg$ is  not due to the
RAVE  data but  instead due  to the  different methods  used to  derive this
parameter in the other works.

For $\logg$, no trend is observed with $\teff$. However a trend with $\logg$
seems to be present, such that the RAVE $\logg$ is slightly overestimated at
the low  end (by  $\sim$0.5~dex).  In addition,  a tendency  to overestimate
$\logg$ at low metallicities is  seen, amounting to the same order.  Because
this  effect is  limited to  the very  low $\logg$  end of  the distribution
($\logg <1 $), which is not highly populated in the RAVE catalog, this leads
to the  conclusion that  our $\logg$ determination  are reliable  within our
quoted uncertainties.

\noindent $\bullet$ $\MH$ :

As  stated in  Paper  II, the  metallicity  indicator obtained  by the  RAVE
pipeline is, due to our medium resolution and limited signal-to-noise ratio,
a  mixture  of  the   real  metallicity,  alpha  enhancement,  and  possibly
rotational  velocity.   To  obtain  an  unbiased estimator,  we  rely  on  a
calibration  relation set  using a  sample of  stars with  known atmospheric
parameters.   Paper  II presented  a  first  calibration  relation using  an
iterative fitting procedure of the relation

$$\MH=c_0 + c_1 . \mh + c_2 . \alp + c_3 . \logg \,.$$

The coefficients of this relation were obtained based on a sample of 45 APO,
24 Asiago, 49  \citet{soubiran}, and 12 M67 cluster  member stars.  With the
larger number of  reference stars available for this release  and due to the
new version of the processing pipeline, modified to increase the reliability
of  the atmospheric  parameters,  we recompute  and  extend the  calibration
relation.  However,  we now  restrict the analysis  to the  reference sample
consisting of  echelle data.  This sample  was selected to  evenly cover the
($\logg$,$\teff$) plane of the RAVE  survey and was processed using the same
technique and reduction algorithm, therefore providing an homogeneous set of
reference data.   Also, with the knowledge  gained from the  analysis of the
correlation between parameters, the  proposed calibration relation now takes
the form

\begin{equation}
\MH=c_0 + c_1 . \mh + c_2 . \alp + c_3 . \frac{\teff}{5040} +c_4 . \logg +
c_5 .  \mathrm{STN} \,,
\label{e:calib}
\end{equation}

\noindent
where  we  added $\teff$  to  the calibration  relation  due  to the  strong
correlation  observed in  Fig.~\ref{f:reobs_param_correl}  and discussed  in
Section~\ref{s:correl}. \snr\ is  also included as one expects  an impact of
the  noise at  the low  \snr\ regime  where the  pipeline may  mistake noise
spikes  for enhanced  metallicity. Since  $\teff$  seems to  be the  primary
source of  error for $\mh$, we  computed four calibration  relations for the
various cases with and without \snr\ or $\logg$. As for the DR2 calibration,
we see no evidence for higher  order terms and therefore restrict our search
for the best calibration to first order (linear) relations.

\begin{figure}
\centering
\includegraphics[width=7cm]{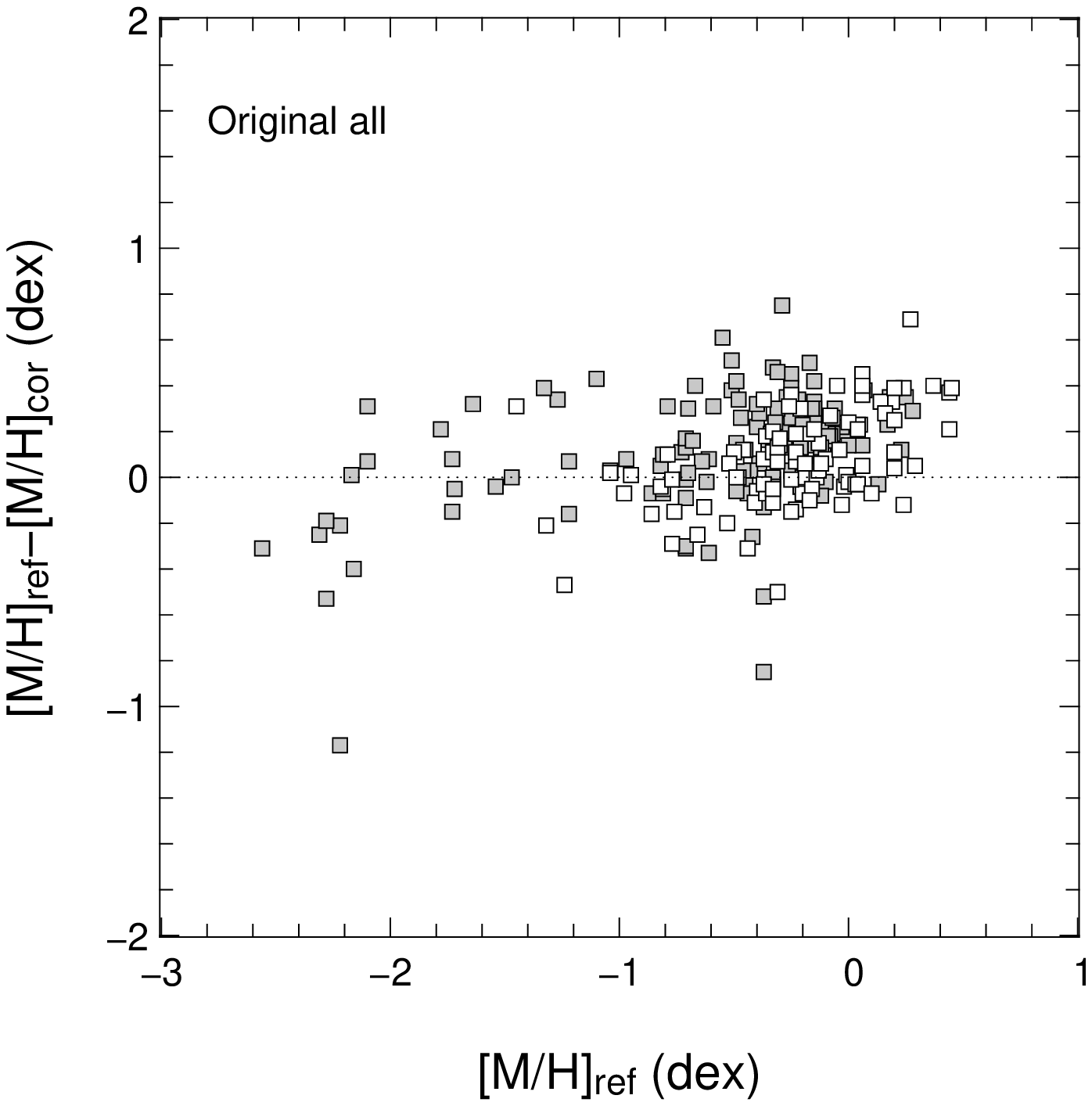}
\includegraphics[width=7cm]{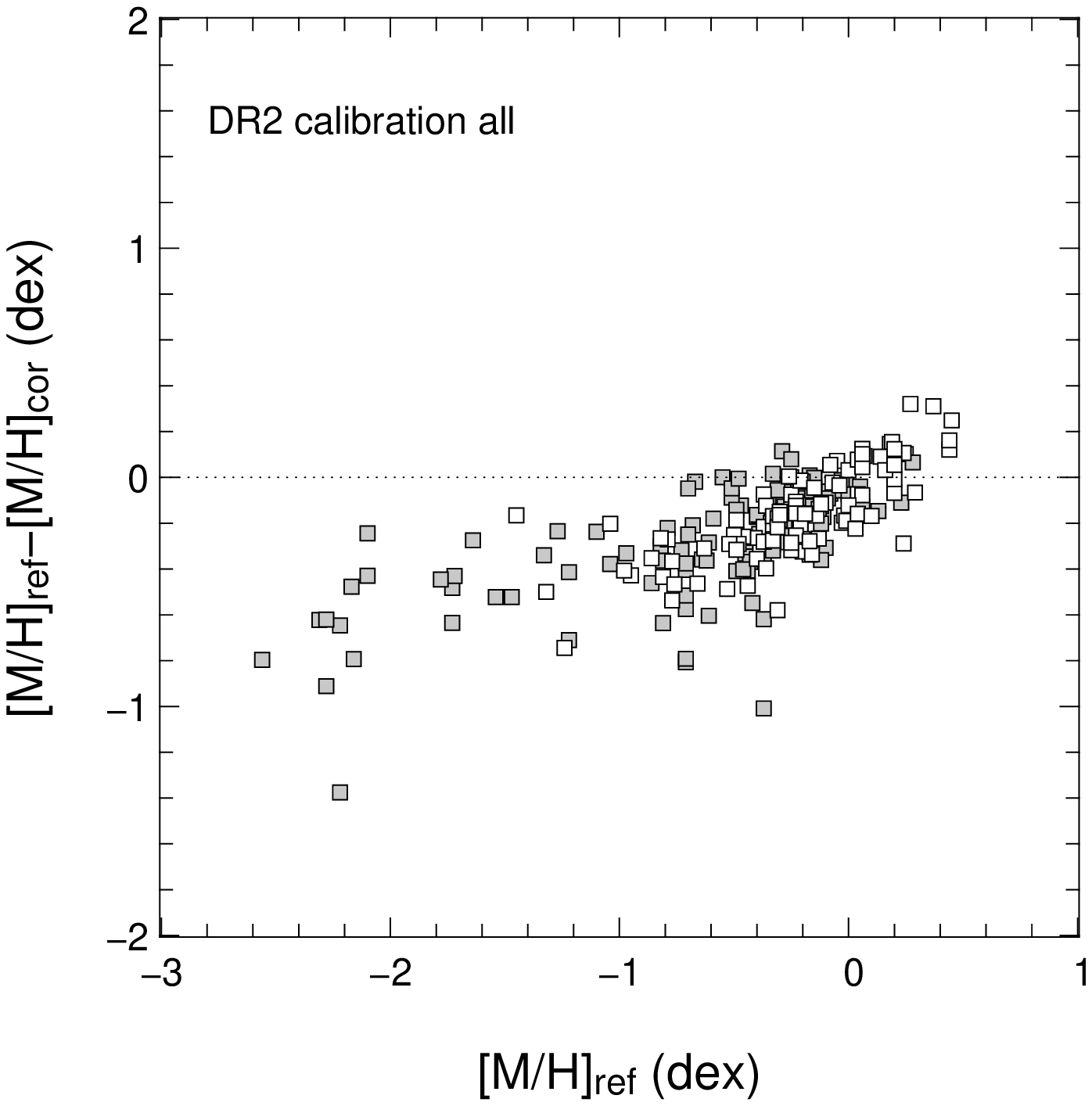}
\includegraphics[width=7cm]{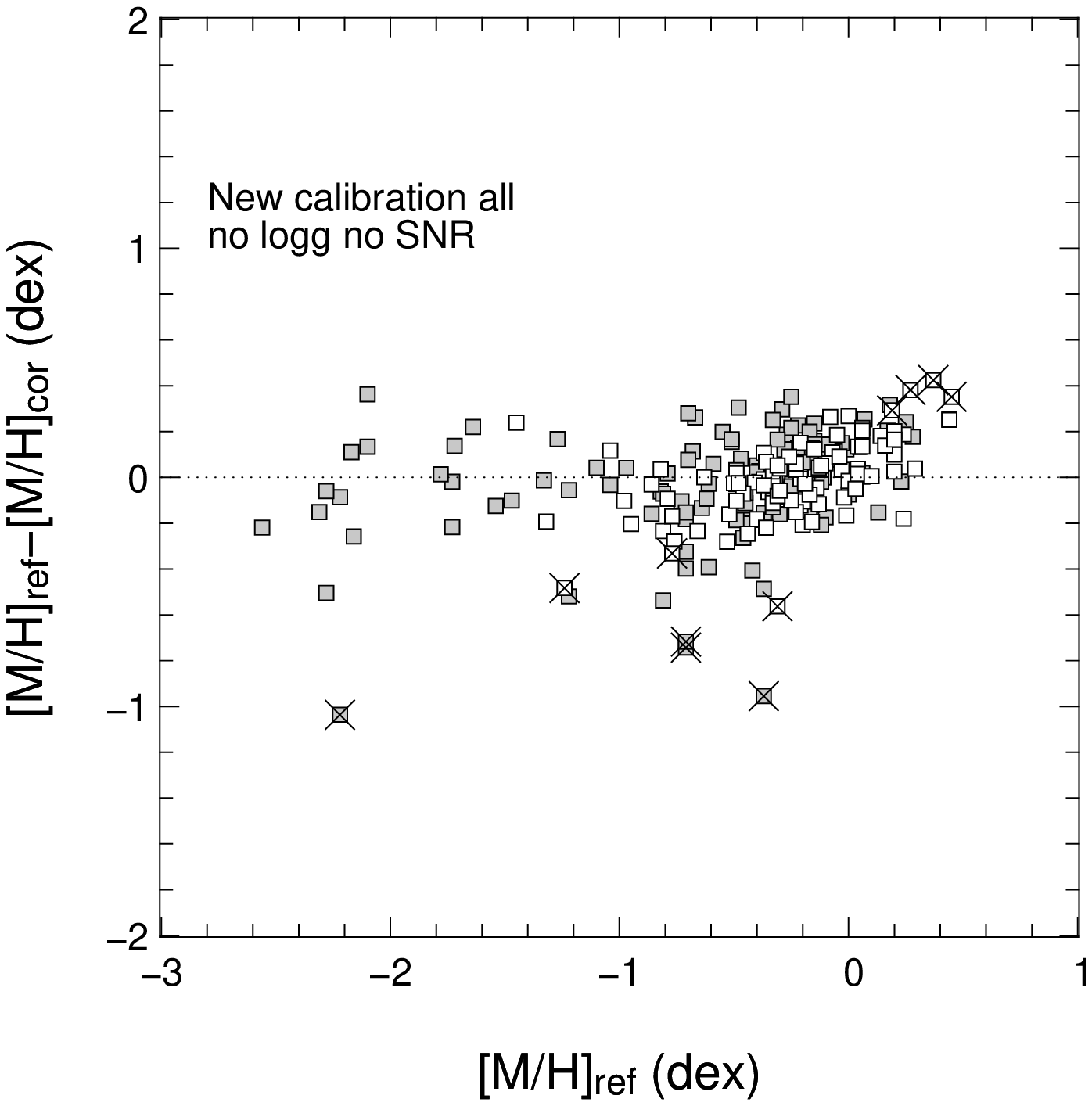}
\includegraphics[width=7cm]{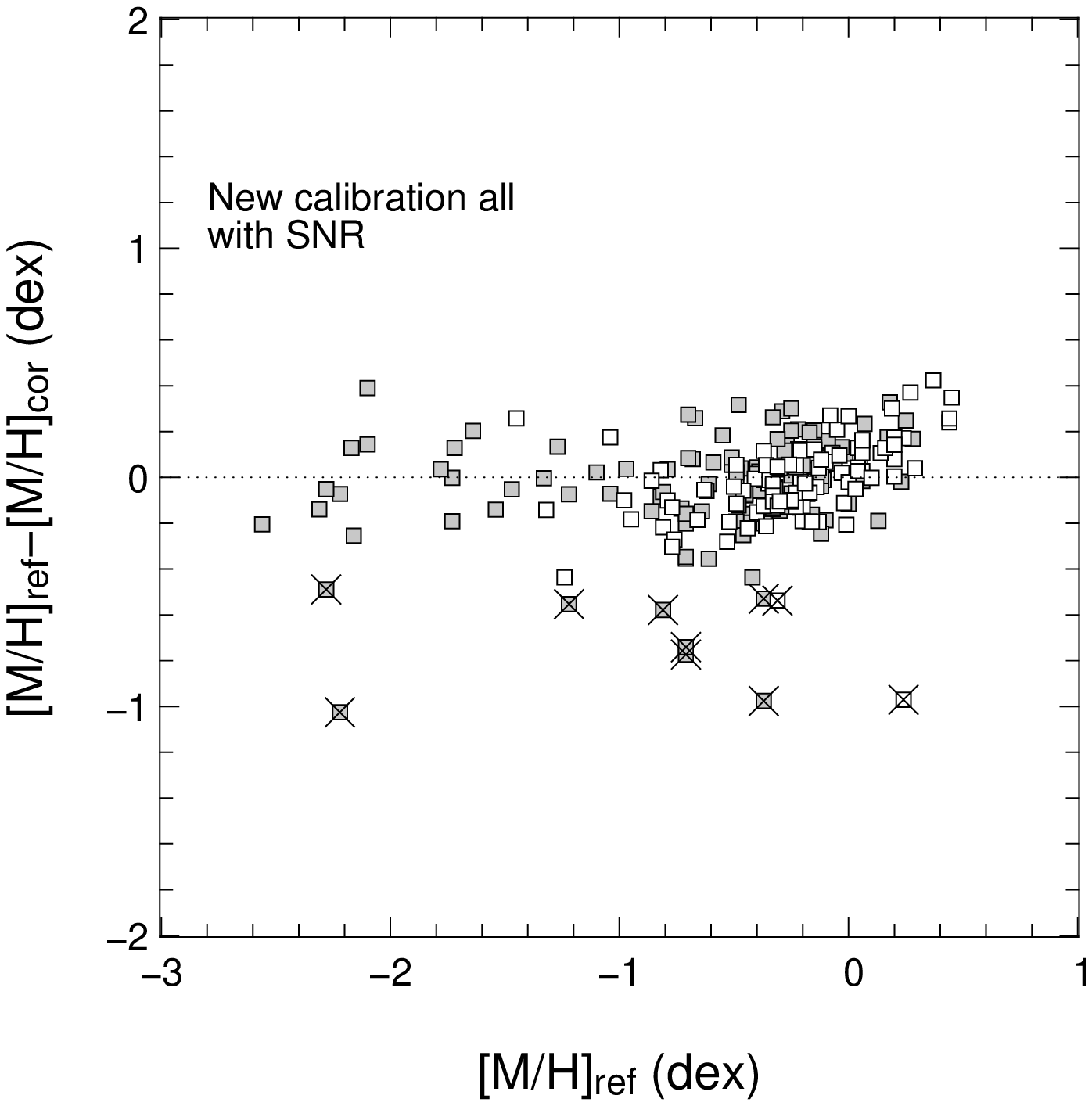}
\includegraphics[width=7cm]{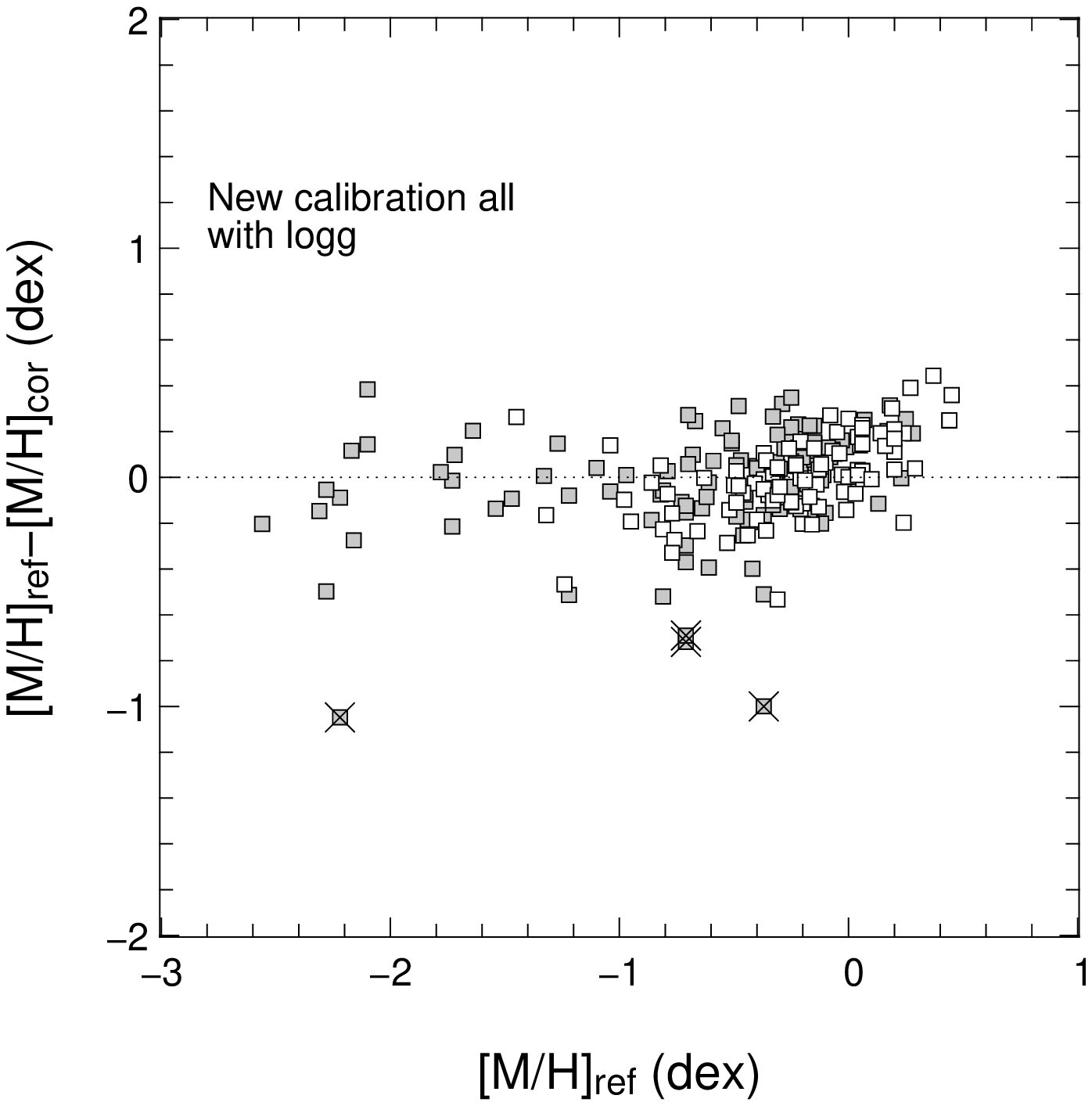}
\includegraphics[width=7cm]{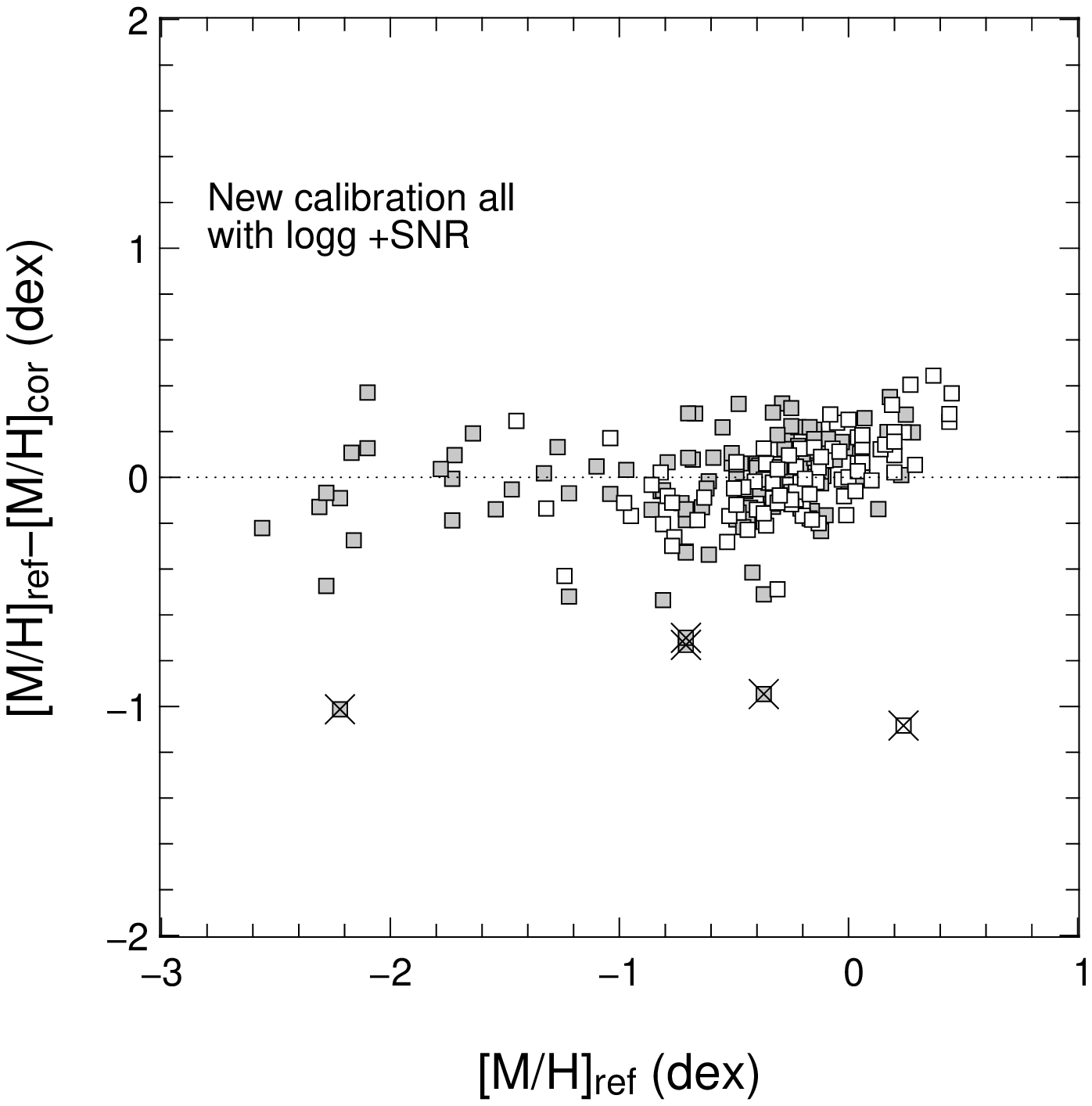}
\caption{Difference  between the reference  $\MH$ and  RAVE $\MH$  using the
 different  calibration relations  as a  function of  reference  $\MH$. The
 crosses indicate the  observations rejected from the fit  by the iterative
 procedure.}
\label{f:calib_MH}
\end{figure}

The coefficients  for the calibration  relations are obtained  by minimizing
the difference between the calibrated $\MH$ and the reference $\MH$ using an
iterative procedure to reject outliers.  The resulting calibration relations
are  summarized in  Table~\ref{t:calib}  where $N_{\rm  tot}$  is the  total
number of  observations used to  compute the calibration relation.   A blank
value in a  column indicates that the calibration  relation does not include
the corresponding parameter.  The residuals between the calibrated $\MH$ and
the reference  $\MH$ as a function  of the reference $\MH$  are presented in
Figure~\ref{f:calib_MH} where the  top panels present the raw  output of the
DR3 pipeline  (panel marked original)  and the residuals obtained  using the
DR2  calibration relation  on the  DR3 atmospheric  parameters  values.  The
following four panels are for the different calibration relations considered
here.   Finally,  Table~\ref{t:calib_stat}  presents  the  mean  offset  and
standard deviation computed from the residuals in the different cases.

\begin{sidewaystable}\footnotesize
\caption{Coefficients in the calibration relation for the RAVE metallicities
 using different sets of parameters for the fit. $N_{\rm tot}$ is the total
 number  of data  points  used to  derive  the calibration,  $c_i$ are  the
 coefficients from  Eq.~\ref{e:calib}.  The first line  presents the output
 of  the new  RAVE  pipeline while  the  second line  presents the  results
 obtained  when one  applies  the  calibration relation  of  Paper II.  The
 following  lines are  the  calibration relations  obtained  using the  new
 pipeline outputs.}
\label{t:calib}
\begin{tabular}{c c c c c c c c}
\hline
\hline
Calibration & $N_{\rm tot}$ & $c_0$ & $c_1$ & $c_2$ & $c_3$ & $c_4$ & $c_5$\\
\hline
\multicolumn{8}{c}{Full sample}\\
\hline
DR2 calibration & - & 0.404 & 0.938 & 0.767 & - & -0.064 & - \\
DR3 no \snr\ no $\logg$ & 223 & $0.578\pm0.098$ & $1.095\pm0.022$ &
$1.246\pm0.143$ & $-0.520\pm0.089$ & - & - \\
DR3 with \snr\ & 217 & $0.587\pm0.091$ & $1.106\pm0.024$ & $1.261\pm0.140$ &
$-0.579\pm0.078$ & - & $0.001\pm0.0004$ \\
DR3 with $\logg$ & 223 & $0.518\pm0.127$ & $1.111\pm0.031$ & $1.252\pm0.144$
& $-0.399\pm0.187$ & $-0.019\pm0.026$ & - \\ 
DR3 with \snr\ and $\logg$ & 222 & $0.429\pm0.132$ & $1.101\pm0.032$ &
$1.171\pm0.147$ & $-0.391\pm0.186$ & $-0.018\pm0.026$ & $0.001\pm0.0004$ \\ 
\hline
\multicolumn{8}{c}{Dwarfs only}\\
\hline
DR3 no \snr\ no $\logg$ & 89 & $0.612\pm0.236$ & $1.081\pm0.045$ &
$1.215\pm0.203$ & $-0.546\pm0.196$ & - & - \\
DR3 with \snr\ & 75 & $0.706\pm0.199$ & $1.250\pm0.055$ & $1.491\pm0.184$ &
$-0.683\pm0.165$ & - & $0.001\pm0.0004$ \\
DR3 with $\logg$ & 82 & $-0.174\pm0.222$ & $1.061\pm0.047$ & $1.621\pm0.158$
& $-0.751\pm0.160$ & $0.232\pm0.038$ & - \\ 
DR3 with \snr\ and $\logg$ & 81 & $-0.170\pm0.217$ & $1.063\pm0.047$ &
$1.586\pm0.155$ & $-0.751\pm0.155$ & $0.219\pm0.037$ & $0.001\pm0.0003$ \\ 
\hline
\multicolumn{8}{c}{Giants only}\\
\hline
DR3 no \snr\ no $\logg$ & 127 & $0.763\pm0.197$ & $1.094\pm0.027$ &
$1.210\pm0.193$ & $-0.711\pm0.207$ & - & - \\
DR3 with \snr\ & 119 & $0.399\pm0.178$ & $1.087\pm0.027$ & $1.300\pm0.185$ &
$-0.383\pm0.179$ & - & $0.001\pm0.0005$ \\
DR3 with $\logg$ & 127 & $0.354\pm0.287$ & $1.162\pm0.044$ & $1.285\pm0.194$
& $-0.049\pm0.398$ & $-0.078\pm0.040$ & - \\ 
DR3 with \snr\ and $\logg$ & 127 & $0.239\pm0.297$ & $1.154\pm0.045$ &
$1.217\pm0.200$ & $-0.006\pm0.398$ & $-0.080\pm0.040$ & $0.001\pm0.0007$ \\ 
\hline
\end{tabular}
\end{sidewaystable}

\begin{figure}
\centering
\includegraphics[width=7cm]{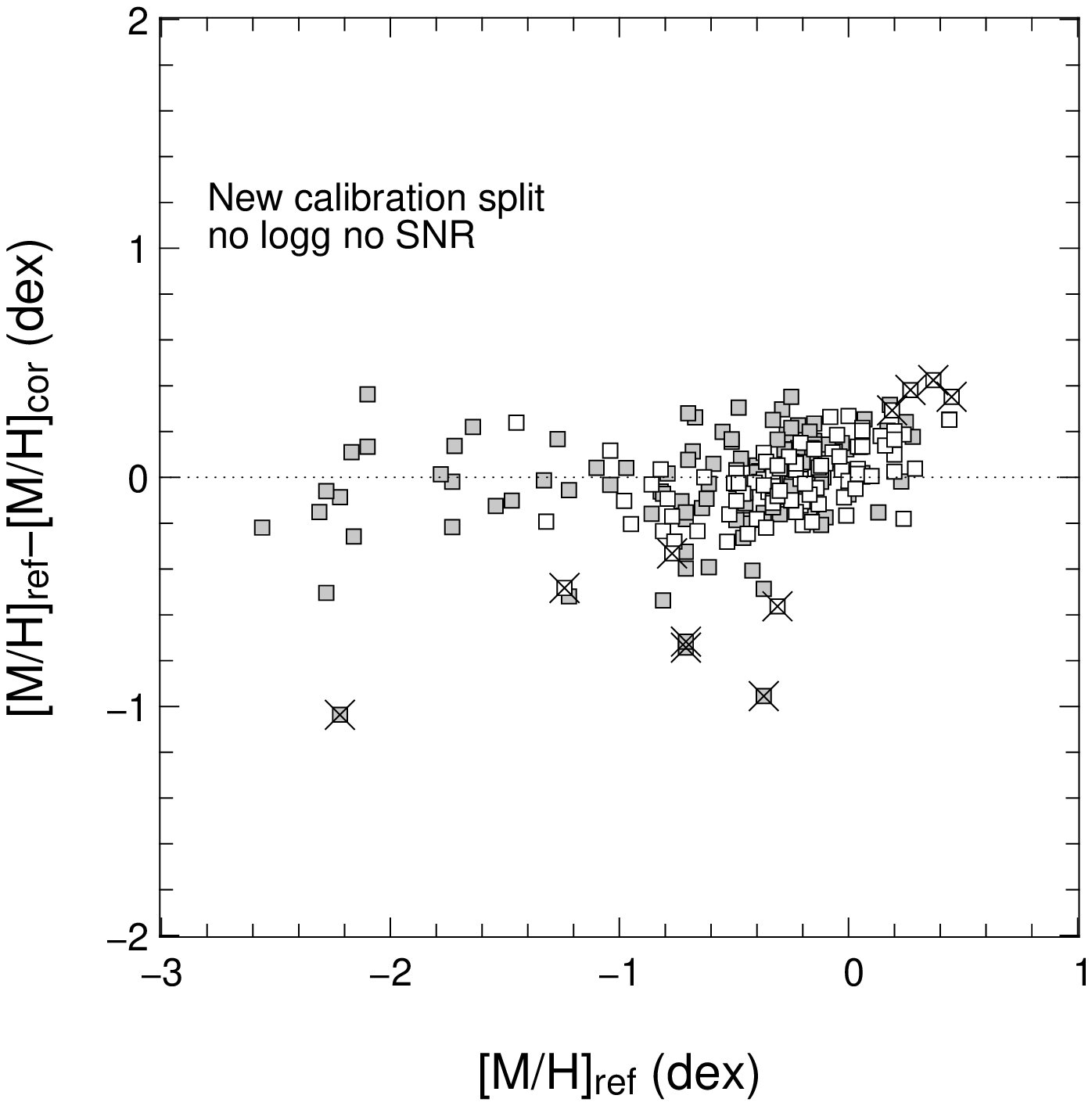}
\includegraphics[width=7cm]{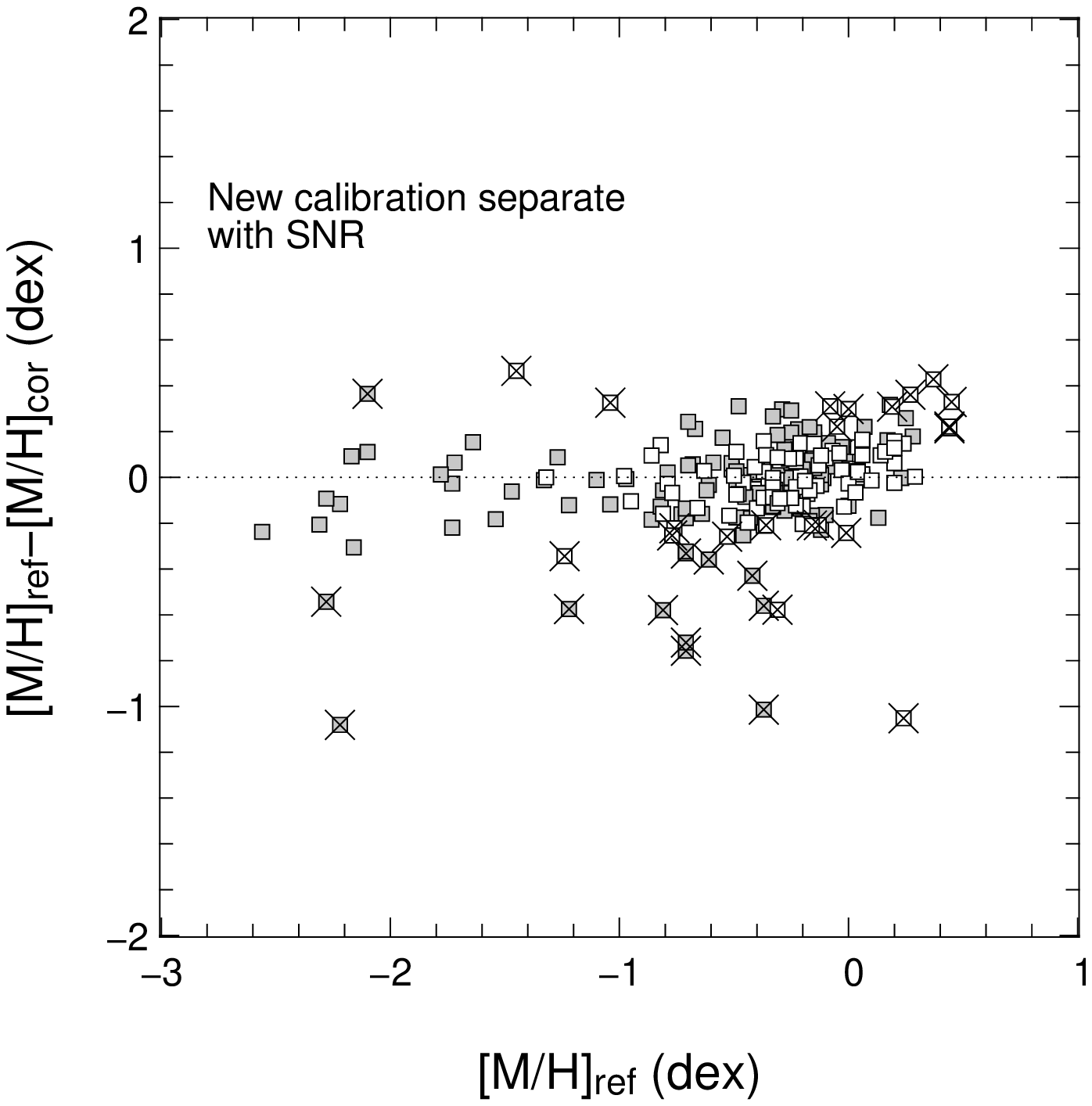}
\includegraphics[width=7cm]{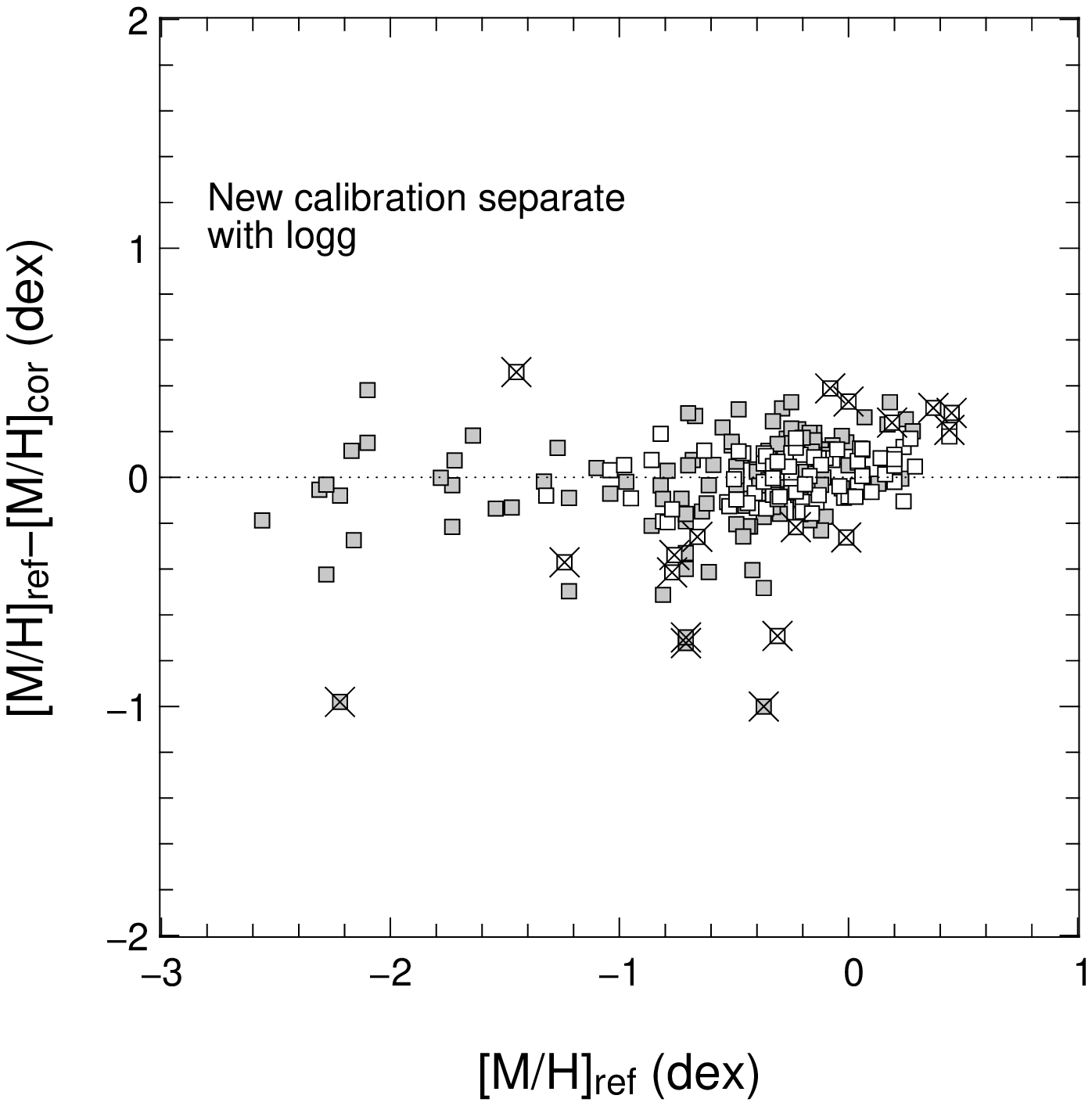}
\includegraphics[width=7cm]{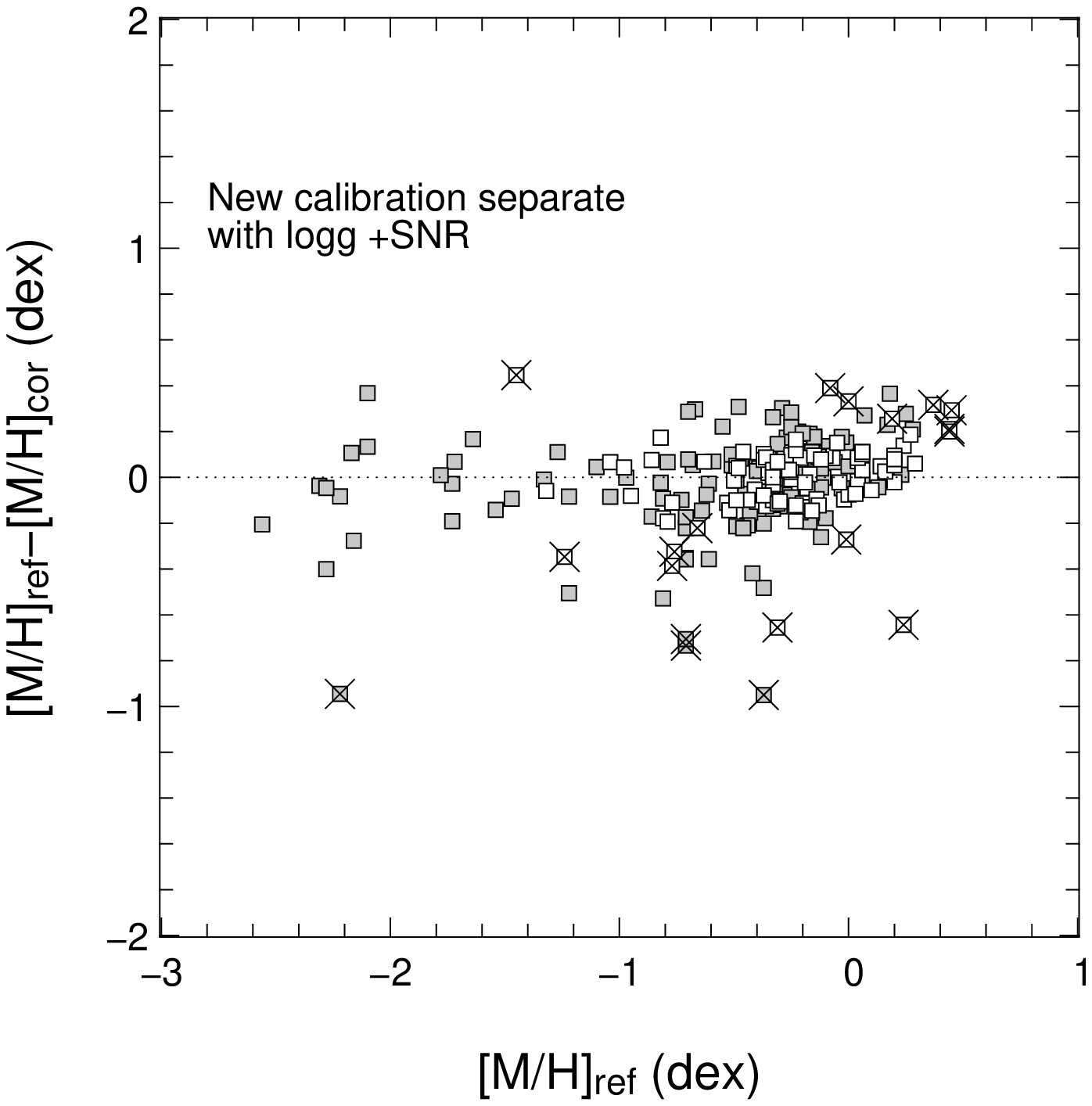}
\caption{Same as Fig.~\ref{f:calib_MH} but using separate calibration
 relations for dwarfs and giants.}
\label{f:calib_MH_split}
\end{figure}

\begin{table}
\centering
\caption{General properties of the different calibration relations presented
 in  Tab.~\ref{t:calib}.  $\Delta  \MH$  is the  mean difference  $\MH_{\rm
   ref}-\MH_{\rm corrected}$  and $\sigma_\MH$ is  the dispersion.  $N_{\rm
   rej}$ is the number of  observations rejected by the iterative procedure
 as  outliers.  For  each calibration  relation, we  also  provide separate
 statistics for dwarfs and  giants obtained using the calibration relations
 derived specifically for each sample.}
\label{t:calib_stat}
\begin{tabular}{c c c c c}
\hline
\hline
Calibration & $\Delta \MH$ & $\sigma_{\MH}$ & $N_{\rm rej}$\\
\hline
No calibration & 0.10 & 0.24 & - \\
DR2 calibration & -0.22 & 0.23 & - \\
DR3 no SNR no $\logg$ & +0.00 & 0.18 & 4 \\
\indent [-] Dwarfs & -0.01 & 0.14 & 7 \\
\indent [-] Giants & 0.00 & 0.18 & 4 \\
DR3 with SNR & 0.00 & 0.16 & 10 \\
\indent [-] Dwarfs & 0.01 & 0.10 & 21\\
\indent [-] Giants & 0.00 & 0.14 & 12\\
DR3 with $\logg$ & 0.00 & 0.18 & 4 \\ 
\indent [-] Dwarfs & 0.00 & 0.10 & 14\\
\indent [-] Giants & 0.00 & 0.18 & 4\\
DR3 with SNR and $\logg$ & 0.00 & 0.17 & 5 \\ 
\indent [-] Dwarfs & 0.00 & 0.10 & 15\\
\indent [-] Giants & 0.00 & 0.18 & 4\\
\hline
\end{tabular}
\end{table}

From Figure~\ref{f:calib_MH}, it is  clear that applying the DR2 calibration
to the DR3  pipeline outputs is not satisfactory and produces  a bias at low
metallicity.  This behavior is expected as the pipeline has been modified to
produce a  better agreement to  the metallicity distribution which,  for DR2
showed a reduced tail at the low metallicity end. As the correlation between
the  parameters  is  significant  (see \S~\ref{s:correl})  and  because  the
calibration  relation is  built upon  the  output parameters  (with a  large
contribution from $\mh$ which is modified compared to the DR2 pipeline), one
therefore  expects the  DR2 calibration  relation not  to hold  for  the DR3
parameters.   Ideally,  the DR3  parameters  would  not  need a  calibration
relation.  However the  raw output of the DR3 pipeline  still suffers from a
small   systematic   effect,  underestimating   the   true  metallicity   by
$\sim$0.1~dex with some systematic dependency on $\teff$.

Applying  the calibration  relations proposed,  the RAVE  metallicties agree
with the  echelle values (see Table~\ref{t:calib_stat}). However,  as can be
seen from  Fig.~\ref{f:calib_MH}, a systematic trend is  observed for dwarfs
at high metallicity, where the difference between RAVE and the echelle value
reaches 0.4~dex for the highest  metallicity stars.  At low metallicity, the
dispersion is significantly reduced and when applying any of the calibration
relations, the two  determinations agree well.  Adding $\logg$  or $\snr$ to
the calibration relation does  not improve the residuals significantly.  For
$\logg$  this is  understood as  it is  the atmospheric  parameter  with the
largest  uncertainty.    Hence  its  dispersion  prevents   it  from  having
significant  weight in  the calibration  relation, even  though we  know the
error  on this  parameter is  strongly correlated  to errors  in  $\MH$ (see
Section~\ref{s:correl}).  For  $\snr$, the situation is less  clear but part
of its low weight in the calibration relation is linked to the fact that, in
order to observe RAVE targets at high resolution, we selected targets in the
bright part of  the catalog to ensure enough $\snr$ in  the spectra to allow
precise measurements of the atmospheric parameters. Hence, the region of the
$\snr$ space where this parameter plays an important role ($\snr<20$) is not
properly sampled,  lowering its weight  on the calibration  relation whereas
above this threshold, no correlation with $\snr$ is observed.

Finally, to  improve on the  situation for the  dwarfs, we split  the sample
between dwarfs and giants (see Fig.~\ref{f:calib_HR} for the criterion used)
and applied the  same procedure to each sub-population.  The result of these
calibration relations is presented in Fig.~\ref{f:calib_MH_split}, the basic
statistics being  reported in Table~\ref{t:calib_stat}  for each calibration
relation.

Using separate calibration relations for dwarfs and giants does help improve
the    dispersion    for    dwarfs,     but,    as    we    can    see    in
Fig.~\ref{f:calib_MH_split},  the calibration relation  is unable  to remove
the bias  at high  $\MH$, the  most discrepant stars  being rejected  by the
fit. Only  a mild improvement is  obtained.  Separating the  dwarfs from the
giants changes the calibrated metallicity  for these stars by only 0.02~dex,
or 0.05~dex if one also uses $\logg$ in the calibration.

\section{Catalog presentation}
\label{s:catalog}

\begin{figure}[hbtp]
\centering
\includegraphics[width=14.7cm,angle=0]{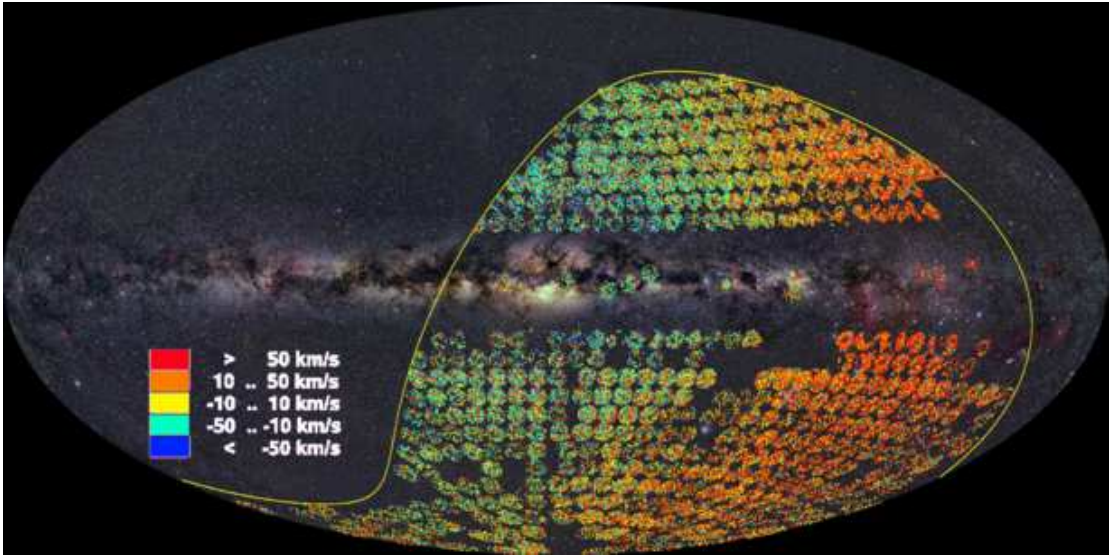}
\caption{Aitoff projection in Galactic coordinates of RAVE $3^{\rm rd}$ Data
 Release fields.  The yellow line  represents the celestial equator and the
 background is from Axel Mellinger's all-sky panorama.}
\label{f:aitoffRVs}
\end{figure}

The  DR3 release  of the  RAVE  catalog contains  $83\,072$ radial  velocity
measurements  for $77\,461$  individual stars.   Atmospheric  parameters are
provided for  $41\,672$ spectra ($39\,833$ stars). These  data were acquired
over $257$ observing nights, spanning  the time interval April $11^{\rm th}$
2003 to  March $12^{\rm th}$  2006, and $976$  fields. The data new  to this
release cover the  time interval March $31^{\rm st}$  2005 to March $12^{\rm
  th}$ 2006 where $32\,477$ new  spectra were collected.  The total coverage
of    the    pilot    survey    is   then    $11\,500$    square    degrees.
Figure~\ref{f:aitoffRVs} plots the  general pattern of (heliocentric) radial
velocities,  where  the dipole  distribution  is  due  to a  combination  of
asymmetric drift and the Solar motion  with respect to the Local Standard of
Rest.

The DR3  release is split  into two catalogs:  Catalog A and Catalog  B. The
first  catalog  contains  the  higher  signal-to-noise  data,  which  yields
reliable  values  for  the  stellar  parameters, and  includes  both  radial
velocities    and    stellar    parameters   (temperature,    gravity    and
metallicity). The second catalog contains the lower signal-to-noise data and
does not include stellar parameters.  The criterion for dividing between the
two  catalogues  was  based on  the  STN  values,  where available,  with  a
threshold  value  of $\mathrm{STN}=20$  between  Catalogs  A  and B.   Table
\ref{t:DR3_catalogs} summarizes  the catalogues, where  we see that  70\% of
the data are in Catalog A.

The DR3 release can be queried  or retrieved from the Vizier database at the
CDS,   as   well   as    from   the   RAVE   collaboration   website   ({\tt
 www.rave-survey.org}).  Table~\ref{t:A1}  describes  its  column  entries,
where the same format is used  for both catalogs for ease-of-use even though
the stellar parameter columns are NULL in Catalog B.  Catalog A contains the
measured stellar  parameters from  the RAVE pipeline  and includes  also the
inferred value of  the $\alpha$-enhancement. As explained in  the DR2 paper,
this is  provided strictly for calibration purposes  only and \textbf{cannot
 be used to infer the $\alpha$-enhancement of individual objects}.

\begin{deluxetable}{llp{3.5cm}l}
\centering
\tablecaption{The two DR3 catalogs
\label{t:DR3_catalogs}}
\tablewidth{0pt}
\tablecolumns{7}
\tablehead{
Catalog& Number of &Selection criteria & Results included\\\
name&Entries& &\
}
\startdata
\hline
Catalog A& $57\,272$& $20 < \mathrm{STN}$ or $20< \mathrm{SNRatio}$&Radial velocities, stellar parameters\\
Catalog B&$25\,800$&$6\le \mathrm{STN}<20$ or $6\le \mathrm{SNRatio}<20$&Radial velocities\\
\enddata
\end{deluxetable}

\begin{figure}[hbtp]
\centering
\includegraphics[width=14.7cm,angle=0]{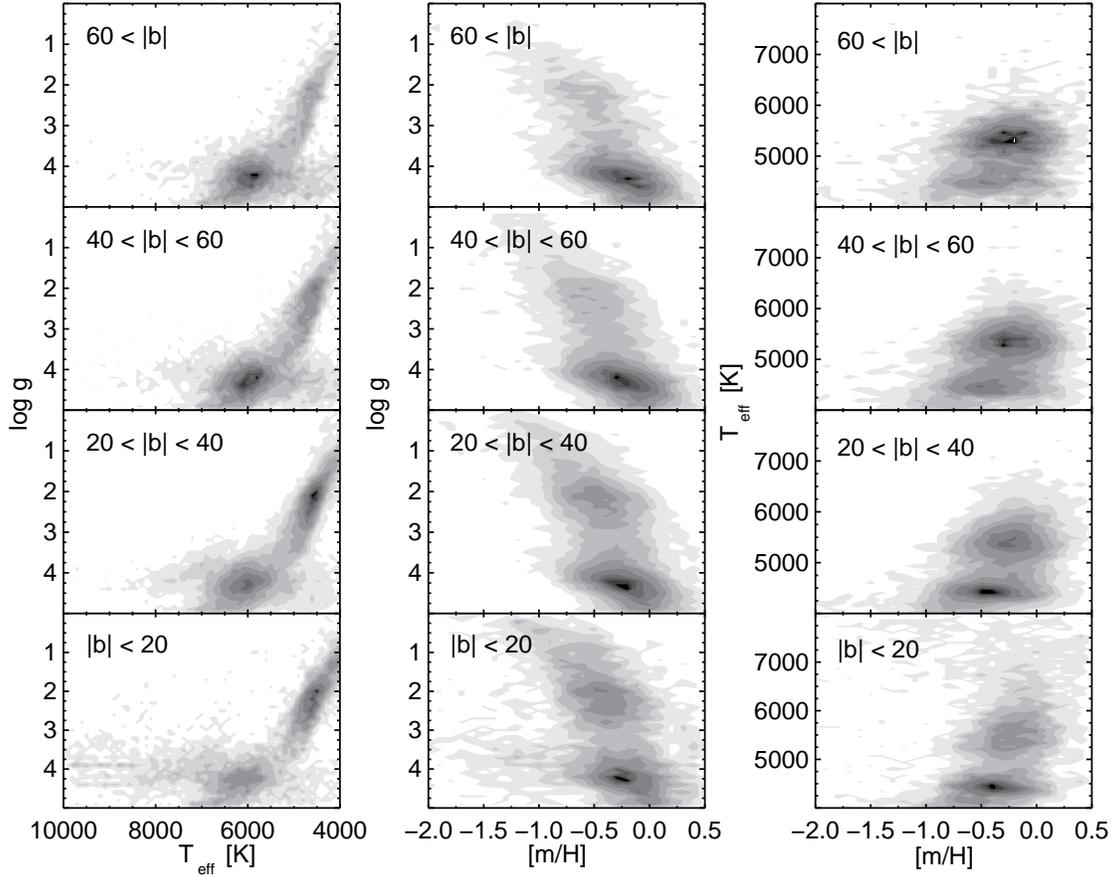}
\caption{The temperature-gravity-metallicity  plane for different  wedges in
 Galactic latitude.  }
\label{f:paramscalib}
\end{figure}

\begin{figure}[hbtp]
\centering
\epsscale{1.0}\includegraphics{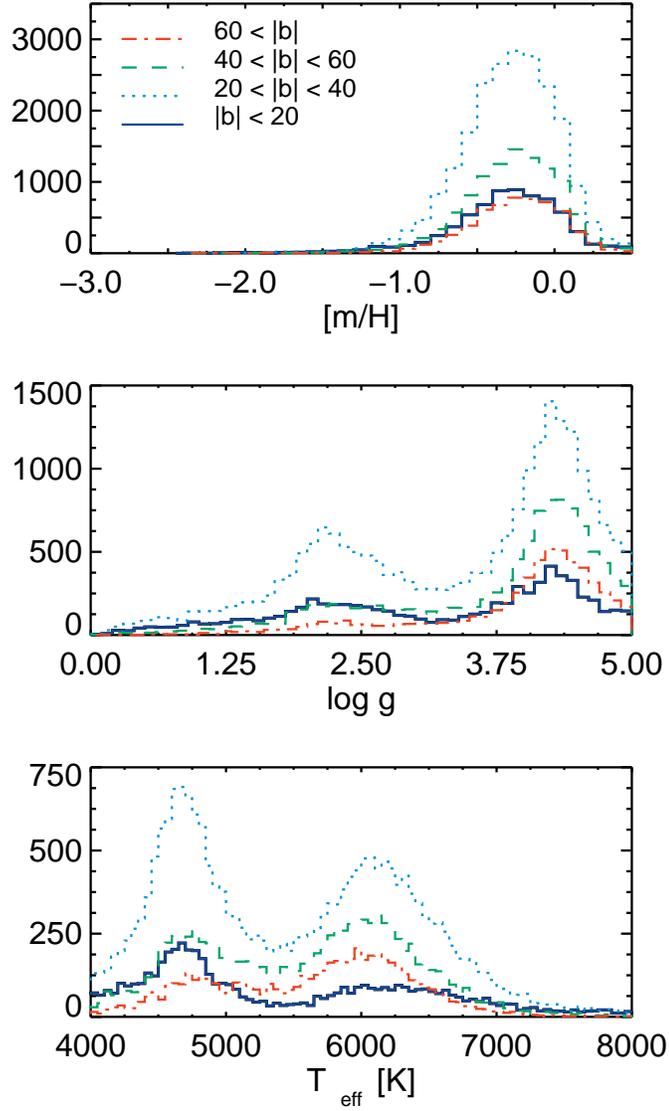}
\caption{Temperature, gravity,  and metallicity histograms  for spectra with
 published stellar parameters.  Histograms for individual Galactic latitude
 bands are plotted separately with the  key given in the top panel. Spectra
 with $|b| \le 20^\circ$ include calibration fields.  }
\label{f:parameterhisto}
\end{figure}

Following Paper~II, in Fig.~\ref{f:paramscalib}  we plot the location of all
spectra on the temperature-gravity-metallicity wedge for different slices in
Galactic latitude. The main-sequence and giant-star groups (particularly the
red-clump  branch) are clearly  visible, with  their relative  frequency and
metallicity  distribution  varying  with  latitude.  For  the  hotter  stars
($\teff>9\,000\,$K) there is significant discretization in $\log g$. This is
caused  by  the  combination  of  a  degeneracy  in  metallicity  for  these
Paschen-line dominated  spectra and  a smaller range  in possible  $\log g$,
which leads to  the penalization algorithm having a  tendency to converge on
the  same solution.  Figure~\ref{f:parameterhisto}  plots histograms  of the
parameters  for different  latitudes.  The  fraction of  main-sequence stars
increases  with the  distance from  the Galactic  plane (see  Paper~2  for a
discussion).  The metallicity distribution  function becomes more metal poor
for the higher-latitude  fields as well.  Also the  shift of the temperature
distribution  towards  higher  temperature  turn-off stars  with  decreasing
Galactic latitude is clearly visible.

\begin{deluxetable}{lcccccc}
\tablecaption{Number and fraction of RAVE database entries with a counterpart 
in the photometric catalogs
\label{t:photometricquality}}
\tablewidth{0pt}
\tablecolumns{7}
\tablehead{
\colhead{Catalog name} & Number of & \%\ of entries &\multicolumn{4}{c}{\%\ with quality flag}\\
                      & entries   & with counterpart& A & B & C & D \\
}
\startdata
\hline
Catalog A&&&&&&\\
2MASS &$ 57\,184$&$ 99.9\%$&$ 99.9\%$&$  0.0\%$&$  0.0\%$&$  0.1\%$\\
DENIS &$ 43\,178$&$ 75.4\%$&$ 73.4\%$&$ 24.2\%$&$  2.3\%$&$  0.2\%$\\
USNO-B &$ 55\,686$&$ 97.2\%$&$ 99.3\%$&$  0.5\%$&$  0.0\%$&$  0.1\%$\\
\hline
\hline
Catalog B&&&&&&\\
2MASS &$ 25\,699$&$ 99.6\%$&$ 99.4\%$&$  0.0\%$&$  0.0\%$&$  0.6\%$\\
DENIS &$ 19\,433$&$ 75.3\%$&$ 74.5\%$&$ 22.6\%$&$  2.2\%$&$  0.7\%$\\
USNO-B &$ 25\,094$&$ 97.3\%$&$ 98.8\%$&$  0.8\%$&$  0.0\%$&$  0.4\%$\\
\enddata
\end{deluxetable}

\begin{deluxetable}{rlrrrr}
\tablecaption{Summary of proper-motion sources and their 
average and 90\%\ errors
\label{t:PMquality}}
\tablewidth{0pt}
\tablecolumns{6}
\tablehead{
SPM & Catalog & Number of & Fraction  & Average & 90\% \\
Flag & Name    & Entries   & of entries& PM error & PM error\\
             &         &           &           & [\masyr]&[\masyr]
}
\startdata
&Catalog A&&&\\
0&No proper motion &$   595$&$  1.0\%$&&\\
1&Tycho-2 &$  3\,517$&$  6.1\%$&$ 3.2$&$ 4.3$\\
2&SSS &$  1\,427$&$  2.5\%$&$24.2$&$30.1$\\
3&PPMX &$ 24\,554$&$ 42.9\%$&$ 3.5$&$ 4.7$\\
4&2MASS + GSC 1.2 &$    30$&$  0.0\%$&$18.8$&$27.3$\\
5&UCAC2 &$ 24\,498$&$ 42.8\%$&$ 4.8$&$ 8.5$\\
6&USNO-B &$  2\,651$&$  4.6\%$&$ 5.7$&$ 8.6$\\
\hline
1-5&All with proper motion &$ 56\,677$&$ 99.0\%$&$ 4.7$&$ 7.4$\\
\hline
\hline
&Catalog B&&&\\
0&No proper motion &$   341$&$  1.3\%$&&\\
1&Tycho-2 &$   300$&$  1.2\%$&$ 3.4$&$ 4.7$\\
2&SSS &$  2\,519$&$  9.8\%$&$25.8$&$34.2$\\
3&PPMX &$  6\,481$&$ 25.1\%$&$ 4.2$&$ 5.7$\\
4&2MASS + GSC 1.2 &$    32$&$  0.1\%$&$22.0$&$26.4$\\
5&UCAC2 &$ 13\,451$&$ 52.1\%$&$ 7.7$&$12.4$\\
6&USNO-B &$  2\,676$&$  10.4\%$&$ 5.3$&$ 8.5$\\
\hline
1-5&All with proper motion &$ 25459$&$ 98.7\%$&$ 8.3$&$13.6$\\
\enddata
\end{deluxetable}

\subsection{Photometry}
\label{photometry}

As in the previous releases, DR3 includes cross-identifications with optical
and near-IR catalogs (USNO-B: $B1$,  $R1$, $B2$, $R2$; DENIS: $I$, $J$, $K$;
2MASS: $J$, $H$, $K$).  The nearest-neighbor criterion was used for matching
and we  provide the distance to the  nearest neighbor and a  quality flag on
the  reliability of the  match.  Table~\ref{t:photometricquality}  shows the
completeness and  flag statistics  for the two  catalogs, where we  see that
Catalog A's coverage  and quality are slightly better  than those of Catalog
B. This is  because Catalog A is dominated by  lower-magnitude objects while
Catalog B contains mainly the  higher-magnitude objects.  For both, however,
nearly  all  stars were  successfully  matched  with  the 2MASS  and  USNO-B
catalogs.  About 3/4 of  the stars lie in the sky area  covered by the DENIS
catalog.

\begin{figure}[hbtp]
\centering
\includegraphics[width=14.7cm,angle=0]{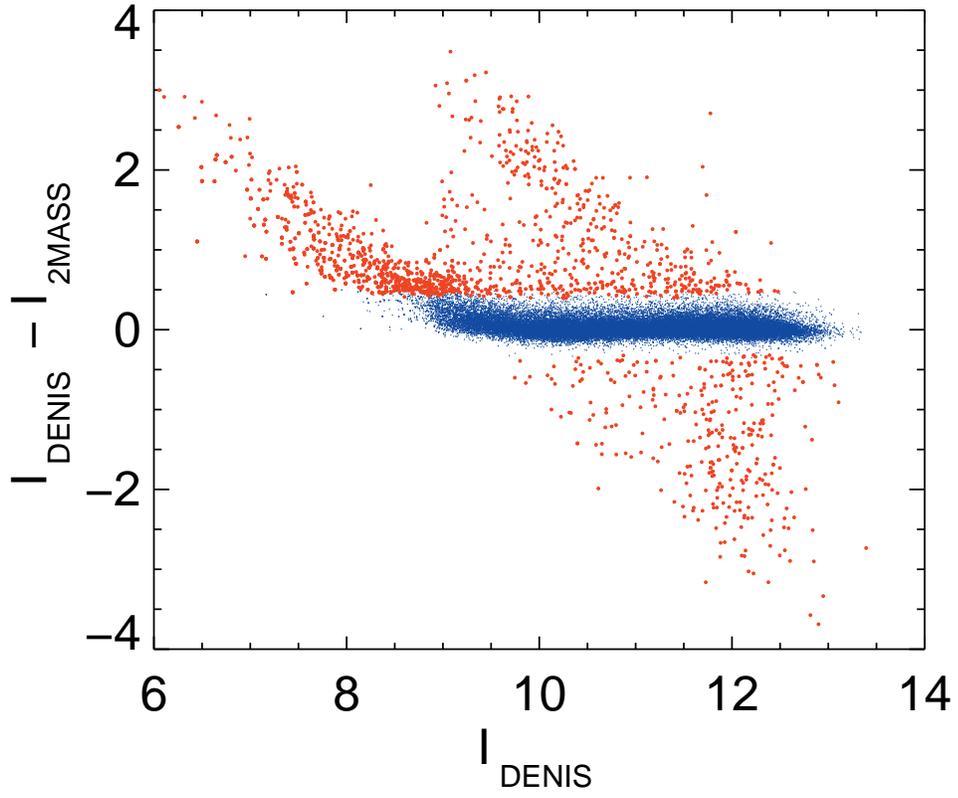}
\caption{Difference between DENIS and  jury-rigged 2MASS $I$ magnitudes as a
 function  of  $I_\mathrm{DENIS}$. The  blue  points  are one  satisfying
 equation~\ref{e:I_condition}, while  the larger red points are ones that
 do not. }
\label{f:DENISI_problem}
\end{figure}

Our wavelength range is best represented  by the $I$ filter. As discussed in
detail in Paper~II, there are some problems with a fraction of the DENIS $I$
magnitudes,   particular  for   $I_\mathrm{DENIS}<10$,  due   to  saturation
effects. Following the  methodology of DR2, we compare  the DENIS magnitudes
against an approximate one calculated from 2MASS J and K (see Equation 24 in
Paper~2).    Fig.~\ref{f:DENISI_problem}   compares   the  DENIS   and   the
``jury-rigged''  2MASS $I-$magnitudes  for  all stars  in  the current  data
release. We see  that the two magnitudes agree for  the majority of objects,
but   a    significant   fraction    have   large   errors:    $10\%$   have
$|(I_\mathrm{DENIS} - I_\mathrm{2MASS})| > 0.2$, with differences of up to 4
magnitudes. It  was proposed in Paper~II  that $I_\mathrm{DENIS}$ magnitudes
should be avoided when the condition
\begin{equation}
-0.2 < (I_\mathrm{DENIS} - J_\mathrm{2MASS}) - 
       (J_\mathrm{2MASS} - K_\mathrm{2MASS}) < 0.6 \,,
       \label{e:I_condition}
\end{equation}
is not  met.  In Figure~\ref{f:DENISI_problem} we  differentiate between the
stars that do and do not satisfy this condition, where we see how it selects
out the problematic $I_\mathrm{DENIS}$ magnitudes.

\subsection{Proper motions}
\label{s:propermotion}

As in  DR2, the proper motions are  sourced from the PPMX,  Tycho-2, SSS and
UCAC2 catalogs. As described in Paper~II, the most accurate available proper
motion is  chosen for  each object.  Table~\ref{t:PMquality}  summarizes for
both  Catalogs  A  and B  the  proper-motion  sources  and the  average  and
$90^\mathrm{th}$ percentile  errors.  The quality  of the proper  motions is
slightly worse  for Catalog B, because  the fainter objects  in this catalog
include a higher proportion of the more distant objects in the survey.

\section{Conclusions}
\label{s:conclusions}

This third data release of the RAVE survey reports $83\,072$ radial-velocity
measurements  for  $77\,461$  stars,  covering more  than  $11\,500$  square
degrees in the  southern hemisphere. The sample is  randomly selected in the
magnitude interval  $9 \leq I \leq  12$. This release  also provides stellar
atmospheric   parameters  for   $41\,672$  spectra   representing  $39\,833$
individual stars.

Since DR2,  we modified the RAVE  processing pipeline to  account better for
defects  in the observed  spectra due  to bad  pixels, fringing,  or locally
inaccurate continuum normalization.  The main driver of the modification was
to improve on  the known limitation of our  estimates of stellar atmospheric
parameters.  Also,  the algorithm to  correct for the zero-point  offset has
been revised, enabling a better control of our radial-velocity accuracy.

The accuracy for  the radial velocities is marginally  improved with the new
pipeline, the distribution  of internal errors in the  radial velocities has
mode  $0.8\kms$ and  median  $1.2\kms$, and  95\%  of the  sample having  an
internal  error better  than  $5\kms$,  which is  the  primary objective  of
RAVE. Comparing our radial velocities  to independent estimates based on 373
measurements from  five data sources we find  no evidence for a  bias in our
radial velocities, our mean radial velocity error being $\sim2\kms$.

A significant effort has been  spent in improving the quality and validation
of our  stellar atmospheric parameters  with respect to the  external errors
and  biases.  The  internal errors  due to  the method  and the  sampling of
synthetic spectra grid remain unchanged  and are presented in Paper II.  The
new calibration  sample consists  of 362 stars  from four  different sources
(either custom  observations or literature)  and cover the full  HR diagram.
Comparing our measured parameters to these reference measurements, we find a
good agreement for $\teff$ and $\logg$  with a mean offset and dispersion of
(-63,  250)~K  for $\teff$  and  (-0.1,  0.43)~dex  for $\logg$,  which  are
consistent  with  DR2. The  $\mh$  distribution  is  improved but  the  true
metallicity  $\MH$ remains  a  combination of  $\mh$,  $\alp$, and  $\teff$.
Taking  $\logg$ or  \snr\  into  the calibration  of  [M/H] only  marginally
improves the situation and the simplest calibration relation is preferred.

This data release  is the last one based on  the pilot-survey input catalog.
Further  releases will  be  based on  an  input catalog  built upon  DENIS-I
magnitudes.  This catalog, supplemented  by the  catalog of  distances, make
this release an unprecedented tool to study the Milky Way.


\acknowledgments {\bf  Acknowledgments} Funding  for RAVE has  been provided
by: the  Anglo-Australian Observatory; the  Leibniz-Institut f\"ur Astrophysik
Potsdam (AIP);  the Australian National University;  the Australian Research
Council;  the   French  National   Research  Agency;  the   German  Research
foundation;  the Istituto  Nazionale  di Astrofisica  at  Padova; The  Johns
Hopkins   University;   the  National   Science   Foundation   of  the   USA
(AST-0908326);  the  W.M. Keck  foundation;  the  Macquarie University;  the
Netherlands  Research  School  for   Astronomy;  the  Natural  Sciences  and
Engineering Research  Council of Canada; the Slovenian  Research Agency; the
Swiss  National Science  Foundation;  the Science  \& Technology  Facilities
Council of the UK; Opticon;  Strasbourg Observatory; and the Universities of
Groningen,  Heidelberg   and  Sydney.   The   RAVE  web  site  is   at  {\tt
 http://www.rave-survey.org}.  The  European Research Council  has provided
financial support through ERC-StG 240271(Galactica).

\newpage

\section*{Appendix A}

Table~\ref{t:A1} describes the contents  of individual columns of the Third
Data   Release    catalog.   The    catalog   is   accessible    online   at
{\tt www.rave-survey.org} and  via the Strasbourg astronomical  Data Center (CDS)
services.

\begin{deluxetable}{cclcll}
\tabletypesize{\footnotesize}
\tablecaption{Catalog description \label{t:A1}}
\tablehead{
Column & Character & Format & Units & Symbol & Description\\ 
number & range     &        &      &        &     }
\startdata
1&  1- 16&A16   &---   & Name            &Target designation\\
2& 18- 34&A16   &---   & RAVEID          &RAVE target designation\\
3& 36- 48&F12.8 &deg   & RAdeg           &Right ascension (J2000.0)\\
4& 50- 62&F12.8 &deg   & DEdeg           &Declination (J2000.0)\\
5& 64- 73&F9.5  &deg   & GLON            &Galactic longitude\\
6& 75- 84&F9.5  &deg   & GLAT            &Galactic latitude\\
7& 86- 93&F7.1  &$\!\kms$  & HRV             &Heliocentric radial velocity\\
8& 95-101&F6.1  &$\!\kms$  & eHRV            &HRV error\\
9&103-109&F6.1  &mas yr$^{-1}$& pmRA            &proper motion RA\\
10&111-117&F6.1  &mas yr$^{-1}$& epmRA           &error proper motion RA\\
11&119-125&F6.1  &mas yr$^{-1}$& pmDE            &proper motion DE\\
12&127-133&F6.1  &mas yr$^{-1}$& epmDE           &error proper motion DE\\
13&135-136&I1    &---   & Spm             &source of proper motion (1)\\
14&138-143&F5.2  &mag   & Imag            &Input catalog I magnitude\\
15&145-153&A8    &---   & Obsdate         &Date of observation yyyymmdd\\
16&155-165&A10   &---   & FieldName       &Name of RAVE field\\
17&167-168&I1    &---   & PlateNumber     &Plate number used\\
18&170-173&I3    &---   & FiberNumber     &Fiber number [1,150]\\
19&175-180&I5    &K     & Teff            &Effective Temperature\\
20&182-186&F4.2  &dex   & logg            &Gravity\\
21&188-193&F5.2  &dex   & Met             &[m/H]\\
22&195-199&F4.2  &dex   & alpha           &[Alpha/Fe]\\
23&201-209&F8.1  &---   & CHISQ           &chi square\\
24&211-216&F5.1  &---   & S2N             &DR2 signal to noise S2N\\
25&218-223&F5.1  &---   & STN             &Pre-flux calibration signal to noise STN\\
26&225-230&F5.1  &---   & CorrelationCoeff&Tonry-Davis $R$ correlation coefficient\\
27&232-236&F4.2  &---   & PeakHeight      &Height of correlation peak\\
28&238-244&F6.1  &$\!\kms$  & PeakWidth       &Width of correlation peak\\
29&246-252&F6.1  &$\!\kms$  & CorrectionRV    &Zero point correction applied\\
30&254-260&F6.1  &$\!\kms$  & SkyRV           &Measured HRV of sky\\
31&262-268&F6.1  &$\!\kms$  & SkyeRV          &error HRV of sky\\
32&270-275&F5.1  &---   & SkyCorrelation  &Sky Tonry-Davis correl. coefficient\\
33&277-282&F5.1  &---   & SNRatio         &Spectra signal to noise ratio\\
34&284-290&F6.3  &mag   & BT              &Tycho-2 BT magnitude\\
35&292-298&F6.3  &mag   & eBT             &error BT\\
36&300-306&F6.3  &mag   & VT              &Tycho-2 VT magnitude\\
37&308-314&F6.3  &mag   & eVT             &error VT\\
38&316-328&A12   &---   & USNOID          &USNO-B designation\\
39&330-336&F6.3  &mas   & DisUSNO         &Distance to USNO-B source\\
40&338-343&F5.2  &mag   & B1              &USNO-B B1 magnitude\\
41&345-350&F5.2  &mag   & R1              &USNO-B R1 magnitude\\
42&352-357&F5.2  &mag   & B2              &USNO-B B2 magnitude\\
43&359-364&F5.2  &mag   & R2              &USNO-B R2 magnitude\\
44&366-371&F5.2  &mag   & IUSNO           &USNO-B I magnitude\\
45&373-374&A1    &---   & XidQualityUSNO  &Cross-identification flag (2)\\
46&376-392&A16   &---   & DENISID         &DENIS designation\\
47&394-400&F6.3  &mas   & DisDENIS        &Distance to DENIS source\\
48&402-408&F6.3  &mag   & IDENIS          &DENIS I magnitude\\
49&410-414&F4.2  &mag   & eIDENIS         &error DENIS I magnitude\\
50&416-422&F6.3  &mag   & JDENIS          &DENIS J magnitude\\
51&424-428&F4.2  &mag   & eJDENIS         &error DENIS J magnitude\\
52&430-436&F6.3  &mag   & KDENIS          &DENIS K magnitude\\
53&438-442&F4.2  &mag   & eKDENIS         &error DENIS K magnitude\\
54&444-445&A1    &---   & XidQualityDENIS &Cross-identification flag (2)\\
55&447-463&A16   &---   & TWOMASSID       &2MASS designation\\
56&465-471&F6.3  &mas   & Dis2MASS        &Distance to 2MASS source\\
57&473-479&F6.3  &mag   & J2MASS          &2MASS J magnitude\\
58&481-485&F4.2  &mag   & eJ2MASS         &error 2MASS J magnitude\\
59&487-493&F6.3  &mag   & H2MASS          &2MASS H magnitude\\
60&495-499&F4.2  &mag   & eH2MASS         &error 2MASS H magnitude\\
61&501-507&F6.3  &mag   & K2MASS          &2MASS K magnitude\\
62&509-513&F4.2  &mag   & eK2MASS         &error 2MASS K magnitude\\
63&515-518&A3    &---   & TWOMASSphotFLAG &2MASS photometric flag\\
64&520-521&A1    &---   & XidQuality2MASS &Cross-identification flag (2)\\
65&523-526&A3    &---   & ZeroPointFLAG   &Zero point correction flag (3)\\
66&528-536&A8    &---   & SpectraFLAG     &Spectra quality flag (4)\\
67&538-542&F4.2  &---   & MaskFLAG        &MASK flag (5)\\
\enddata
\tablecomments{
(1): Flag value between 0 and 4:
   0- no proper motion,
   1- Tycho-2 proper motion,
   2- Supercosmos Sky Survey proper motion,
   3- PPMX proper motion,
   4- GSC1.2 x 2MASS proper motion,
   5- UCAC-2 proper motions. 
\\
(2): Flag value is A,B,C,D or X:
   A- good association,
   B- 2 solutions within 1 arcsec,
   C- more than two solutions within 1 arcsec,
   D- nearest neighbor more than 2 arcsec away,
   X- no possible counterpart found. 
\\
(3): Flag value of the form FGSH, F being for the entire plate, G for the 50
fibers group to  which the fiber belongs. S flags  the zero point correction
used: C for cubic  and S for a constant shift.If H is set  to * the fiber is
close to a 15 fiber gap.
   For F and G the values can be A, B, C, D or E:
   A- dispersion around correction lower than $1\kms$,
   B- dispersion between 1 and $2 \kms$,
   C- dispersion between 2 and $3 \kms$,
   D- dispersion larger than $3 \kms$,
   E- less than 15 fibers available for the fit. 
\\    
(4): Flag identifying possible problem in the spectra (values can be combined): a- asymmetric Ca lines,
c- cosmic ray pollution, 
e- emission line spectra, 
n- noise dominated spectra, 
l- no lines visible, 
w- weak lines, 
g- strong ghost, 
t- bad template fit, 
s- strong residual sky emission, 
cc- bad continuum, 
r- red part of the spectra shows problem, 
b- blue part of the spectra shows problem, 
p- possible binary/doubled lined, 
x- peculiar object.
\\
(5): Flag identifying the fraction of the spectrum unaffected by continuum
problems from the MASK program. Spectrum with MaskFLAG lower than 0.70 must
be used with caution. \\
}
\end{deluxetable}

\end{document}